\RequirePackage{fix-cm}
\documentclass[smallextended,natbib]{svjour3}    
\smartqed  
\usepackage{graphicx}
\usepackage{color}
\usepackage{hyperref}
\begin{document}

\title{The distribution of dark matter in galaxies}

\author{Paolo Salucci}

\institute{P. Salucci
\at 
SISSA, 34100 Trieste, Italy\\
\email{salucci@sissa.it}
}

\date{Received: date / Accepted: date}

\maketitle

\begin{abstract}
The distribution of the non-luminous matter in galaxies of different luminosity and Hubble type is much more than a proof  of the existence of dark particles governing the structures of the Universe. Here, we will review the complex but well-ordered scenario
of the properties of the dark halos also in relation with those of the baryonic components they host. Moreover, we will
present a number of tight and unexpected correlations between selected properties of the dark and the luminous matter.
Such entanglement evolves across the varying properties of the luminous component and it
seems to unequivocally lead to a dark particle able to interact with the Standard Model particles over cosmological times.
This
review will also focus on whether we need a paradigm shift, from pure collisionless dark particles emerging from ``first
principles'', to particles 
that we can discover only by looking to how they have designed the structure of the galaxies. 
\end{abstract}

\newpage
\setcounter{tocdepth}{3}
\tableofcontents

\section{Introduction}

\begin{figure}[htb]
\includegraphics[width=\textwidth]{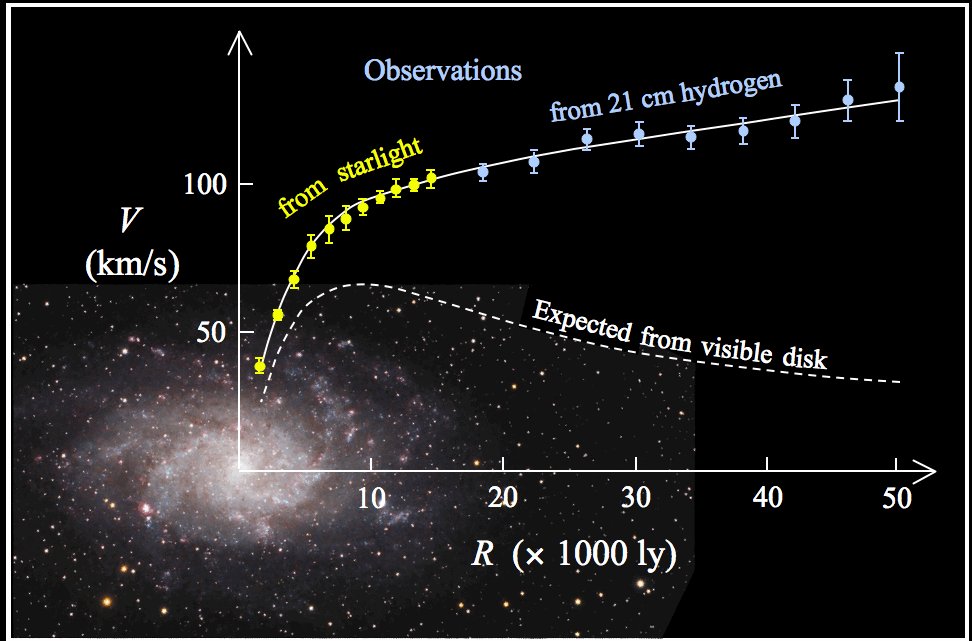}
\caption{The image of M33 and the corresponding rotation curve (\citealt{cs}). What exactly does this large anomaly of the
gravitational field indicate? The presence of {\it i)} a (new) non-luminous
massive component around the stellar disk or {\it ii)} new physics of a (new) dark constituent?}
\label{fig:1m33}
\end{figure}

The idea of the presence of large amounts of invisible matter in and around spirals, distributed 
differently from the stellar and gaseous disks, turned up in the 1970s (\citealt{R78, FG, rel80, b1}, see also \citealt{BH}). 
There were, in fact, published optical and 21-cm rotation
curves (RCs) behaving in a
strongly anomalous way. These curves were incompatible with the Keplerian fall-off we would predict from  their 
outer distribution of luminous matter (see Fig.~\ref{fig:1m33}). 

From there, this dark component has started to take a role always more
important in cosmology,
astrophysics and elementary particles physics. On the other hand, the nature and the cosmological history of such dark component 
has always become more mysterious and difficult to be derived from paradigms and first principles. We must remark that a dark
massive 
component in the mass budget of the Universe is necessary to explain: the redshift dependence of the 
expansion of its scale factor, the relative heights of the peaks in the CMB cosmic fluctuations, the
bottom-up growth
of the cosmological structures to their nonlinear phases, the large scale distribution of galaxies and the internal 
mass distribution of the biggest structures of the Universe. These theoretical issues and observational evidences
(that will not be treated in this review) add phenomenal support to the paradigm of a massive dark particle, which, {\it a fortiori},
must lay beyond the zoo of the Standard Model of the elementary particles. This support is not able, however, to determine the
kind, the nature and the mass of such a particle. 
 
There is no doubt that dark matter connects, as no other issue, the different fields of study of cosmology, particle physics
and astrophysics. In the current $\Lambda$ cold dark matter ($\Lambda$ CDM) paradigm, the DM is non-relativistic since its decoupling time and can
be described by a collisionless fluid, 
whose particles interact only gravitationally and very weakly
with the Standard Model particles (\citealt{jel96, BerB}).

In the past 30 years, in the preferred $\Lambda$CDM scenario, the complementary approach of detecting messengers of the
dark particle and creating it
at colliders has brought over an extraordinary theoretical and experimental effort that, however, has not reached a
positive result. 
Moreover, on the scales  $< 50\ {\rm kpc}$, where great part of the DM resides, there is a growing
evidence of increasingly quizzical properties of
the latter are, so that, a complex and surprising scenario, of very difficult understanding,
is emerging.
 
\subsection{Scope of the review}

 The distribution of matter in galaxies does not seem to be the final act of a simple and well understood history which has 
developed itself over the whole age of the Universe. It seems, instead, to lead to 
 one of the two following possibilities: 1) the dark particle is a WIMP, however, baryons enter, 
heavily and in a very tuned way, into the process of galaxy formation, modifying, rather than following, the original DM 
distribution  2) the dark particle is something else, likely interacting with SM particle(s) and very likely lying
beyond our current ideas of physics. 

In both cases, investigating deeply the distribution of dark matter in galaxies is necessary and worthwhile. In the first
case, the peculiar imprint that baryons leave on the original distribution of the dark particles can serve us as an
indirect, but telling, investigation of the latter.
In the second case, with no guidance from first principles, a most complete investigation of the dark matter distribution in galaxies is
essential to grasp its nature. 

In any case, it is now possible to investigate such issue in galaxies of various morphological types and luminosities. We
are sure that this will help us to shed light on the
unknown physics underlying the dark matter mystery. 

There are no doubts that the topic of this review is related and, in some case, even entangled with other main topics of
cosmology and astroparticle physics. However, this work will be kept focused on the
properties of dark matter where it mostly resides. Then, a number of issues, yet linked to the dark matter in galaxies, will
not be dealt here or will be dealt in a very schematic way.
 This, both because we sense that looking for the ``naked
truth'' of the galactic dark matter is the best way to approach the related mystery and because there are recent
excellent reviews, suitable to complete the whole picture of dark matter in galaxies. 
These include: ``The Standard Cosmological Model: Achievements and Issues'' (\citealt{e18}), standard
and exotic dark-matter candidate particles and their related searches and productions (\citealt{RST, L17}), the 
$\Lambda$CDM scenario and its observational challenges (\citealt{NO, SD, BB,T18}), 
``The Connection Between Galaxies and Their Dark Matter Halos'' (\citealt{WT}), ``Status of dark matter in the universe'' (\citealt{Free}), ``Galaxy
Disks'' (\citealt{vkf}) and ``Chemical Evolution of Galaxies''
(\citealt{MATT2}). In addition, in the next sections, when needed, I will indicate the readers the papers that extend and deepen
the content
here presented. 

Let us stress that, although in this review one can find several observational evidences that can be played in
disfavor of 
the $\Lambda$CDM scenario, this review is not meant to be a collection of observational challenges to such scenario
and several issues at such regard, e.g., \cite{mel}, will not be considered here.

It is worth pointing out that here we do not consider the theories alternative to the DM, that is, theories that dispose of
the dark particle. The main reasons are 1) space: an honest account of them will require to add 
many more pages to this longish review and 2) my personal bias: no success in explaining the observations at 
galactic scale can compensate the intrinsic inability that these theories have in conceiving the galaxy formation process and interpreting the {\it corpus} of the cosmological observations.

\subsection{The presence of dark matter in galaxies}

Let us introduce the ``phenomenon'' of dark matter in galaxies as it follows: be $M(r)$ the mass distribution of the gravitating
matter and $M_L(r)$ that of the sum of all the luminous components. Let us notice that the radial logarithmic derivative  of both mass profiles can be obtained from
observations. Then, we realize
that in spirals, for $r> r_T$, they do not match, in detail: 
$d\,\log M/d\,\log r >d\,\log M_L/d\,\log r$ (see Fig.~\ref{fig:1m33} where the transition radius $r_T\simeq 4\ {\rm kpc}$). Then, we
introduce a non luminous component whose mass
profile
$M_H(r)$ accounts
for the disagreement:

\begin{equation}
\frac{d\,\log M(r)}{d\,\log r} = \frac{M_L(r)}{M(r)} \frac{d\,\log M_L}{d\,\log r} +
\frac{M_H(r)}{M(r)} \frac{d\,\log M_H}{d\,\log r} \,.
\label{eq:1}
\end{equation}

The above immediately shows that the phenomenon of the mass discrepancy in galaxies emerges from the discordance between the value
of the radial
logarithmic derivative of the total mass profile and that of the luminous mass profile.
We need to insert in the r.h.s.\ of Eq.~(\ref{eq:1}) an additional (dark) term. This also implies that the DM phenomenon
emerges observationally and can be investigated only if
we
are able to accurately measure the distribution of luminous and gravitating matter.  In fact, the rotation curves $V(r)
\propto (M(r)/r)^{1/2}$
have a property which is rarely found in astrophysics. We start with the fact that a good determination of the logarithmic
derivative $\nabla \equiv d\,\log V/ d\,\log r$ is essential to successfully mass model a galaxy. Now, the analysis of $N$
individual 
RCs with the same value of $\nabla =\nabla_0$ and with a large uncertainty, e.g., $\delta \nabla_0= 0.2$ gives much
\emph{less}
information on the mass distribution than one single
RC with $\delta\nabla_0= \pm 0.2/\sqrt N$.  In short a RC with large uncertainties gives no information on the underlying galaxy mass distribution. 

There is, however, a way to exploit the
information carried by the low quality RCs, namely, to properly {\it stack} them in coadded curves, killing so large part of their random uncertainties. 

The luminous components of galaxies show a striking variety in morphology and in the values of their structural quantities.
The range in magnitudes and central surface brightness are 15 mag and 16 mag/arcsec$^2$. The distribution of the luminous
matter in spirals 
is given by a stellar disk\,+\,a stellar central bulge and an extended HI disk and in ellipticals and dSphs by a
stellar spheroid.

How will the variety of the properties of the luminous matter contrast with the organized 
uniformity of the dark matter? 
The phenomenological scenario of dark matter in galaxies that we discuss in this review has to be considered as a
privileged way to
understand what dark matter halos are made of and to approach the involved (new) laws or processes of Nature. 

\cite{fre}, in its Appendix~A, firstly drew the attention of the astrophysical community to a discrepancy between the
kinematics and the photometry of 
the 
spiral galaxy NGC 300, that implied the presence of large amounts of non-luminous matter. Then, during the 1970s the contribution of Morton Roberts to the cause of DM  in galaxies has been crucial (\citealt{BB}).  A 
next topical moment was when Vera Rubin published 20 optical RCs, extended out to 2/3 of their optical radii $R_{\rm opt}$, 
that were still rising or flattish at the last measured point (\citealt{rel80}). Decisive kinematics was obtained 
 by means of several 21-cm rotation curves extended out to 2-3 optical radii (\citealt{b1,b2}). Moreover, we have to
mention the \cite{FG} review that played a very important role to spread the idea of a dark halo
component
in galaxies.\footnote{Only much later the universality of the DM phenomenon in spirals did emerge (\citealt{PSS}).}

In this brief historical account of the discovery of dark matter in galaxies, one point should still be made. 
Until to few years ago, the nature of dark matter was not meant to be determined by the properties of
the galaxy gravitational field, but to come from first principles verified by large scales observations. In this
review, instead, we will follow also a reverse-engineering approach: the unknown nature of the DM  is searched
within the (complex) observational properties of the dark halos in galaxies.
 
 \section{The invisible character, dark particles and co.}
 
It is worth starting this review with a brief account of the dark matter candidate particles presently in the ballpark; one
has to keep on mind however, that there are likely to risk not to be ``the'' DM particle. 
 
\subsection{Collisionless and cold dark particles}

\begin{figure}[htb]
\centering
\includegraphics[width=0.6\textwidth]{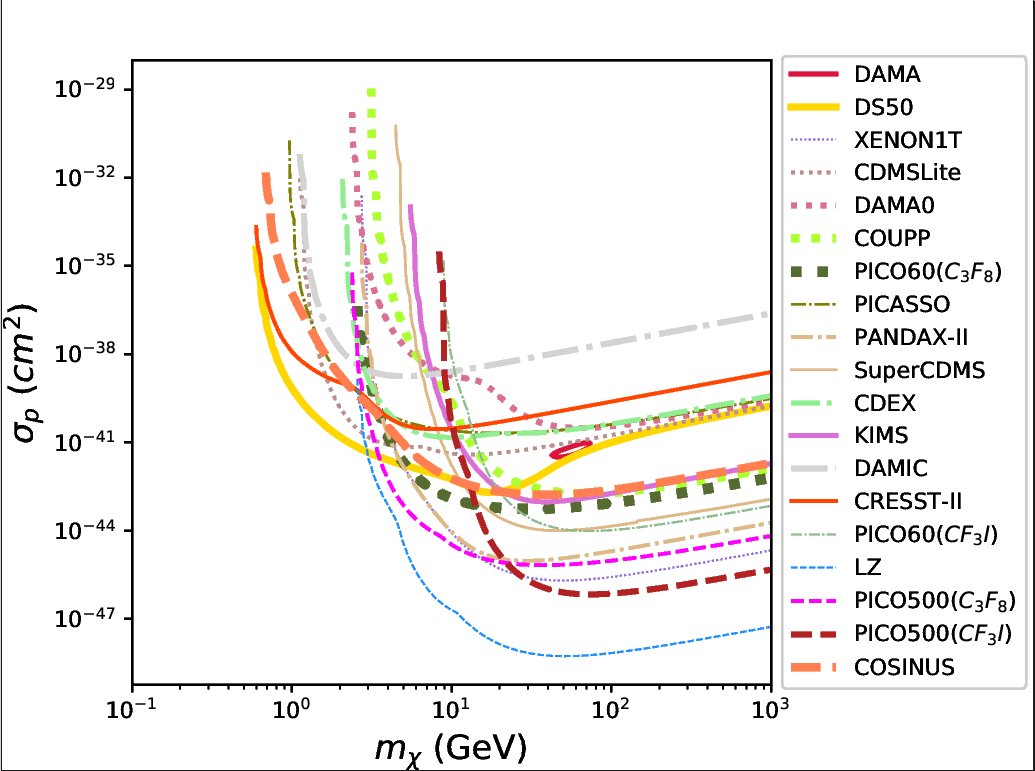}
\caption{{\bf top} Current 90\% C.L. exclusion plots to the effective WIMP--proton cross section, see \cite{Kanel}.}
\label{fig:aa20}
\end{figure}

\begin{figure}[htb]
\centering
 \includegraphics[width=0.7\textwidth]{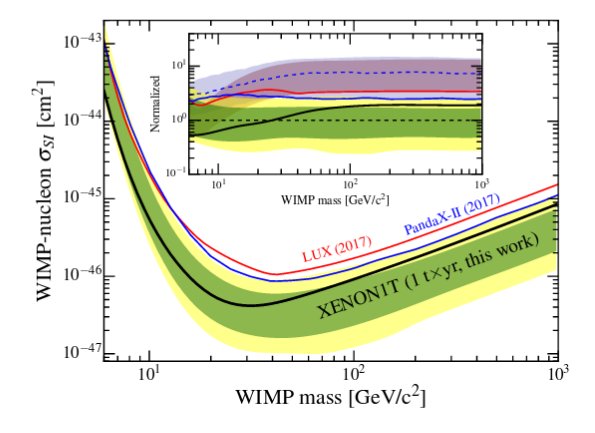}
\caption{Current 90\% C.L.\ exclusion plots to the effective WIMP--nucleon cross section.
Image reproduced with permission from  \cite{apr}, copyright by APS.}
\label{fig:apr}
\end{figure}

Let us start by recalling the motivations that have led to about 30 years of fascination with the
Weakly-Interacting Massive
Particles (WIMPs) and especially with the lightest supersymmetric particle (\citealt{ST}, see also \citealt{KT}). At high
temperatures, ($T \gg m_{\rm 
WIMP}$), WIMPs are thermally created and destroyed. As the temperature
of the Universe decreases due to its expansion, the density is exponentially suppressed ($\propto \exp[{-m_{\rm WIMP}/T}]$) 
and  becomes no longer high enough to pair-create them. When the WIMP mean free path is comparable to the Hubble distance,
the particles also cease to annihilate, leave the thermal equilibrium state and ``freeze-out''. At this point, the co-moving
density remains constant. The temperature for which the freeze-out occurs is about 5\% of the WIMP mass. Therefore, the
(relic) density becomes
constant when the particles are non-relativistic. The value of the relic density $\Omega_{\rm WIMP}$ depends only on the
total annihilation cross-section $\sigma_A$ and the particles' relative velocity $|\vec{v}|$:
\begin{equation}
\Omega_{\rm WIMP}\simeq \frac{6 \cdot 10^{-27}\, {\rm cm^3 s^{-1}}}{<\sigma_A |\vec{v}|>} ,
\label{eq:relicdensity}
\end{equation}
The scale of weak interaction strength ($\sim \alpha^2/m_{\rm WIMP}^2$) implies that $<\sigma_A |\vec{v}|>10^{-25}\, {\rm cm^3 s^{-1}}$, 
where $\sigma_A$ is the cross section and the WIMP mass is taken to be 100~GeV. The resulting relic density for such a
particle
would be within a factor 3 of the 
measured value of the dark matter density $\Omega_m$ (e.g., \citealt{P16}). This remarkable coincidence is referred to
as the ``WIMP miracle.'' This particle, today, should interact with ordinary matter only through weak
interaction, in addition to the gravitational one. The former should occur via the exchange of a scalar particle, or a
vector boson
interaction. These interactions together with the particle-particle annihilations ongoing in the densest region of the
Universe, would make the particle detectable.

It is known that this scenario reproduces a wealth of 
cosmological observations, particularly on scales $> 10$~Mpc. On the other hand, WIMPs have so far escaped detection (see
Figs.~\ref{fig:aa20}--\ref{fig:apr}) and, furthermore, there is a number of small-scale issues that put in question their being the dark particle in galaxies.

\subsection{An unexpected new candidate for cold dark particles}

There might be a connection between the dark matter in galaxies, in particular the cold DM and the gravitational waves
produced by the merging of stellar-mass black holes and possibly detectable by LIGO-Virgo experiments. This is due to the intriguing
possibility
that DM consists of black holes created in the very early Universe. In this case, the detection of primordial black
hole binaries could provide an unambiguous observational window to pin down the nature of dark matter (\citealt{grn}). These
objects are also detectable as effect of their continuous merging since recombination. This violent process can have
generated a stochastic background of gravitational waves that could be detected by LISA and PTA (see also \cite{gb}). 

It is known that massive primordial black holes form at rest with respect to the flow of the expanding 
Universe and then with zero spin. Moreover, they have negligible cross-section with the ordinary matter and constitute
a
right candidate for the $\Lambda$CDM scenario (see, however, \citealt{kl17}). Of course, just substituting
WIMPs with primordial BHs does not immediately relieve the severe tension with the observations at galactic scales that these 
 particles have. It is, however open the question whether these primordial BHs could have some sort of interaction
with baryons which is instead forbidden to WIMPs. 
 
\subsection{Self-interacting DM particles}

Self-interacting dark matter (SIDM) particles were proposed by \cite{spst} (see also \citealt{Bo,bd}) to solve the
core-cusp and missing satellites problems (see also \citealt{TY,Bz}). DM particles scatter elastically with each other
through 2-2 interactions and, as low-entropy particles,
are heated by elastic collisions within the dense inner halo and leave the region: the central and nearby densities are then
reduced, turning an
original cusp into a core. The collision
rate is:
\begin{equation}
R_{\rm scatt} = \sigma v_{\rm rel} \, \rho_{\rm DM} /m \approx 0.1\ {\rm Gyr}^{-1} \times \Big(\frac{\rho_{\rm DM}}{0.1 \,
 M_\odot /{\rm pc}^3} \Big) \Big(\frac{v_{\rm rel}}{50\ {\rm km/s}} \Big) \Big(\frac{\sigma /m}{1\ {\rm cm^2/g}} \Big),
\end{equation}
where $m$ is the DM particle mass, $\sigma, v_{\rm rel}$ are the cross section and relative velocity for
scattering. Within the central region of a typical dwarf galaxy we have: $\rho_{\rm DM} \sim 0.1 \, M_\odot/{ pc}^3$ and
$v_{\rm rel} \sim 50\ {\rm km/s}$. Therefore, the cross section per unit mass ($\sigma/m$) must be at least:
\begin{equation}
\sigma/m \sim 1\ {\rm cm^2/g} \approx 2 \times 10^{-24}\ {\rm cm^2/GeV}
\end{equation}
to have an effect; this corresponds to about one scattering per particle over 10 Gyr galactic timescales. With the above
value of $\sigma/m$, $R_{\rm scatt}$ is negligible during the early Universe when structures form. SIDM, therefore retains the
success of large-scale structure formation of the $\Lambda$CDM scenario, and affects the dark structures on small scales
only once they are already virialized. 

The self-interacting dark matter is then a cusp-core density
profile transformer (e.g., \citealt{vog14b, za, ka}). As result of the annihilation among these particles in the denser 
inner regions of the galactic halos, the originally cuspy DM density  becomes constant with radius. Outside the
core region, the number of annihilations rapidly falls as $\rho_{\rm DM}^2(r)$ and the halo profile remains identical to the
original one.

\subsection{FUZZY dark particles}

The idea is that the dark matter is a scalar dark particle of mass $m_a\sim 10^{-22}\ {\rm eV}$. At large scales its 
coherent macroscopic
excitations can mimic the behavior of the cold dark matter (CDM). At the scale of
galaxies, however, this particle has macroscopic wave-like properties that may explain the classic ``discrepancies'' of the
standard DM scenario (\citealt{W,HO, Be+17,Ri}).
 
Once in galaxies, these particles behave as Bose--Einstein condensate (BEC); in this model, the inter-particle distance
is much smaller than their de Broglie wave length. The particles move
collectively as a wave: their equation of state can lead to cored configuration like those observed.
The capability to detect such Bose--Einstein-condensed scalar
field dark matter with the LIGO experiment is under analysis (\citealt{lsrd17}).

 \subsection{Warm dark matter particles}

\begin{figure}[htb]
\centering
\includegraphics[width=0.82\textwidth]{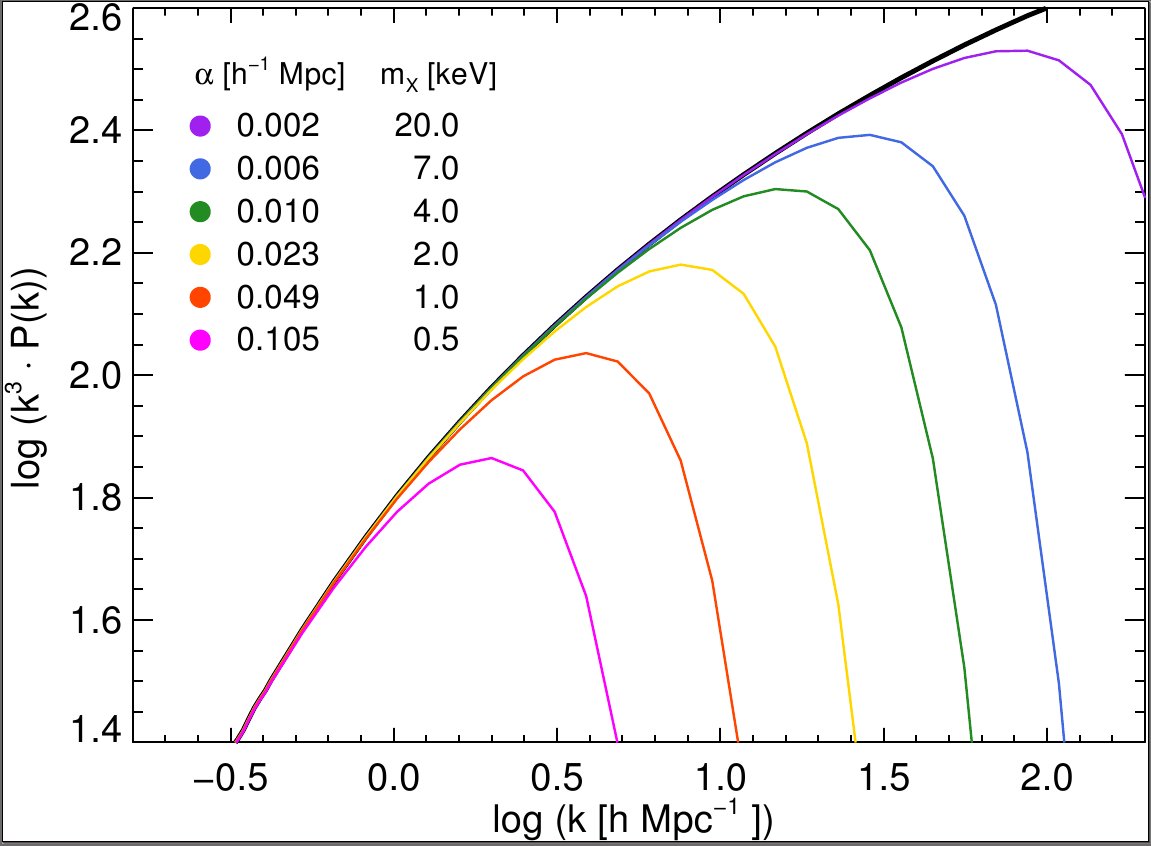}
\caption{Linear power spectra in  $\Lambda$CDM ({\it black line})  and $\Lambda$WDM ({\it coloured lines)} scenarios. 
$\Lambda$WDM models are labelled by their thermal relic mass and value
of the damping scale $\alpha$. We have
 $(\frac{P_{\rm WDM}} {P_{\rm CDM}})^{1/2} =[ 1 + ({\alpha}k) ^{2 /1.1}] ^ {-5 / 1.1}$,
$k$ is the wave-number. 
Image reproduced with permission from \cite{ke+14}, copyright by the authors.}
\label{fig:WDM}
\end{figure}

Warm dark matter (WDM) particle decouples from the cosmological plasma when it is still mildly relativistic. 
These particles can be created in
the early Universe in a variety of ways (\citealt{DW, SeF, Ku}). In the case where the WDM consists of thermal
relics, the suppression of small-scale power in the linear power spectrum (e.g., \citealt{Br})
$P_{\rm WDM}$, can be conveniently parametrized by reference to
the CDM power spectrum $P_{\rm CDM}$, see Fig.~\ref{fig:WDM}. 
In the more likely cases in which the WDM particle is a non-resonantly produced
sterile neutrino, its mass $m_{\rm sterile}$, can be related to
the mass of the equivalent thermal relic
(\citealt{v05}).This conversion depends on the specific particle production mechanism. 

Given the mass of this particle being about 2~keV, its de~Broglie length-scale is of the order of 30~kpc, so
that, inside the optical region of galaxies a quantum pressure emerges (\citealt{dds, dvs}) and plays a role in the
equilibrium
of
the structures.
The DM particles follow, then, a Fermi--Dirac distribution:
\begin{equation}
f_{\rm FD}(p;T,\mu)= \frac{g}{(2\pi \hbar)^3}\frac{1}{\exp[(E-\mu)/T]+1} \,,
\end{equation} 
where $p$ and $E=p^2/(2m)$ are the momentum and the single-particle kinetic energy; $T(r)$, expressed in terms of energy, 
is the average temperature 
of DM particles at a radius $r$: $T(r) \propto V^2(r)$ in spirals and $T(r) \propto \sigma_{\rm l.o.s.}^2(r)$ in
pressure dominated systems. Noticeably, $f(p)$ has an upper limit: $ f(p) \le \frac{g}{(2\pi \hbar)^3}$, where $g$ is the
number of internal degrees of freedom. We have, in this case, that the quantum pressure and not the Gravity Force shapes the inner DM density
profile.  WDM particles can be detected: they can 
produce a monochromatic gamma ray line at $ 2m_{\rm WDM}\ {\rm keV}$, which is constrained by X-ray measurements, e.g., \citealt{Boy07}.

The properties of WDM
particles, their scientific case and cosmological role and the various strategies to detect them, have
 recently been presented in a White Paper (\citealt{all}).

\subsection{In search for dark matter}

For 30 years, WIMPs have been the first target in our attempt to detect and identify the dark particle. During the past
decades, the
sensitivity of the experiments involved has improved by three to four orders of magnitude, but
an evidence for
their existence is yet to come. On the other hand, searches at hadron colliders (which attempts to produce WIMPs 
through the collision of high energy protons and the subsequent 
formation of stable dark matter particles that can be identified through the production of quarks and gluons), have given no
result (see \citealt{But}).

It is agreed that no conclusive detection signal of the particle has yet arrived as result of a many
year-long
extensive search program that combined, in a complementary way, direct, indirect, and collider probes (see \citealt{amel} for
a detailed review).

However, it is worth discussing astrophysical aspects, related to the above searches, that have an intrinsic importance
and
that are valid also for any particle investigation. In direct searches, the differential event rate $R_{\rm scatt}$
\begin{equation}
\frac{{d} R_{\rm scatt}}{{d} E} \propto g(v_{\min}) \rho(R_\odot) \,,
\end{equation}
is proportional to $\rho(R_\odot)$ the local (i.e., at the solar radius) dark matter density and to the function 
$g(v_{\min}) = \int_{v> v_{\min}}^{v_{\rm esc}} \frac{f({\bf v})}{v} {d}^3 {\bf v}$.
$v_{\min}$ is the minimum particle speed that can cause in the detector a recoil of energy $E$ (\citealt{go02}).

$v_{\rm esc}$
is the escape velocity from the Milky Way: $v_{\rm esc}=(570 \pm 120){\rm km /s}$ 
(\citealt{ns13}). A reference value of $\rho(R_\odot)=0.3\ {\rm GeV/cm^3}$ is often adopted however recent accurate  determinations indicate  a
rather higher value: $\rho(R_\odot)=(0.43\pm 0.06)\ {\rm GeV/cm^3}$ (\citealt{sng, cu10}).

To obtain $g(v_{\min})$, one needs the whole DM density distribution, however, for the Milky Way, we can
consider the galaxy halo as an isotropic isothermal sphere with density profile $\rho(r) \propto r^{-2}$. Then 
$f({\bf v}) = \frac{N}{2 \pi \sigma_{v}^2} \exp{\left(- \frac{{\bf v}^2}{2 \sigma_{v}^2} \right)}$,
where $N$ is a normalization constant and $\sigma_{v}$ is the DM particles one-dimensional velocity dispersion, which in the
present model is related to the circular velocity $V(r)$ by: $\sigma_{v} = V(r) / \sqrt{2}$.

The indirect searches of DM are based on astrophysical observations of the products of the DM particles self-annihilation 
(or
decay) able to climb up the emissions coming from the likely astrophysical mechanisms also producing antiprotons and
positrons. The
photon spectrum $\frac{dN^f_{\gamma}}{dE_{\gamma}}$, with $E_{\gamma}$ the photon energy, is expected to be proportional
to
$\int_{\rm l.o.s.} dl ~\rho^2 (r)$ for annihilations and $\int_{\rm l.o.s.} dl ~\rho (r)$ for decays; as usual, $\rho(r)$ is the
DM
density within the galaxy and the integrals are performed over 
the line of sight $l$. The dependence of $\rho(r)$ on the above fluxes leads to a dependence of the signal on the inner
distribution of DM in galaxies, modulo the fraction between the size of the dark halo and that of the telescope beam both projected on the plane of the 
sky (for details including the application to the Galactic Center, see \citealt{ga16}). As consequence of that, indirect searches require
an accurate knowledge of the
halo density profiles and, in this perspective, one should also consider cored dark matter halo distributions,
in performing the analysis on the $\gamma$ flux.
Here, we do not further enter in this (important) issue (see, e.g., \citealt{ga16}).

\section{Baryons in galaxies}

The luminous components in galaxies show a striking variety in morphology and in dimensions. Noticeably, the total
luminosity and the radius $R_{1/2}$ enclosing half of the latter are good tags of the objects.

\subsection{Spirals, LSB and UDG} 

Caveat some occasional cases not relevant for the present topic, the stars are distributed in a thin disk with surface
luminosity (\citealt{fre}, for a study on 967 late type spirals, see \citealt{PSS})
\begin{equation}
I(R) =I_0 e^{-R/R_D}=\frac{M_{D}}{2 \pi R_{D}^{2}}\: e^{-R/R_{D}}
\Big(\frac{M_D}{L}\Big)^{-1} \,,
\label{eq:fre}
\end{equation}
where $R_D= 1/1.67 ~ R_{1/2}$ is the disk length scale, $I_0$ is the central value of the surface luminosity and $M_D$ is the
disk mass. The
light profile of late spirals does not depend on galaxy luminosity and the length scale
$R_D$ sets a consistent reference spatial scale.\footnote{We take $R_{\rm opt}\equiv 3.2 \, R_D$ as the reference stellar
disk edge.}

The contribution to the circular velocity from this stellar component is: 
\begin{equation}
V_{\rm disk}^{2}(r)=\frac{G M_{D}}{2R_{D}}
x^{2}B\left(\frac{x}{2}\right),
\label{eq:freeman}
\end{equation}
where $x\equiv R/R_{D}$ and $B=I_{0}K_{0}-I_{1}K_{1}$, a combination of
known Bessel functions.

Classical LSB galaxies usually have central surface brightness down to $\mu_B(0) \sim 22$--$23$ mag arcsec$^{-2}$
(\citealt{I88}). Extremely low surface brightness (LSB) galaxies with unexpectedly large sizes, namely ultra-diffuse galaxies
(UDGs), are found in nearby galaxy clusters (\citealt{Bo91,tl18}). UDGs have much lower central
surface brightness ($\mu(0)=24$--$26$ mag arcsec$^{-2}$ in $g$ band and half-light radii $R_{1/2}>1.5\ {\rm kpc}$ that, in spirals,
are found in objects with stellar masses more than 10 times higher (\citealt{vad15, din}).
In LSBs/UDGs the stellar disks follow the Freeman exponential profile as in normal spirals, but their two structural parameters ($I_0$ and $R_D$)
do not
correlate as in the latter, where, approximately: $L_I \propto R_D^2$. 

\subsubsection{HI distribution in disk systems}

Spirals have a gaseous HI disk which usually is important only as tracer of the
galaxy gravitational field. Only at the outer radii ($R>R_{\rm opt}$) of low
luminosity objects, such disk becomes the major baryonic component of the circular velocity and must be included in the galaxy velocity model. 

The HI disks show, very approximately, a Freeman distribution with a scale length about three times larger than that of the
stellar
disc (\citealt{e11, wael14}). 
\begin{equation}
\mu_{\rm HI}(R)=\mu_{\rm HI,0}~ e^{-\frac{R}{3\,R_D}}
\label{eq:muhi} 
\end{equation}

A rough estimate of the contribution of the gaseous disc to the circular velocity is 
\begin{equation}
V_{\rm HI}(R)^2=1.3 \Big( \frac {M_{\rm HI}}{9\,M_{D}} \Big) V_{\rm disk}^2 \Big(\frac {R} {3\,R_D}\Big)
\label{eq:muhi2} 
\end{equation}
where the coefficient 1.3 is due to the He contribution. Of course when the resolved HI surface density is available, one 
derives $V_{\rm HI}(R)^2$ directly from the latter. Inner H$_2$ and CO disks are also present, but they are negligible with respect to the stellar and HI ones
(\citealt{G+10,cs}).

\subsection{Ellipticals} 

Ellipticals are more compact objects than spirals so that, in objects with same stellar mass $M_\star$, they
probe inner regions of the DM halo than spirals. Their profiles are well represented by the Sersic Law: 
\begin{equation}
 \ln \Big[\frac{\Sigma_S(R)}{\Sigma_{R_e}}\Big] = -q\Big[\Big(\frac{R}{R_e}\Big)^{\frac{1}{m}} -1 \Big] \,,
\end{equation}
$\Sigma_S(0)=\Sigma_{R_e}e^q$, where $R$ is the projected radial coordinate in the plane of the sky, $\Sigma_{R_e}$ is
the line of sight (l.o.s.) projected surface brightness at a projected scale radius $R_e\simeq R_{1/2}$ and $q=2m-1/3$ with $m$ a free parameter. 
 By deprojecting the surface density $\Sigma_S(R/R_e,m)$, we obtain the luminosity density $j(r)$ and by assuming a radially
constant stellar
mass-to-light ratio $(M/L)_\star$ we obtain the spheroid stellar density $\rho_\star (r)$. 
 
\subsection{Dwarf spheroids}

The distribution of stars in dSph plays a major role in the analysis of their internal kinematics. The
information we have comes from the bright stars detected by dedicated imaging or spectroscopy and, more
recently, by surveys like the Sloan Sky Digital Survey and Gaia. The 3D stellar density is obtained from the deprojection of the 2D luminosity profile and an assumed mass-to-light ratio. The former is well reproduced by the 
 Plummer density profile (\citealt{pl}), characterized by a length scale $R_e$ and a central density $\nu_0=3\, M_{\rm sph}
/(4\pi R_e^3)$ with $M_{\rm sph}$ the total stellar mass. The projected mass (luminosity) distribution is given by:
$\Sigma(R)=\frac{M_{\rm sph}}{\pi R_e^2} \left(1+x^2\right)^{-2}, x= R/R_e $. Then, the 3D stellar density is given by
\begin{equation}
 \nu(x)=\nu_0 \left(1+x^2\right)^{-5/2} \,.
\label{eq:plu}
\end{equation}

\section{Probing the gravitational potential in galaxies} 

\subsection{Rotation curves}

The rotation curves (RCs) of spirals are an accurate proxy of their gravitational potential. We measure recessional velocities by
Doppler shifts, and from these (often 2D) data, 
we construct the RC $V(R)$. This process estimates also the sky coordinates of the galaxy kinematical center, its systemic
velocity, the degree of symmetry and, often, the inclination angle. 

Notice that the effectiveness of the RC is proved in many ways: e.g., in systems with $M_I<-18 $ in the innermost luminous matter dominated regions
 the gravitating mass (measured by $V(R)$) agrees with the predictions from the light distribution (\citealt{ra}).

The rotation curves in disk systems have 3 different components: the relationship with the total gravitational potentials 
$\phi_{\rm tot}=\phi_b+ \phi_{\rm H}+\phi_{\rm disk}+\phi_{\rm HI}$ is
\begin{equation}
V^2_{\rm tot}(r)=r\frac{d}{dr}\phi_{\rm tot}=V^2_b +V^2_{\rm H}+V^2_{\rm disk}+V^2_{\rm HI} \,.
\end{equation}

Then, the velocity fields $V_i$ are the solutions of the four separated equations:
 $\nabla^2 \Phi_i= 4 \pi G \rho_i $
where $\rho_i $ are dark matter, stellar disk, stellar bulge, HI disk surface/volume densities ($\rho_h(r),
\rho_{bu}(r), 
\mu_{d}(r) \delta (z), \mu_{\rm HI}(r) \delta (z)$ with $\delta(z)$ the Kronecker function, $z$ the cylindrical coordinate)
and $\phi_i$ the gravitational potential.

\begin{figure}[htb]
\centering
\includegraphics[width=1\textwidth]{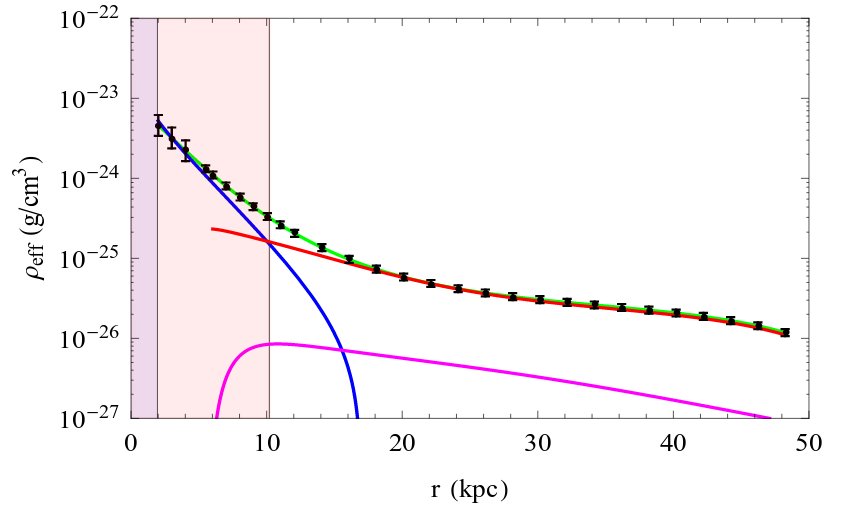}
\caption{A test case: NGC 3198. Effective total density ({\it points with errorbars}). Contributions: stellar
disk ({\it blu }), HI
disk 
({\it magenta}), dark matter ({\it green line}), all components ({\it green}). Regions in which the method: 1) is
not applicable
({\it pink}), 2) provides us with a) the value of disk mass
({\it green}) b) the halo density profile ({\it white}). 
Image reproduced with permission from \cite{ks15}, copyright by ESO.}
\label{fig:newrho}
\end{figure}

Recently, a new way to exploit the RC to obtain the DM halo density distribution has been devised (\citealt{sng}). We assume that spirals are
composed by a stellar disk (\citealt{fre}), a HI disk and an unspecified spherical DM halo with density profile $\rho_H(r)$. Other baryonic
components can be added, if needed.\footnote{The HI component is obtained directly from observations, however, it is 
always negligible because $dV^2_{\rm HI}/dr\simeq 0$.} From
the radial derivative of the equation of
centrifugal equilibrium we obtain
\begin{equation}
\rho_H(r)=\frac{1}{4 \pi G r^2} \frac{d}{dr}\left[r^2\left(\frac{V^2(r)}{r}-a_{D}(r)\right)\right],
\label{eq:rhoh}
\end{equation}
We have $a_{D}(r)= \frac{GM_{D}r}{R_
 D^3}(I_0 K_0 - I_1 K_1)$, where $I_n$ and $K_n$ are the
modified Bessel functions computed at $\frac{r}{2 R_ D}$. 
Noticeably, the second term  of the r.h.s.\ of Eq.~(\ref{eq:rhoh}) goes exponentially to zero for $r/R_D> 2 $ (see Fig.~\ref{fig:newrho}).
Then, for $R>2 \ R_D$, we can determine the DM
density profile (see Fig.~\ref{fig:newrho}). On the other hand, for  $R< R_D$, the DM distribution 
is negligible, so that, if we have a good spatial coverage of the inner RC, we can use Eq.~(\ref{eq:rhoh}) also to obtain the disk
mass
with
good
precision.

\subsection{A reference velocity for disk systems}
 
In spite of the fact that $V(R)$, the circular velocity, is a function of radius we often require
a meaningful reference velocity to tag each disk system. In the literature there is no shortage of
proposed reference velocities, among those: $V_{\rm flat}$, $V_{\rm last}$, the linewidths
$W_{20}$, $W_{50}$ and the maximum velocity $V_{\max}$. Obviously, if the RC of an object is not available, we are
forced to choose one
of 
these kinematical measurements as a reference velocity, however, we must stress that they are very biased: a)~a
flat part of
the RC occurs only a limited number of objects and only over a limited radial region (\citealt{PSS}); b)~$V_{\rm last}$
depends on the
distribution of HI
in the galaxies and on the sensitivity of radio telescope used; c)~the linewidths are similar to the case b) and
furthermore they depend on the full RC profiles; d)~the significance of $V_{\max}$ changes as galaxy
luminosity changes, sometimes coinciding with the outermost available velocity, in other cases, with the innermost
one.  The best unbiased reference velocity for spirals is the quantity: $V(k R_D)$ that also involves  the  stellar disks length scale. We have  $k=2.2$ or $3.2$, according
whether we are investigating the properties of the luminous or of the dark matter.

\subsection{Vertical motions}

The main goal of the DiskMass Survey (\citealt{dms1,dms2}) was to determine the dynamical mass-to-light ratio of the
galaxy disks $(M/L)_{\rm dyn}$ by a suitable  use
of the stellar and gas kinematics. At a radius
$R$, for a locally isothermal disk, we have  
\begin{equation}
(M/L)_{\rm dyn} = \frac{\sigma_z^2}{\pi \,G \,b \,h_z I(R)} \,,
\label{eq:ML}
\end{equation}
where the value $b=1.5$ is a reasonable approximation for the
composite (gas+stars) density distribution (\citealt{vk88}), 
$I$ the surface luminosity obtained from the photometry, $\sigma_z$ the vertical component of the stellar velocity
dispersion. Noticeably, with the advent of 2-dimensional spectroscopy using integral field units
(IFU), the accuracy and the $z$-extension of the measurements of $\sigma_{z}$ has been dramatically increased;
$h_z$ is the disk scale height (\citealt{vs81,bah84}) that can be directly measured, and that well correlates
with the disk scale length $R_D$ (\citealt{kel, dms1}).

Let us stress that this approach leading to Eq.~(\ref{eq:ML}) is certainly a new avenue for investigating dark matter
in
galaxies, but some warning must be raised in that it can be subject to relevant biases (\citealt{H17}). 
 
\subsection{Dispersion velocities}
 
It is well known that in spheroids the kinematics is complex, the stars are in gravitational equilibrium by balancing the
gravitational potential,
they are subject to, with the pressure arisen from the r.m.s.\ of their 3D motions. Moreover, we cannot directly measure the
radial/tangential velocity
dispersions linked to the mass profile, but only their projected values (e.g., \citealt{cga}). 

The SAURON (\citealt{bac01}) Integral Field Spectroscopy survey (\citealt{dz02}) was
the
first project to map the two-dimensional stellar kinematics of a sample of 48 nearby ellipticals with $M_B<18$. This survey
was followed by the ATLAS$^{\rm 3D}$ project (\citealt{ca11}), a multiwavelength survey of
260 ETGs galaxies. In \cite{ca16} one finds the details of these observations.
 
The dispersion velocity is related to the gravitational potential of a galaxy by the Jeans equation that we express as
(\citealt{BT})
\begin{equation}
\frac{\partial\ln \sigma^2_r}{\partial\ln r} = -\frac{1}{\sigma^2_r} \frac{GM}{r} - \gamma_\star -2 \beta \,.
\label{eq:jeans0}
\end{equation} 
Here, $G$ is the gravitational constant and $M(r)$ is the enclosed mass. The velocity
anisotropy $ \beta = 1 - \frac{\sigma_{\theta}^2 + \sigma_{\phi}^2}{2 \sigma_{r}^2}$,
where $\sigma_{\theta, \phi, r}$ are the velocity dispersions in the $r, \theta$ and $\phi$ directions, can be a function of radius $r$ (\citealt{BT}). It is useful to define:  $\alpha= d\,\log \sigma_r/d\,\log r$. Almost always the motions in the $\theta$ and $\phi$ directions are assumed to coincide. 

$\nu_\star$ is the 3D stellar density distribution, $\gamma_\star= d\,\log \nu_\star/d\,\log r$. Under the assumption of constant $\beta$, the radial velocity dispersion $\sigma_{r}(r)$ can be
expressed as
\begin{equation}
\sigma^{2}_{r}(r) = \frac{1}{\nu_\star (r)} \int^{\infty}_{r} \nu_\star (r') \left(\frac{r'}{r}\right)^{2 \beta} \frac{G
M(r')}{r'^2}\, dr' \,.
\label{eq:jeans}
\end{equation}

We can then determine the galaxy mass profile by means of 
Eq.~(\ref{eq:jeans}) and the line-of-sight velocity dispersion $\sigma_{\rm l.o.s.}$ when the anisotropy factor $\beta(r)$ is
known or assumed: 
\begin{equation}
\sigma_{\rm l.o.s.}^2(R)=\frac{1}{I(R)}\int_{R^2}^\infty { d}r^2 
\frac{\nu_\star}{\sqrt{r^2-R^2}} \sigma_r^2
\left[1-\beta\frac{R^2}{r^2}\right],
\label{eq:jeans1}
\end{equation}
where $I(R)$ and $\nu_\star(r)$ are related by $I(R) = 2 \int_R^{+\infty} \frac{\nu_\star(r)
r \, dr}{\sqrt{r^2-R^2}}$.
$I(R)$ and $\sigma_{\rm l.o.s.}(R)$ are directly measured. 

The Schwarzschild method can be seen as a (complex) extension of the Jeans method and it is especially applied to dSph
galaxies where the stellar component is totally negligible (\citealt{cr99, br13}). It is based, fixed a specific gravitational potential, on the
integration of test particle orbits drawn from a grid of integrals of motions, i.e., the energy and the angular momentum.
The main feature of this method is that, differently from the Jeans method, it can successfully use
the observed second
and fourth velocity moment profiles to break the mass-anisotropy degeneracy (\citealt{br13}).

\subsection{Fast spheroidal rotators}

In the case of objects (e.g., S0 galaxies) in which the dispersion velocity combine with the rotation motions to balance the
galaxy self-gravity, 
there is a simple and efficient anisotropic generalization of the axisymmetric Jeans formalism
which is used to model the stellar kinematics of galaxies (see \cite{ca16} for details). The following is assumed: (i) a
constant mass-to-light ratio $M/L$
and (ii) a velocity ellipsoid that is aligned with cylindrical coordinates $(R,z)$ and characterized by the classic anisotropy
parameter $\beta_z=1- \sigma_z^2/\sigma_R^2$. These simple models are fit to integral-field observations
of the stellar kinematics of fast-rotator early-type galaxies.
With only two free parameters ($\beta_z$ and the inclination) the models generally provide remarkably good descriptions of the
shape of the first ($V$) and second ($V_{\rm rms}\equiv\sqrt{V^2+\sigma^2}$) velocity moments. 
The technique can be used to determine the
dynamical mass-to-light ratios and angular momenta of early-type fast-rotators and it allows for the inclusion of
dark
matter, supermassive central black holes, spatially varying anisotropy, and multiple kinematic components.

 \subsection{Dispersion velocities versus rotation curves}

Here, it is worth making a comparison between the circular velocity
$V(r)$ and the radial (or line-of-sight) velocity dispersion of an irrotational gravitational tracer with distribution
$\nu_\star(r)$ and
with anisotropy $\beta(r)$. From Eq.~(\ref{eq:jeans0}) we get:
 \begin{equation} 
(-\gamma_\star (r) + 2 (\beta(r)+\alpha(r)))~ \sigma_r^2(r) = V^2(r)
\label{eq:vsigma} 
\end{equation} 
$\alpha(r)$ and $\gamma_\star(r)$ are the logarithmic derivatives of $\sigma_{\rm l.o.s.}$ and $\nu_\star$.
Let us notice that, in dispersion velocity supported
systems, even in the case of isotropic orbits: $\beta(r)=0$, it is
necessary to know the spatial distribution of the tracers in order to make any inference on the DM distribution. Flat RC
and flat dispersion velocity profiles do not necessarily indicate the same gravitational field.

\subsection{Masses in spheroids within half-light radii}

We can measure the \emph{total} mass enclosed within the half-light radius $R_{1/2}$ 
by measuring $\sigma_{\rm l.o.s.}(R_{1/2})$ the line of sight velocity dispersion at this radius (\citealt{wo10}). Since
$\sigma_{\rm tot}^2 = \sigma_r^2 +
\sigma_\theta^2 + \sigma_\phi^2 = (3 - 2 \beta)\sigma_r^2$ we can write the Jeans equation as $G \ M(r) r^{-1} = \sigma_{\rm tot}^2(r) + \sigma_r^2(r) \left (-\gamma_\star+\alpha -3 \right)$. Let us define $R_3$ as $\gamma_\star(R_3)=3$\footnote{$R_3\simeq 1.1 \ R_{1/2}$} since $\alpha(R_3) \ll 3$
 from the observed $\sigma_{\rm los}(r)$ profiles, then, at $R=R_3$,
we have, independently of the value of the anisotropy:
\begin{equation}
M(R_{1/2})\approx 3 G^{-1} \sigma_{\rm l.o.s.}^2 (R_{1/2}) R_{1/2}
\label{eq:Wolf}
\end{equation}

APOSTLE cosmological hydro dynamical simulations have tested the validity and accuracy of this mass estimator and found that the resulting measurements are, at most, biased by 20\% (\citealt{cel17}). 

\subsection{Tracer mass estimator}
 
Given a number of $N$ of tracers in dynamical pressure supported
equilibrium with no systematic rotation and moving with l.o.s.\ velocities within a dark halo of mass profile $M(r)$, the TME is expressed as 
\begin{equation}
 M(r_{\rm out}) = \frac{C}{GN}\sum_{i=1}^N V_{{\rm l.o.s.,i}}^2 R^{\epsilon}_i \,.
\label{eq:TME}
\end{equation}
 
The prefactor $C$ depends on i) the slope $\epsilon$ of the gravitational potential, assumed to be: 
$
\Phi(r) \propto \frac{v_0^2}{\epsilon} \left(\frac{a}{r} \right)^{\epsilon}; 
\medskip\
v_0^2\, \log \left(\frac{a}{r} \right) (\epsilon = 0).
$ ii) the ``slope'' $\gamma_\star$ of the de-projected density profile of the tracers ($\rho_{\rm trac}(r)\propto r^{-\gamma_\star}$) iii)
the
orbital anisotropy $\beta$ of the tracers.

We then have $C = \frac{(\epsilon+\gamma_\star-2\beta)}{I_{\epsilon,\beta}} r_{\rm out}^{1-\epsilon}$
 with $r_{\rm out}$ the distance of the outermost tracer and 
$I_{\epsilon,\beta}=\frac{\pi^{1/2}\Gamma(\epsilon/2+1)}{4\Gamma((\epsilon/2+5)/2)} \left[\epsilon+3-\beta(\epsilon+2)\right]$, 
where $\Gamma$ is the Gamma function (\citealt{wa10, ae11}).

The mass estimator in Eq.~(\ref{eq:TME}) 
performs very well, especially in the case in which the tracers are in random orbits, so that $\beta=0$ and for ellipticals where we have $\alpha =0 \pm 0.1$. In these cases, the
uncertainties on the two latter quantities do not bias the mass estimate.

\subsection{Weak lensing}

We briefly recall here that weak gravitational lensing is a powerful tool for probing the dark matter distribution in
galaxies (\citealt{sch,hj08,mv, b16}). 
It is known that observed images of distant galaxies are coherently deformed by weak lensing effects caused by foreground
matter distributions. These distortions enable the measurement of the mean mass profiles of foreground lensing galaxy through the
stacking of the background shear fields (\citealt{ZM}).
To determine halo mass, we measure the excess surface mass
density $\Delta\Sigma(R)=\overline{\Sigma(<R)}-\overline{\Sigma(R)}$, which is the difference between the projected average
surface mass within a circle of radius $R$ and the surface density at
that radius. The tangential shear $\gamma_{t}$ is directly related to the above quantities 
through $\Delta\Sigma(R)=\Sigma_{\rm crit}\langle\gamma_{t}(R)\rangle$,
where $\Sigma_{{c}}$ is the critical surface density defining the Einstein radius of the lens
\begin{equation}
 \Sigma_{c}=\frac{c^2}{4\pi G}\frac{D_{s}}{D_{l} D_{ls}} \,,
\end{equation}
where $D_{s}$, $D_{l}$, and $D_{ls}$ are the distances to the source,  to the lens and the lens-source one, respectively. 
The lens equation relates $\gamma_t$ with the distribution of matter in the lensing galaxy:
\begin{equation}
\gamma_t(R)= (\bar \Sigma(R)- \Sigma(R))/\Sigma_c \,,
\end{equation}
where $\Sigma(R)=2 \int_{0}^{\infty} \rho(R,z)\, dz $ is the projected mass density
of the object distorting the galaxy image, at projected radius $R$ and $\bar{\Sigma}(R)= \frac{2}{R^2} \int_{0}^{R} x
\Sigma(x)\,
dx$ is the mean projected mass density interior to the radius $R$.

\subsection{Strong lensing}

Gravitational lensing occurring in very aligned galaxy-galaxy-observer structures magnifies and distorts the images of 
a distant galaxy providing us with relevant information on the mass structure of the intervening galaxy so as of 
the background source (see \citealt{T}).

\begin{figure}[htb]
\includegraphics[width=\textwidth]{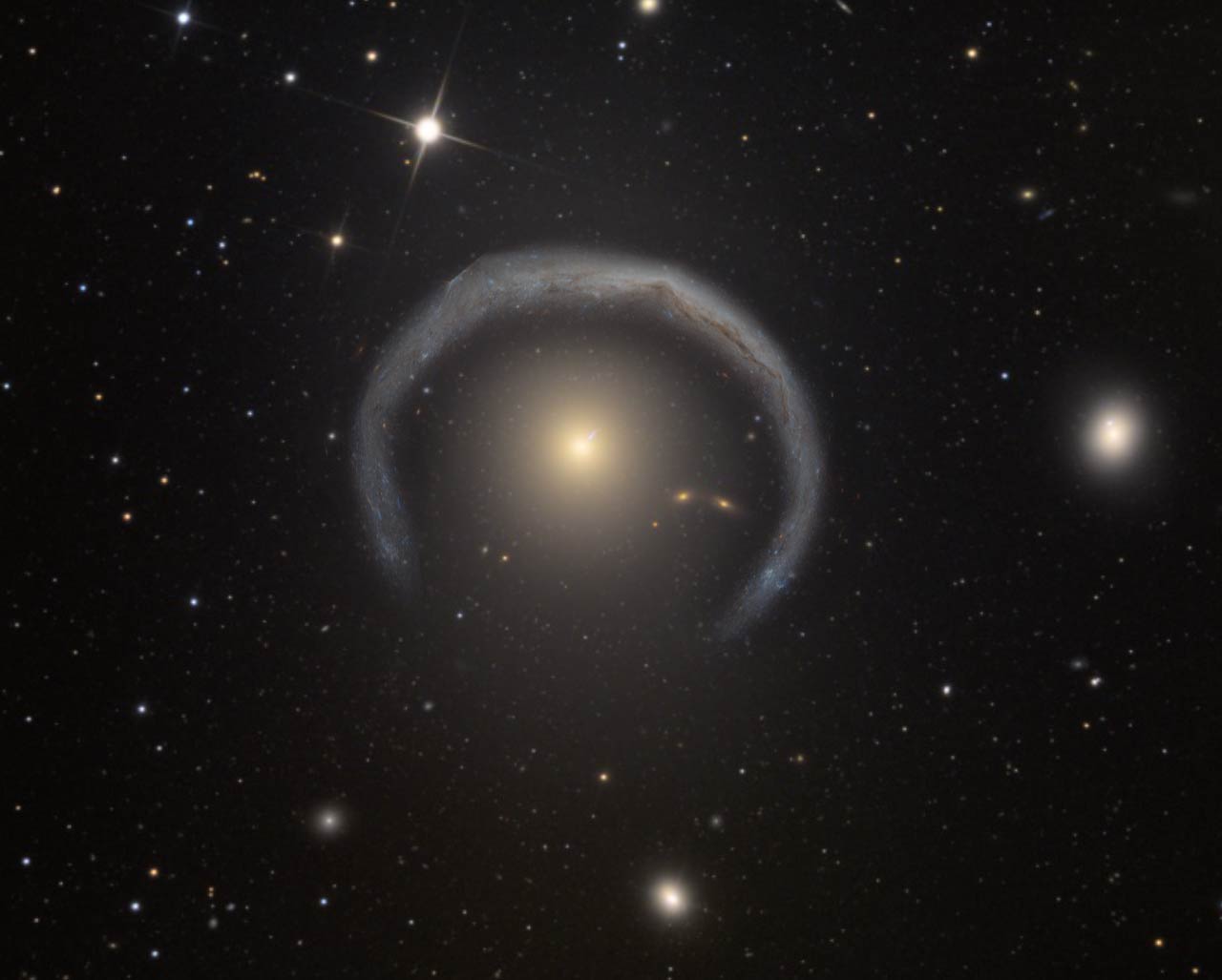}
\caption{Einstein ring (artist's concept). This extraordinary GR effect provides us with the value of the projected
mass of the galaxy lens inside $R_E$.}
\label{fig:x2}
\end{figure}

The lens system is axially symmetric, the radial coordinate $r$ is related to cylindrical polar coordinates by
$r=\sqrt{\xi^{2}+z^{2}}$ where $\xi$ is the impact parameter measured from the center of the lens. 
The mean surface density inside
the radius $\xi$ is
\begin{equation} 
\bar{\Sigma}(\xi)=\frac{1}{\pi \xi^{2}}\int_{0}^{\xi}2\pi \xi' \Sigma(\xi')\,d\xi'. 
\end{equation}
The presence of an Einstein ring of radius $R_E$, at projected galactocentric
distance
$\xi$ (see Fig.~\ref{fig:x2}), 
allows us to obtain the projected total mass
inside $\xi$: 
\begin{equation}
(M_{\rm halo}(\xi)+ M_{\rm stars}(\xi))= \pi R_E^2 \Sigma_c \,.
\label{eq:ME}	
\end{equation}
 
\subsection{X-ray emission \& hydrostatic equilibrium}

Isolated ellipticals have an X-ray emitting halo of regular morphology, that, extends out to very large radii. The gravitating
mass inside a radius $r$, $M(r)$ can be derived from their X-ray flux if the emitting gas is in hydrostatic
equilibrium. From its density and the temperature profiles we obtain the total mass profile (\citealt{frg, ef}:
\begin{equation}
M(<r) = {kT_{g}(r) r \over {G \mu~ m_p}}\Big({d\,\log {\rho_g(r)}\over{d\,\log r}} + 
{d\,\log T_{g}(r)\over{d\,\log r}}\Big),
\end{equation}
where $T_{g}$ is the (measured) ionised gas temperature, $\rho_g$ the gas density, $k$ is the Boltzmann's
constant, $\mu=0.62$ is the mean molecular weight and $m_p$ is the mass of the proton.

\section{The mass of the stellar component in galaxies}

 We can assume that the stellar mass surface density $\Sigma_\star(r)$ is proportional to the luminosity 
surface density, which in galaxies is well measured by CCD infrared photometry. Radial variations of the $M_\star/L$ ratio 
exist and often are astrophysically relevant, but rarely they play a role in the determination of the mass profile of
galaxies. 

The total galaxy luminosity is related to its stellar content, hence, the direct approach to derive 
the galactic mass in stars by modelling their spectral energy distribution in terms of age, metallicity, initial mass
function of the stellar component. This modelling, pioneered by \cite{tin}, is performed by the well-known stellar population
synthesis technique. The SED of a galaxy, selected colour indices and absorption lines are all reproduced by a theoretical
models calculated under
different assumptions regarding the above physical quantities. In practice, the exercise is not straightforward
because 
degeneracies among age, metallicities, IMF and dust content, to name some, do arise and 
different combinations of the former quantities yield to very similar SEDs.

\cite{bdj} found rather
simple relationships between mass-to-light ratios and certain colour indices. In detail, they investigated a suite of 
spectrophotometric spiral galaxy evolution models that assumed a Salpeter Initial Mass Function, an exponentially
declining star formation rate and a current age of 12 Gyr and found that the stellar mass to light ratios correlate
tightly with galaxy colours (see also \citealt{Bel3}).

The important  stellar mass-to-light ratios in the Spitzer $3.6\ \mu{\rm m}$ band
(${{\Upsilon}_\star}^{3.6\ \mu{\rm m}}$) and in the K-band (${{\Upsilon}_\star}^{K}$) have also been derived by constructing
stellar population synthesis models,
with various sets of metallicity and star-formation histories (see \citealt{oh,db08}).

We have 
\begin{equation}
\log({{\Upsilon}_*}^{K}) = 1.43 \times (J-K) -1.38 \  \  \ {\Upsilon}_*^{3.6\ \mu{\rm m}}=0.92 \ {\Upsilon}_*^{K}-0.05 \,.
\end{equation}

The values of the galaxy stellar masses as derived from their SEDs have
been compared with those obtained by other
methods. \cite{gg09} investigated a sample of ellipticals with Einstein rings from which they derived the total
projected mass (dominated by the stellar component) and, from the latter, the total mass of the spheroid.
Then, by using the SDSS multicolour
photometry they fitted the galaxy spectral energy distributions (SEDs) by means of composite stellar-population synthesis models of \cite{BC03} and \cite{mar} and obtained the 
photometric mass of the stellar spheroid.  The two different mass estimates agreed within 0.2~dex (see also \cite{TS10}). 

\cite{sya} have estimated kinematically the disk mass from the rotation curve of 18 spirals of 
different
luminosity and Hubble types and have compared them with the values obtained by fitting their SED with
spectro-photometric models. They found 
$M_{\rm pho}\propto M_{\rm kin}^{1.0 \pm 0.1} $ with a r.m.s.\ of 40\% suggesting that photometric and kinematical estimate of the masses of the stellar galaxy disks are statistically consistent. 

We have to caution about one consequence of the found disagreement of about 0.15 dex among the dynamical and the spectro-photometric estimates. This value is small to affect existing color stellar mass relationships, but it is large if we want to use it for mass modelling purposes. In
fact, in spirals, for $R< R_D$, 
the dark and the luminous components of the circular velocity are of the same order of magnitude $V_h \simeq V_d (M_{\rm D,true}/M_{\rm D,phot})^{-0.5}$ and therefore an uncertainty of $(10^{0.15}-1)
~100\% \sim 40\% $ on the value of $M_{\rm D,phot}$ jeopardizes the derivation of the DM velocity contribution and even more that of the subsequent DM halo density.

For spiral galaxies there is a reliable method to estimate the disk mass which is immune from the latter uncertainty. We start from the gravitating
mass inside $R_{\rm opt} $: $M_g(R_{\rm opt})\equiv G^{-1} V_{\rm opt}^2 R_{\rm opt}$ and $\nabla$, the rotation curve logarithmic slope
measured at $R_{\rm opt}$: $ \nabla\simeq 3.2 (1-V(2.2~R_D)/V(R_{\rm opt}))$. From \cite{PS90} we
have:
\begin{equation}
M_D = (0.72 - 0.85 \nabla)~M_g(R_{\rm opt}) \,,
\label{eq:nablad}
\end{equation}
where the disk mass has uncertainty of 20\%.

\section{DM halo profiles}

In this section, we will introduce the DM halo profiles that are presently adopted: the empirical ones and those emerging
from specific theoretical scenarios, see Fig.~\ref{fig:haloden}. It is useful to remind that $M_h(r)= G^{-1} V^2_h(r)
r=\int^r_0
4\pi r^2 \rho_h(r) \, dr $ with $V_h(r)$ the halo contribution to the circular velocity $V(R)$. 

\subsection*{BT-URC}

The empirical DM halo density profile, adopted for the URC of \cite{PSS}, takes the form
(see also \citealt{BT})
\begin{equation}
\rho_{\rm BT-URC}(r)= \frac{1}{G} \frac{v_0 ^2 (r^2+3r_0^2)} {(r_0 ^2 + r^2)^2}, \ \ \ \ M_{\rm BT-URC}(r) = \frac{1}{G} \frac{v_0
^2
r^3}{r_0 ^2 + r^2} \,,
\label{eq:LOG}
\end{equation} 
where $r_0$ and $v_0$ are the core radius and the asymptotic circular velocity
of the halo, respectively.

\subsection*{Navarro--Frenk--White} 

In $\Lambda$CDM the structure of virialized DM halos, obtained by $N$-body simulations, have a universal spherically averaged
density profile, $\rho_{\rm NFW} (r)$ (\citealt{nfw}):
\begin{equation}
\rho_{\rm NFW}(r) = \frac{\rho_s}{(r/r_s)(1+r/r_s)^2},
\end{equation}
where $\rho_s$ and $r_s$ are strongly correlated: $r_{s} \simeq 8.8 \left(\frac{M_{\rm vir}}{10^{11}{M}_{\odot}}
\right)^{0.46}\ {\rm kpc}$ (e.g., \citealt{w06}). We define $X\equiv r/R_{\rm vir}$, the concentration parameter $c\equiv r_s/R_{\rm
vie} $ is a weak
function of mass (\citealt{k}): 
\begin{equation}
c= 9.35 ~ \big (\frac {M_{\rm vie}} {10^ {12} M_\odot} \big) ^ {-0.13}
\end{equation}
but a very important quantity in determining the density shape at intermediate radii. The circular velocity for an NFW
dark matter halo is 
given by 
\begin{equation}
V_{\rm NFW}(X)=V_{\rm vie}^2 \frac{1}{X} \frac{{{\ln}} (1 + c X) - \frac{c X}{1 + c X}} {{{\ln}} (1 + c) -\frac{c}{1+c}}\,,
\end{equation}
with $M_{\rm vir}=100~ 4/3 \pi \ \rho_c \ R_{\rm vir}^3$ and $ \rho_c=1.0 \times 10^{-29}\ {\rm g/cm^3}$.

\subsection*{Burkert-URC}

The Burkert empirical profile (\citealt{bu95, sb00}) well reproduces, in cooperation with the velocity components
of
the stellar and gaseous disks, the individual circular velocities of spirals, dwarf disks and low surface brightness systems.
Furthermore, this profile is at the basis of the universal rotation curve of the above systems. The density profile reads as
\begin{equation}
\rho_{\rm B-URC} (r)=\frac{\rho_0 r_0^3}{(r+r_0) (r^2+r_0^2)} \,,
\end{equation}
$r_0$ and $\rho_0$ are the core radius and central density respectively. The velocity profile is: 
\begin{equation}
 V^2_{\rm B-URC} (r) = \frac{G}{r} 2\pi\rho_0r_0^{3} \left[\ln(1+r/r_0) \right] + \frac{1}{2} \ln (1+r^2/r_0^2)- \tan^{-1} (r/r_0) \,.
\label{eq:Burk}
\end{equation}

\begin{figure}[htb]
\centering
\includegraphics[width=10.5cm]{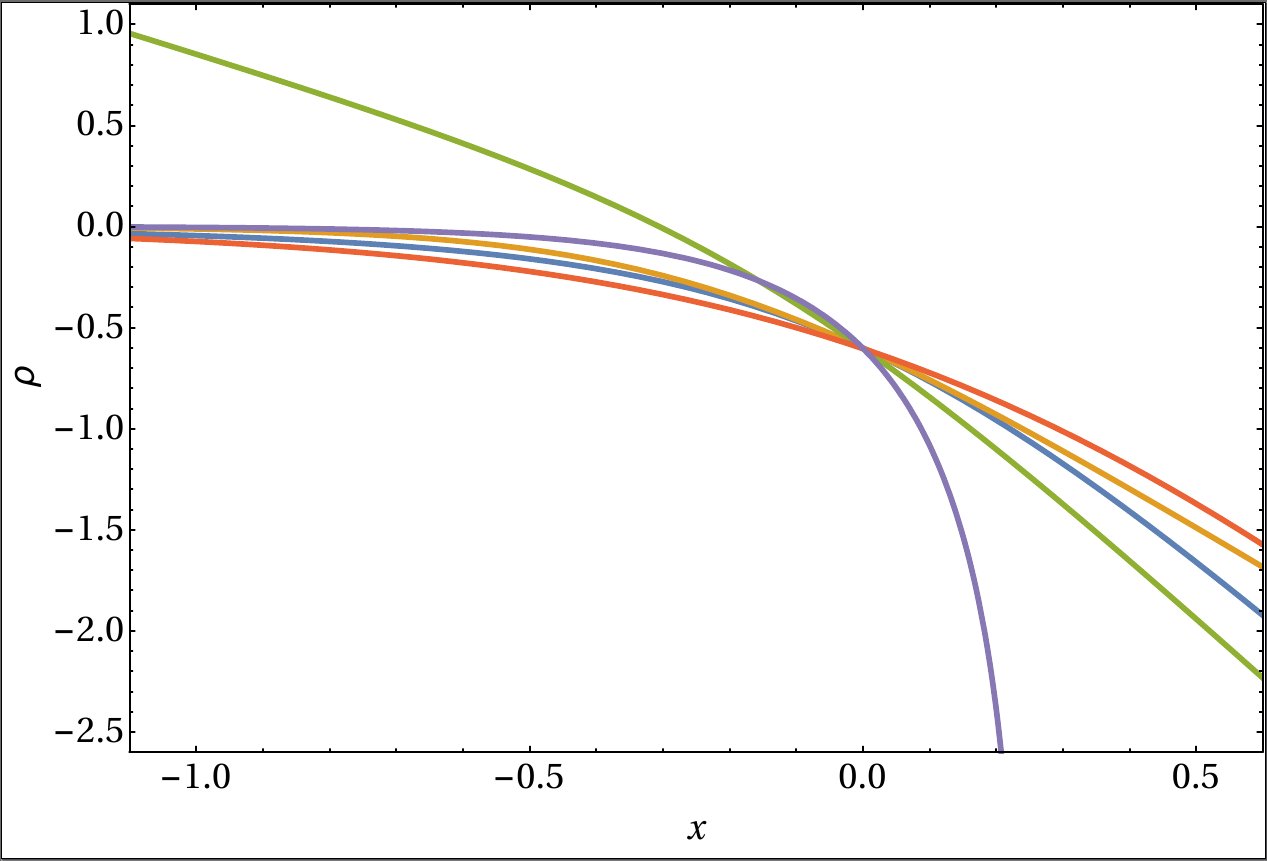}
\caption{DM halos density profiles. NFW {\it (green)}, Burkert-URC {\it (blu)}, fully degenerate fermionic particles {\it (violet)}, Pseudo Isothermal {\it (yellow)} and Binney-URC {\it (red)}.}
\label{fig:haloden}
\end{figure}

This profile represents the (empirical) family of cored distributions (see Fig.~\ref{fig:haloden}). To discriminate 
among them the correct one is, currently, very difficult. It would require a large number of accurate measurements of RCs at
inner radii $r < r_0$. 

\subsection*{Pseudo-isothermal profile}

The PI halo profile $ \rho_{\rm PI}(r)=\frac{\rho_0 r_0^2}{((r^2+r_0^2)} $ is an alternative cored distribution to Eq.~(34).
 This density profile implies that $V_{\rm PI}(r) = {\rm const}$ for $r \gg R_{\rm opt}$, which disagrees with the RC profiles at very
outer radii that show a decline with radius (\citealt{s7}). 

\subsection*{Fermionic halos}

In this scenario there is a strong degeneracy limit for which the DM particles velocity dispersion $\sigma^2_{\rm DM, \min}(\rho)$ has the
minimal value.
This represents the most compact configuration for a self-gravitating fermionic halo (see e.g., \citealt{dnv}). The density profiles of such
fully degenerate halos are universal, depending only on the mass of the configuration: 
\begin{equation}
\rho(x)=\rho_0 \cos^3 \left[\frac{25}{88} \pi x\right], \quad x= r/R_{h} \,,
\end{equation}
where $\rho_0$ is the central DM halo density. This profile is quite peculiar and recognizable in the RCs. 

\subsection*{Zhao halos}

The following density profile (\citealt{Zha}): 
\begin{equation}
\rho(r)=\frac{\rho_{0}}{(\frac{r}{R_{0}})^{\gamma}(1+(\frac{r}{R_{o}})^{\alpha})^{\frac{\beta+\gamma}{\alpha}}} \,,
\label{eq:Zhao}
\end{equation}
where $\rho_{0}$ is the central density and $R_{0}$ the ``core radius'', that, initially, was not proposed for the DM
halo density, is defined by the set of
parameters: $\alpha$, $\beta$, $\gamma$. The case (1, 3, $\gamma$) is sometimes used as a ``cored-NFW'' profile. This is incorrect
because both in the Burkert and in the NFW profiles, the inner regions are not related with the outermost regions, as, instead occurs in the
Zhao model. Moreover, with the latter, we pass from the two free parameters of most of the halo models in the
ballpark, to the five of Eq.~(\ref{eq:Zhao}). This seems in disagreement with observations in spirals, ellipticals and spheroidals that suggest 
that DM halos are one (two)-parameters family. 

\subsection*{Transformed halos}

We want to draw the attention on the profiles which are the outcome of the primordial NFW halos after that these have experienced the effects that it is called baryonic feedback (e.g., \citealt{dc14}). They seem in agreement with those observed around galaxies. However, the collisionless DM paradigm requires that such kind of transformation has occurred in every galaxy of any luminosity and Hubble type and to reach this goal seems extremely difficult. On the other side, the effect of the baryonic feedback to DM halos has to be investigated, no matter what the nature of DM is. In conclusion, a review on this crucial   complex and still on its infancy issue must be a goal future work.

\section{Kinematics of galaxy systems} 
 
A main channel to obtain the DM properties in galaxies is through their kinematics (rotation curves and dispersion
velocities). The analysis could regard individual objects or stacked data of a sample of objects. 

\subsection{The Tully--Fisher and the Baryonic Tully--Fisher} 

\cite{TF} discovered that, in spirals, the neutral hydrogen 21-cm FHWM linewidths $w_{50}$, related, in a disk system,
 to the maximal rotational velocities $V_{\max}$ by: $\log V_{\max} \simeq -0.3 + \log w_{50} -\log \sin i$, with $i$ the
inclination of the galaxy with respect to the l.o.s., correlate with the galaxy
magnitudes $M$
\begin{equation}
M= a \log \Big(\frac{w_{50}}{\sin i} \Big) + b \,,
\label{eq:TF}
\end{equation}
where $a$ is the slope of the relationship and $b$ the zero-point.

 With the availability of a large number of extended RCs, the relation evolved: a radius proportional to the disk length-scale $R_D$ (e.g. 
$R_{\rm opt}$ or $R_{\max}=2.2 \ R_D$) emerged as the reference radius; moreover, the circular velocity 
at this reference radius substituted the linewidth $w$.

It is easy to realize that Eq.~ \ref{eq:TF} just reflects the equilibrium configuration of rotating disks embedded in dark
halos (\citealt{sw}) and that the magnitude $M$ in the relation is the prior for the stellar disk mass. However, it is worth 
going
deeper: in fact, despite that in each spiral the disk and the dark 
components contribute in different proportions to the value of $V(R_{\rm opt}) = (V_d(R_{\rm opt})^2+V_h(R_{\rm opt})^2)^{1/2}$, one finds
that
 $V_d(R_{\rm opt})$ correlates better with magnitudes than $V(R_{\rm opt})$ (\citealt{SFP}). This finding can be understood
in that the latter relationship couples two attributes that pertain exclusively to the stellar disk: its mass, measured kinematically and its luminosity.

\begin{figure}[htb]
\centering
\includegraphics[width=0.8\textwidth]{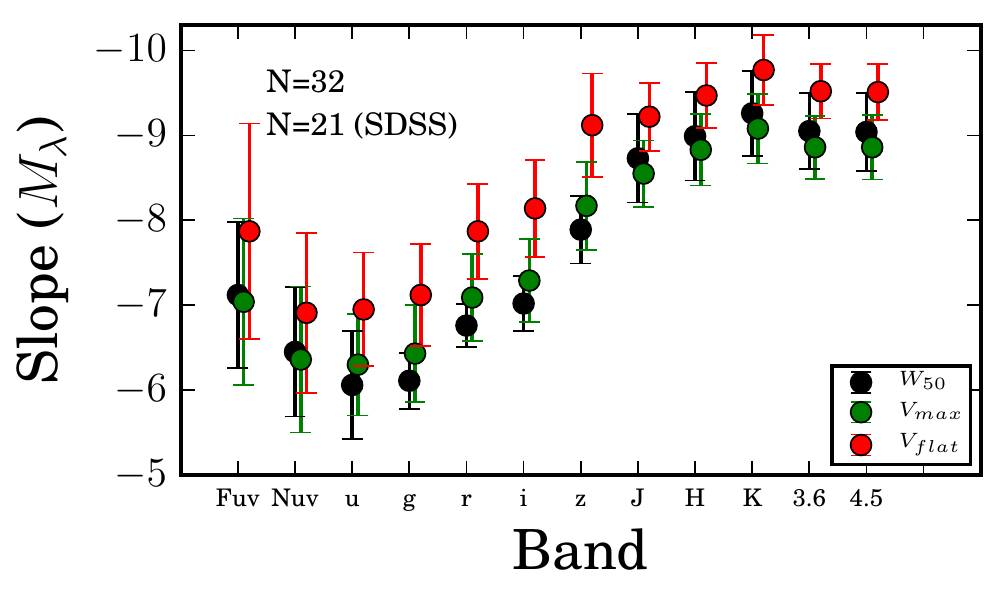}
\includegraphics[width=0.8\textwidth]{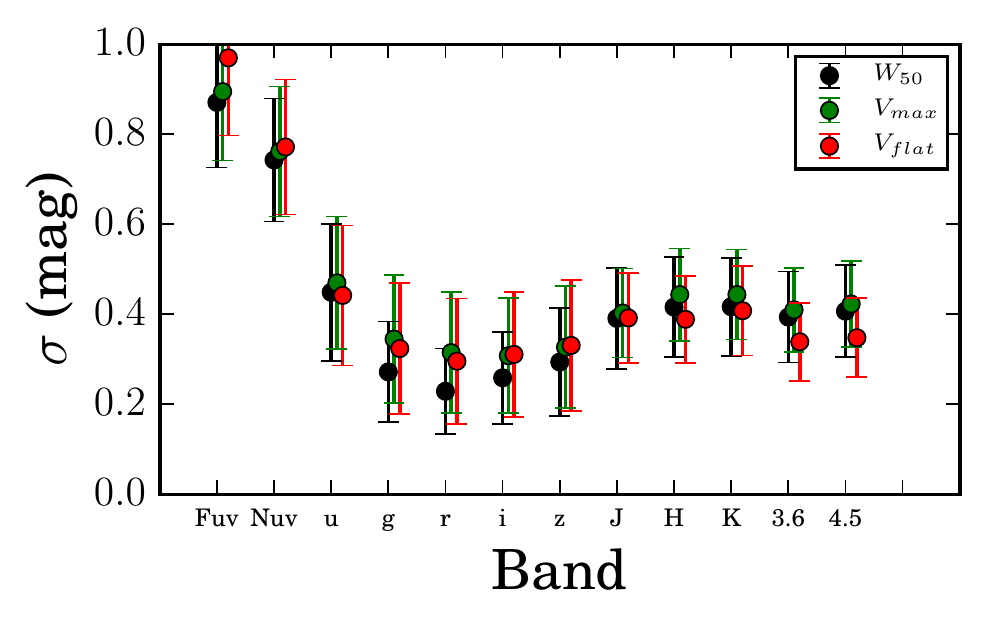}
\caption{The slope and the scatter of the TF relation by adopting different reference velocities and different systems of
magnitude.}
\label{fig:pono}
\end{figure}

The physical meaning of the TF relation as a link between circular velocities and stellar masses has been shown by means of
729 kinematically and morphologically
different galaxies belonging to the SAMI Galaxy Survey 
sample (\citealt{bel}). It has been found:
\begin{equation}
\log V_{2.2} = (0.26\pm 0.017) \log(M_\star/M_\odot) - (0.5 \pm 0.13) \,,
\end{equation}
with $V_{2.2}\equiv V(2.2 R_D)$. Such relationship results in very good agreement with the correspondent one we can derive from
the URC (\citealt{s7}): $\log V_{2.2} = (0.263\pm 0.005) \log(M_\star/M_\odot) - (0.57 \pm 0.05)$.

A recent work (\citealt{pon}) has investigated the statistical properties of the Tully--Fisher relation for a sample of 32
galaxies
with accurately measured distances and with 1) panchromatic photometry in 12 bands: from far ultra-violet to $4.5\,\mu{\rm m}$, and 
2) spatially resolved HI kinematics. For this sample they adopted, in turn, the following reference velocities: the linewidth
$W_{50}$, the maximum velocity $V_{\max}$ and
$V_{\rm flat}$
the average value of the RC in the range ($2$--$5)\, R_D$. With these quantities they constructed 36 correlations, each of them
involving one
magnitude and one
kinematical parameter. They found that the slope of the relationships strongly depends on the band considered and that the
tightest correlation occurs between the $3.6\ \mu{\rm m}$
photometric band magnitude $M_{3.6\, \mu{\rm m}}$ and $V_{\rm flat}$ (see Fig.~\ref{fig:pono}):
\begin{equation}
M_{3.6}= (9.5 \pm 0.3) \log V_{\rm flat} + (3.3 \pm 1.7)
\end{equation}
in good agreement with the value of $8.6 \pm 0.1$ found by \cite{ys} for the slope of the I magnitude of the radial Tully--Fisher relationship at $R=1.2\,R_{\rm opt}$ that becomes $9.6 \pm 0.3$ when translated in the $3.6\ \mu{\rm m}$ band.

\begin{figure}[htb]
\centering
\includegraphics[width=1.0\textwidth]{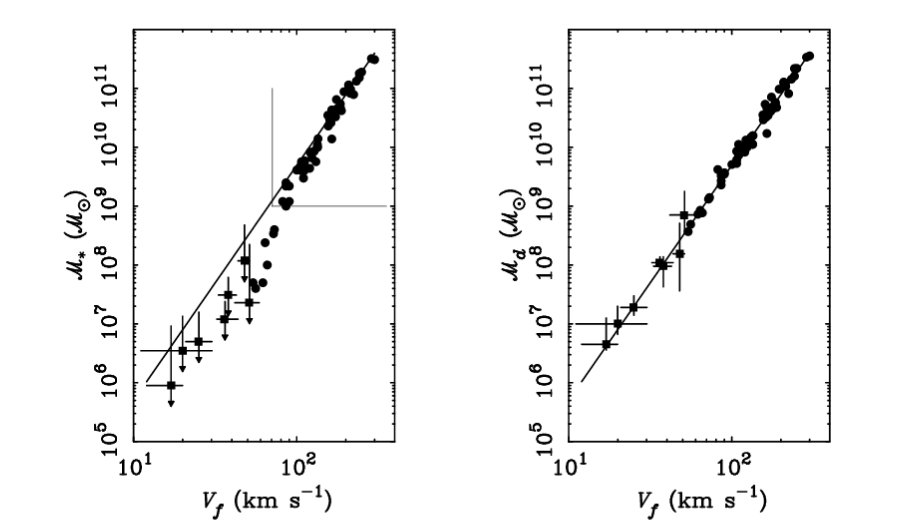}
\includegraphics[width=\textwidth]{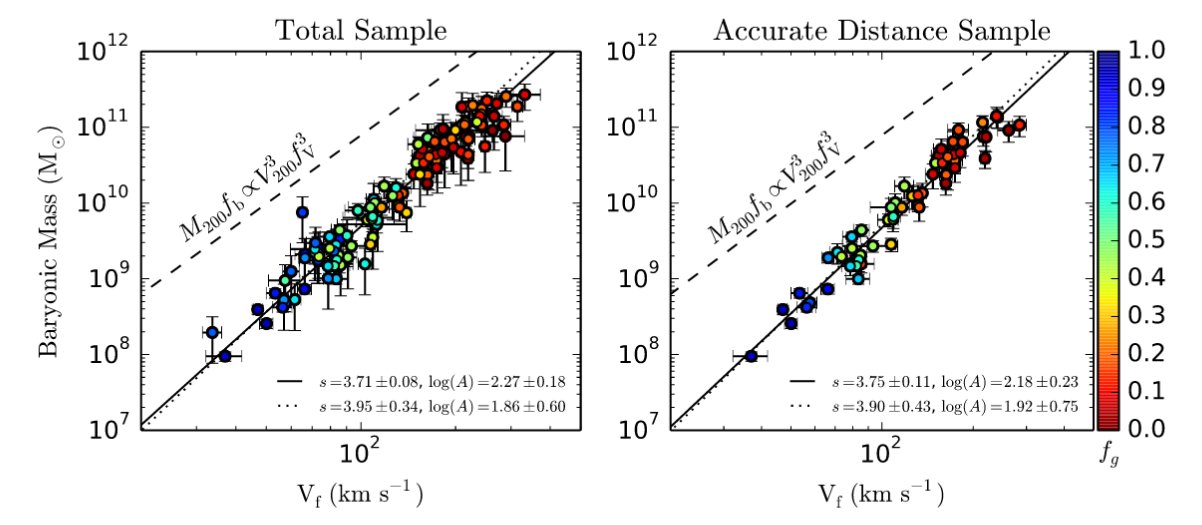}
\caption{ {\bf (top)} The stellar mass {\it (left)} and baryonic {\it (right)} Tully--Fisher relations. 
{\bf (bottom)} The determination of the BTF. 
Images reproduced with permission from [top] \cite{mgc1}, and [bottom] from \cite{LMS}, copyright by AAS.}
\label{fig:BTF}
\end{figure}

\subsection{The Baryonic Tully--Fisher}

\cite{mc00} found a fundamental relationship by correlating the baryonic mass (i.e., the sum of the stellar
and the (HI + He) gas mass) with the reference rotation velocity $V_{\rm flat}$. This Baryonic Tully--Fisher
(BTF) relation has been thorough fully studied and confirmed by several works: (e.g., \citealt{bdj, v01,gel04}). A decisive step
forward in understanding it came from
\cite{mgc1}, who investigated a sample of galaxies with extended 21-cm
rotation curves spanning the range $20\ {\rm km\ s^{-1}} < V_{\rm flat} < 300\ {\rm km\ s^{-1}}$. By using a grid of stellar population
models they estimated the values of the stellar disks masses to which they added those of the HI disks derived by the
observed 21-cm HI fluxes. 
They found: 
\begin{equation}
 M_{\rm bar} = A ~V_{\rm flat}^4 \,; \quad A = 50\, M_\odot\ {\rm km^{-4}\ s^{-4}} 
\end{equation}
(see Fig.~\ref{fig:BTF}). Notice that, by including the HI mass in the galaxy baryonic mass, the BTF becomes log-log
linear and has less intrinsic scatter.

\cite{LMS} investigated the BTF relationship with a sample of 118 disc galaxies (spirals and irregulars) with data of
the highest quality: extended HI high quality rotation curves tracing the total mass distribution and Spitzer photometry at
$3.6\ \mu{\rm m}$ tracing the stellar mass distribution. They
 assumed the stellar mass-to-light ratio ($M_\star /L_{3.6\ \mu{\rm m}}$) to be
constant among spirals and found that the
scatter, slope, and normalization of the relation vary with the adopted $M_\star/L_{3.6\ \mu{\rm m}}$ value, though the
intrinsic scatter is always modest: $\leq 0.1$ dex. The BTF relationship gets minimized for $M_\star/L_{3.6\
\mu{\rm m}} > 0.5$. This result, in
conjunction with the RC profiles of the galaxies in the sample, implies maximal discs in the high-surface-brightness.\footnote{Notice that maximal disks are incompatible with cuspy DM
halos (\citealt{vel85}).}

The BTF relationship slope comes close to 4.0, see Fig.~\ref{fig:BTF}(bottom) and the residuals show no correlation with the galaxy structural parameters (radius or surface
brightness). The above relationship seems to play an important cosmological role, however, the value of its slope
strongly depends on the vagueness in the definition of the reference velocity $V_{\rm flat}$ (\citealt{BSS}).
The DM enters in this relation principally through the value of the dark/ total matter fraction at $R_{\rm flat}$: 
this indicates that the BTF is related more to the disk formation process than to the DM nature.

\subsection{The universal rotation curve and the radial Tully--Fisher}

We can represent all the rotation curves of spirals by means of the universal rotation curve (URC), pioneered in
\cite{rel80}, expressed in \cite{ P91} and set in \cite{PSS} and in \cite{s7}. 
By adopting the normalized radial coordinate $x\equiv r/R_{\rm opt}$, the RCs of spirals are very well described by a universal
profile,
function of $x$ and of $\lambda$, where $\lambda$ is one, at choice, among $M_I$,
the I magnitude, $M_D$, the disk mass and $M_{\rm vir}$, the halo virial mass (\citealt{s7}).

The universal magnitude-dependent profile is evident in the 11 \emph{coadded} rotation curves $V_{\rm coadd}(x, M_I)$
(Fig.~6 of \citealt{PSS} and top of Fig.~\ref{fig:urcmodel}),
built from the individual RCs of a sample of 967 spirals with luminosities spanning their whole I-band range: $-16.3 <
M_I < -23.4$. I-band surface photometry measurements
provided these objects with their stellar disk length scales $R_D$ (\citealt{PS95}).\footnote{See also \cite{lsd} for the
analysis of 24
{\it coadded} RCs obtained from 3500 individual RCs.} 

The coadded RCs are built in a three-step way: \textbf{1)} We start with a large sample of galaxies with RC and suitable
photometry (in the case of \cite {PSS}: 967 objects and suitable I-band measurements). The whole (I) magnitude range is divided in 11 successive
bins centred at $M_I$, as listed in 
Table~1 of \cite{PSS}. \textbf{2)} The RC of each galaxy of the sample is assigned to its corresponding I magnitude bin,
normalized by its
$V(R_{\rm opt})$ value and then expressed in terms of its normalized radial coordinate 
$x $. \textbf{3)} The double-normalized RCs $V(x)/V_{\rm opt}$ curves are coadded in 11 magnitude bins and
in 20 radial bins of length 0.1 and then averaged to get: $V_{\rm coadd}(x, M_I)/ V_{\rm coadd}(1, M_I)$, the points with
errorbars in Fig.~\ref{fig:urcmodel}. The 11 values of $ V_{\rm coadd}(1, M_I)$ are given in Table~1 of \cite{PSS}. The RCs 
are usually increasing or decreasing. Simplifying, they increase when they are dark matter dominated or always for $r< R_D$ and 
decrease for $r>2 R_D$ when they are disk 
dominated.\footnote{We stress that only the RCs with $190\ {\rm km/s} < V_{\rm opt} < 230\ {\rm km/s}$ and in the radial range $1 \, R_D <R<4 \, R_D$ can be considered flattish.} 
The recent finding of RCs of six massive star-forming galaxies that, outside $R_{\rm opt}$, decrease with
radius (\citealt{GEL})
has been considered very surprising. Rightly, it has been proposed that this trend arises because this high-redshift
galaxy population was strongly baryon dominated. However, while the importance of such objects in the cosmological context
is obvious, there is a presence, also in the local Universe, of many baryon dominated decreasing RCs. This was first
drawn to the attention by \cite{P91} and, moreover, it is inbuilt in the URC.

The URC is the analytical function devised to fit the stacked/coadded RCs $V_{\rm coadd}(x,M_I)$.
In principle, it could be any suitable empirical function of ($x,M_I$), the idea of \cite{PSS} was to choose, as 
fitting function, the
sum in quadrature of the velocity components to the circular velocity. Namely, the Freeman stellar disk with one free
parameter, its mass $M_D$
and the dark halo with an assumed profile and two free parameters, the central density $\rho_0$ and the core radius
$r_0$. Then, 
the data $V_{\rm coadd}(x,M_I)$ are fitted by the $V_{\rm URC}$ universal function: 
\begin{equation}
V^2_{\rm URC}(x,M_I) \equiv V^2_{\rm URCd}(x;M_D(M_I)) + V^2_{\rm URCh}(x;\rho_0(M_I), r_0(M_I)) 
\label{eq:urc}
\end{equation}
The first component of the RHS is the standard Freeman disk of Eq.~(\ref{eq:freeman}), the second is the B-URC halo of Eq.~(\ref{eq:Burk}). In dwarf galaxies, a HI term must be included (\citealt{ks17}).

The excellent fit (see Fig.~\ref{fig:urcmodel}) has led us to the validation of the URC idea: there exists a universal function of (normalized) radius and luminosity that well fits the RC of any spiral galaxy
(see \citealt{s7}).\footnote{In short: the 
variance of  $V(x,L)$ is negligible, i.e., the r.m.s.\ of the values of the RCs  in galaxies of same luminosity $L$ and at the same radius $x$ is
negligible.}

\begin{figure}
\centering
\includegraphics[width=1.0\textwidth]{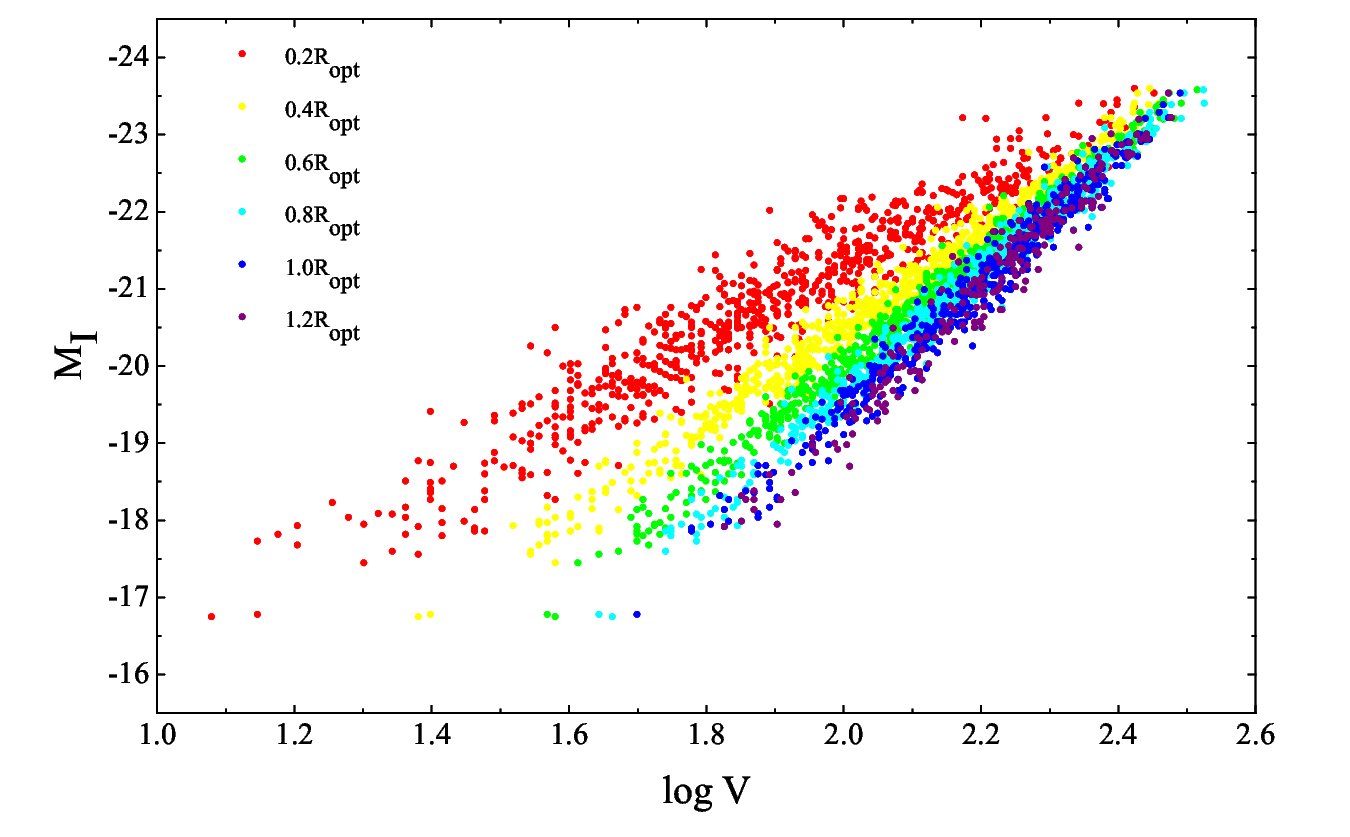}
\caption{The radial TF. The variation of the slopes $a_i$ with $r_i$ is very evident. 
Image reproduced with permission from \cite{ys}, copyright by the authors.}
\label{fig:TFR}
\end{figure}

The radial Tully--Fisher is a relationship \emph{on} the URC surface, orthogonal to the various RCs
(\citealt{ys}; see Fig.~\ref{fig:TFR} top).
At different galactocentric distances, measured in units of the optical size,
$r_i\equiv i~ R_{\rm opt}$ ($i=0.2, 0.3, \dots, 1$), a family of independent Tully--Fisher-like relationships emerges:
\begin{equation}
M_{\rm band} = b_i + a_i \log V(r_i) \,,
\end{equation}
with $M_{\rm band}$ the magnitude in a specific band, often the (R, I)-bands. The RTF
has a very small r.m.s.\ scatter,  at any radius smaller than that of the classical TF. It also shows a large systematic variation of the slopes $a_i$ with $r_i$ that 
range, across the disk, between $-4$ and $-8$. This variation, in cooperation with the smallness of the scatter, indicates that the fractional amount of dark matter inside the optical
radius is luminosity-dependent (\citealt{ys}). 

It is important to stress that, given a sample of RCs, the RTF relationship provides us with an independent method of deriving (if it exists) the underlying coadded RCs and, in turn, the relative URC. \cite {ys}, in fact, have shown that samples with a similar $a_i$ vs  $r_i $ relationship have also  similar  $V_{\rm coadd}(x, {\rm magnitude})$.  This has been applied to the large samples of \cite{co97} and \cite{vel04,vel04b} with the result of finding the same RTF discovered  in the \cite{PSS} sample (see Fig.~8 of \citealt{ys}) and, then, finding very similar coadded RCs.

\section{The dark matter distribution in disk systems} 

The general pattern is the following: spirals show a reference radius $R_T(L_I)$ whose size ranges from $1$ to $3\,R_D$
according to the galaxy luminosity (see Fig.~8 of \cite{PSS} and \cite{pw00}); inside
$R_T(L_I)$ the ordinary baryonic matter fully accounts for the RC, while, for $R>R_T(L_I)$,
is instead unable to justify the profile and the amplitude of the RC. 

\subsection{Dark matter from stacked RCs}

Very extended individual RCs and virial velocities $V_{\rm vir}\equiv (G
M_{\rm vir}/R_{\rm vir})^{1/2}$ obtained in \cite{sh06}, further support the URC paradigm and help determining
the universal velocity function out to the virial radius (\citealt{s7}). It is important to stress that the $V_{\rm URC}$ function 
(and the relative mass model) has, in principle, three free parameters: the
disk mass and two quantities related to the DM distribution
(the halo
central density $\rho_0$ and the
core radius $r_0$). These are obtained by best fitting the $V_{\rm coadd}(x,M_I)$)
and found to be correlated among themselves and
with the luminosity. So, the RCs and the related gravitational potential of spirals belong to a family ruled by
1-parameter that we can choose among many
possibilities, e.g., the halo
mass, which is a combination of $\rho_0$ and $r_0$ and it ranges in spirals as: $3
\times 10^{10} \, M_{\odot} \leq 
M_{\rm vir} \leq 3\times10^{13} \, M_{\odot} $. 
 
\begin{figure}[htb]
\centering
\includegraphics[width=1.0\textwidth]{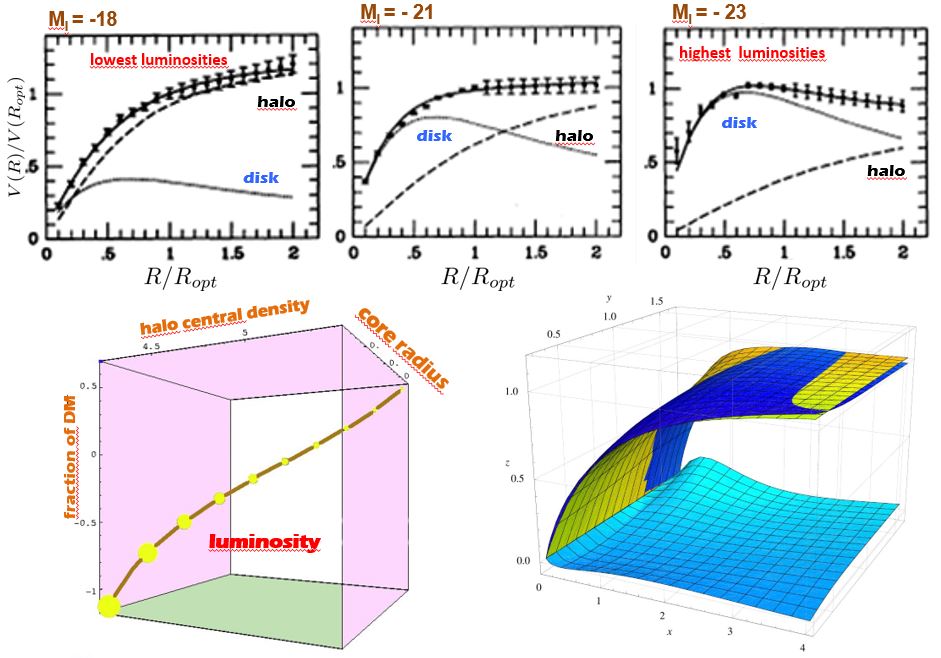}
\caption{ {\bf (top)} The URC best-fit models of the coadded RCs {\it (points with errorbars) } (\citealt{PSS}). 
It is shown: the bin magnitude $M_I$, the disk/halo contributions {\it (dotted/dashed lines)} and
the resulting URC {\it (solid line)}. {\bf (bottom left)}. The 
4-D relationship among the central DM density, its core radius in units of $R_{\rm opt}$, the DM fraction at $R_{\rm opt}$ and the
galaxy I-luminosity (proportional to the area of the circles). {\bf (bottom right)}. The URCs from \cite{PSS},
{\it (yellow)} and from \cite{catel06} {\it (blue)}. Legenda: $x \equiv R/R_D$, $y \equiv \log (M_{\rm vir}/(10^{11}\,M_\odot))$, $z \equiv
V(x)/V(3.2)$. The differences between the two URCs are also indicated.}
\label{fig:urcmodel}
\end{figure}

\subsection{Dark matter from individual RCs}

The study of individual RCs is very similar to that of the stacked ones as regard to their mass modelling, but
it is complementary to it with respect to the data analysis. Moreover, in the core-cusp issue, the individual RCs
have a special role: stacked RCs of spirals, as seen in the previous section, points unambiguously to a cored distribution, but cannot indicate to us whether this is a sort of average property of the entire population of
spirals 
or a property of any single object. Only the analysis of fair number of individual RCs of systems of different
luminosity and Hubble types can answer to this.

It is worth pointing out that, in the first 15 years since the DM discovery from the profiles of the RCs, the latter have always been
reproduced by models including a Freeman disk, a bulge and a dark halo with the {\it cored} Pseudo Isothermal distribution (e.g., \citealt{cf85, vel85}). 
It is well-known that in the current $\Lambda$CDM cosmological scenario the dark matter halos have a
very specific and universal cusped density distribution (\citealt{nfw}). A debate has arisen on the level of the observational
support for such profile (\citealt{deb01, s01, ge04, si05, sp05, kdn, db08, oh11, ad14} to name a few, reviews on this issue:
\citealt{BB, dB10}).\footnote{Let us stress that, in this issue, non circular motions in the RCs play a minor role 
(\citealt{oh,ge05}.}

It is important to remark that the DM cores could come {\it ab initio} from the structural properties of the (exotic?) DM
particles or been created, over all the Hubble time, by dynamical processes occurring inside the galaxies. 

\begin{figure}
\centering
 \includegraphics[width=0.61\textwidth]{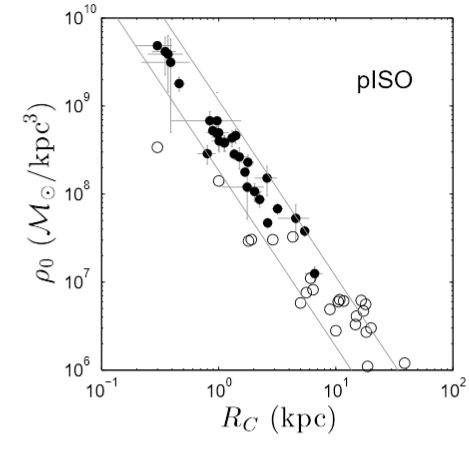}
\caption{The relationship between the size of the DM core radius $R_C$
and the value of the central dark matter density $\rho_0$. 
Image reproduced with permission from \cite{Mel13}, copyright by ESO.}
\label{fig:pih}
\end{figure}

\cite{Mel13} devised and applied to a sample of 30 spirals, a method to decompose the rotation curves in its dark and luminous components. The method exploits the vertical velocity dispersions of the disk stars $\sigma_z$ (see Sect.~6.3).  By reminding that $R_{\max}\equiv 2.2~ R_D$ is the radius
where the disk velocity component has its maximum, they found: $(V_d(R_{\max})/V(R_{\max}))^2= 0.57\pm 0.07$, with a
dependence on galaxy luminosity: in their velocity models, at $R_{\max}$, the disk component prevails over the dark
component in the biggest spirals, while, it is very sub-dominant in the smallest ones. 

They also modeled the dark matter halos with either a PI or a NFW profile and found the former distribution performing
something better and showing a tight $\rho_0$ vs.\ $r_0$ relationship, very similar to that found in spirals by means of a
different analysis (see Fig.~\ref{fig:pih}).

A recent study of NGC5005 (\citealt{Rich15}) can be considered as a test case investigation of the mass
distribution in spirals obtained by means of multi-messenger observations. These included images taken
at $3.6\ \mu{\rm m}$ from the Spitzer Space Telescope, B and R broadband and H$\alpha$ narrowband observations. Very
Large Array (VLA) radio synthesis observations of neutral hydrogen provided the HI surface density and the
kinematics. Spectroscopic integral field unit observations at WIYN 3.5-m telescope provided the ionized gas
kinematics in the inner region. The surface brightness has been carefully decomposed in its disk and bulge
component. The modelling of the composite high resolution rotation curve clearly favors a PI DM halo,
with core radius of $2.5\pm 0.1\ {\rm kpc}$, over the corresponding NFW configuration. 

\begin{figure}[htb]
\centering
 \includegraphics[width=0.8\textwidth]{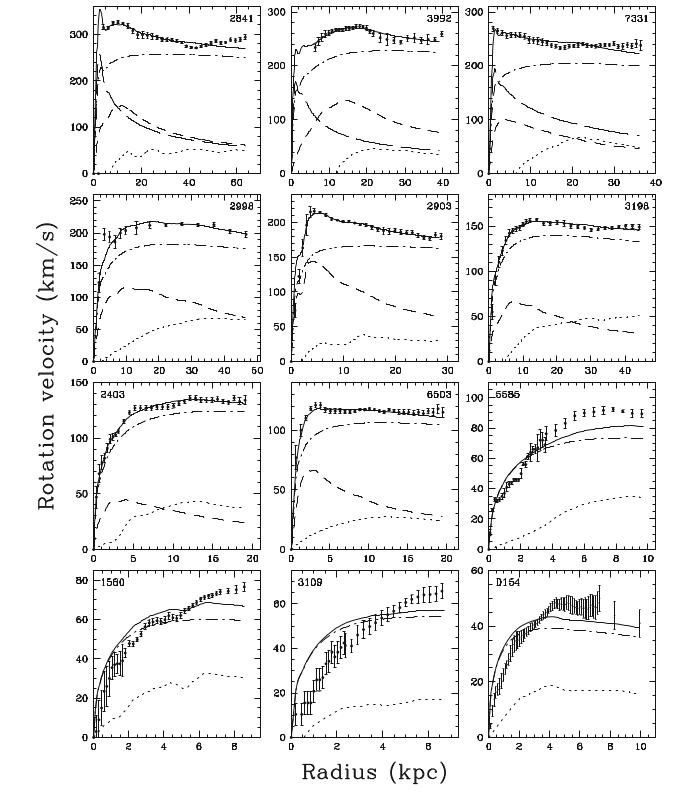} 
\caption{Maximum disc best-fits {\it (solid lines)} to the RCs {\it (dots with errorbars)}. Also shown the contribution of
gas, disc, bulge,
and PI dark halo {\it (dotted, short dashed, long dashed, dash-dot lines)}. 
Image reproduced with permission from \cite{bp15}, copyright by the authors.}
\label{fig:bot}
\end{figure}

\cite{bp15} obtained high resolution kinematics for sample of 12 galaxies, whose luminosities are distributed
regularly
over a
range spanning several orders of magnitude. They found that models with maximum disks, 
 cored DM halos and a unique value of the mass-to-light ratio, i.e. $M_D/L_R=1.0$, fit very well all the RCs, see
Fig.~\ref{fig:bot}. NFW DM halos, independently of the baryonic distribution, cannot fit the RCs of the least massive
galaxies of the
sample, while, for the most massive ones, the best fitting values of the structural parameters of the NFW +stellar/HI disks
models, namely the halo concentration and mass and the mass-to-light ratio of the stellar disk, take often non-physical values.
 
The Spitzer Photometry \& Accurate Rotation Curves sample includes 175 nearby galaxies with 
surface photometry at $3.6\ \mu{\rm m}$ and high-quality rotation curves. This sample spans a broad
range of morphologies (S0 to Irr), luminosities ($\sim 5$~dex), and surface brightness ($\sim 4$~dex). 
 These data have been used by \cite{lms16} in order to build the mass models of the galaxies.
They adopted the specific value of 0.5 for the stellar mass-to light ratio in the $3.6\ \mu{\rm m}$-band as suggested by stellar
population models and found that $V_{\rm bary}/V$ varies with luminosity and surface brightness: the stellar disks in high-mass,
high-surface-brightness galaxies are nearly maximal, while in low-mass, low-surface-brightness galaxies they are very 
submaximal. Moreover, in these galaxies, the cored DM halo + (high mass) stellar disk model, generally, reproduces the sample RCs
very well, differently from the cuspy halo + (low-mass) stellar disk model that often shows a bad fit and/or non-physical
values for the parameters of the mass model.

The mass distribution of 121 nearby
objects with high quality optical rotation curves has been recently derived from the Fabry--P\'erot kinematical GHASP survey of spirals and irregular galaxies (\citealt{kal18}). These galaxies cover all
morphological types of spirals and have infra-red $3.6\ \mu{\rm m}$ emission measurements, good tracers the old stellar population. Combining the
kinematical and the surface brightness data they obtained the mass models once they assumed a specific DM halo density
profile. They considered the PI cored profile and the Navarro--Frenk--White cuspy profile. The value of the 
$M_D/L_{3.6}$ for the stellar disc was obtained for each
objects in two different ways: 1) from the stellar evolutionary models and the WISE W$_1$-W$_2$
colours, 2) from fitting the RC. Both approaches found that: (i) the rotation curves of most galaxies are
better fitted with a cored rather than with a cuspy profile, (ii)
there are luminosity/Hubble type dependent relationships between the parameters of the DM and those of the luminous
matter. In detail, in the PI halos framework they found that core radius $\propto$ (central DM halo density)$^{-1}$, in very good
agreement with \cite{KF04, Doet}. In the NFW framework they found a very strong dependence of the concentration 
on the halo virial mass, in disagreement with the outcome of N-body simulations (e.g., \citealt{k}).

\subsubsection{The Galaxy}

The investigation of DM distribution in our Galaxy is clearly important under many aspects, although
 it is made difficult by our location inside it. The stellar component can be modeled as a Freeman
exponential thin disk of length
scale $R_ D=(2.5\pm 0.2)\ {\rm kpc}$ (e.g., \citealt{ju08}). 

\begin{figure}[htb]
\centering
\includegraphics[width=0.93\textwidth]{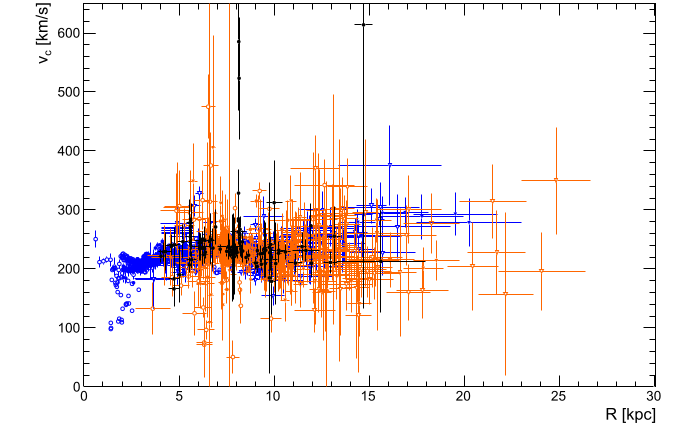}
\caption{Rotational velocities in the Milky Way derived from gas and stellar kinematics {\it (blue, orange)} and masers measurements
{\it (black)}). Notice measurements with huge uncertainty. 
Image reproduced with permission from \cite{pi17}, copyright by the authors.}
 \label {fig:fig14.png}
\end{figure}

Very precise measurements of position
and proper motion of maser sources (\citealt{ho12}) provide us with a reliable solar galactocentric
distance of $R_\odot =
8.29 \pm 0.16\ {\rm kpc}$ and a circular speed, at $R_{\odot}$, of $V(R_\odot)=(239 \pm 5){\rm km/s}$. Adopting these
values, for
$R<R_\odot$, we can transform the available HI disk terminal velocities $V_T$ into circular velocities $V(R)$:
$
V(R/R_\odot)=V_{T} (R/R_\odot)+\frac{R}{R_\odot} V_\odot
$
(see \citealt{mm11, ns13} and references inside). For $R>R_\odot$ out to
$\sim 100\ {\rm kpc}$, the MW circular motions
are inferred from the kinematics of tracer stars in combination with the Jeans equation (\citealt{Xet, bet}).\footnote{The raw kinematical data needed to build the Galaxy RC can be found \cite{pi17}, see Fig.~\ref{fig:fig14.png}.} In \cite{sof17} the issue of the RC of the MW compared  with those of spirals of similar luminosity is discussed. 

The mass 
model of the MW is that of any other spiral: it includes a central bulge, a stellar disk, an
extended
gaseous disk and all these components are embedded in a spherical dark halo (see \citealt{cao81, cu10, ns13, so13}). 
 As regard to the latter, in a number of studies,  the available kinematics is not able to discriminate between the cored
and a cusped DM halo profiles (e.g., \citealt{cu10,cu12}). 

\cite{ns13} have alternatively assumed a B-URC
and a NFW DM halo profile. They fitted the resulting velocity models to the available kinematical data: HI terminal
velocities,
circular velocities as recently estimated from maser star forming regions and velocity dispersions of stellar halo tracers in
the outermost Galactic regions. They found, for the first model, the following best fit values: $\rho_0 = 4 \times 10^{7}\,
M_\odot/{\rm kpc}^3$, $r_0 =10\ {\rm kpc}$ and $M_D= 6\times 10^{10}\, M_\odot $, $M_{\rm vir}=1.2 \times 10^{12}\, M_\odot$ that coincide with
those of the URC with the same virial mass and optical radius. The mass model with NFW halo
profile fits quite well the dynamical data, however, the resulting best fit value for the concentration parameter $c$ is: 
$c=20
\pm 2$, higher
than the
predicted value from only dark matter $\Lambda$ CDM simulations. Similar findings were obtained also by
\cite{cu10,cu12,dea12}.


\subsection{Low surface brightness galaxies}

\begin{figure}[htb]
\centering
\includegraphics[width=0.88\textwidth]{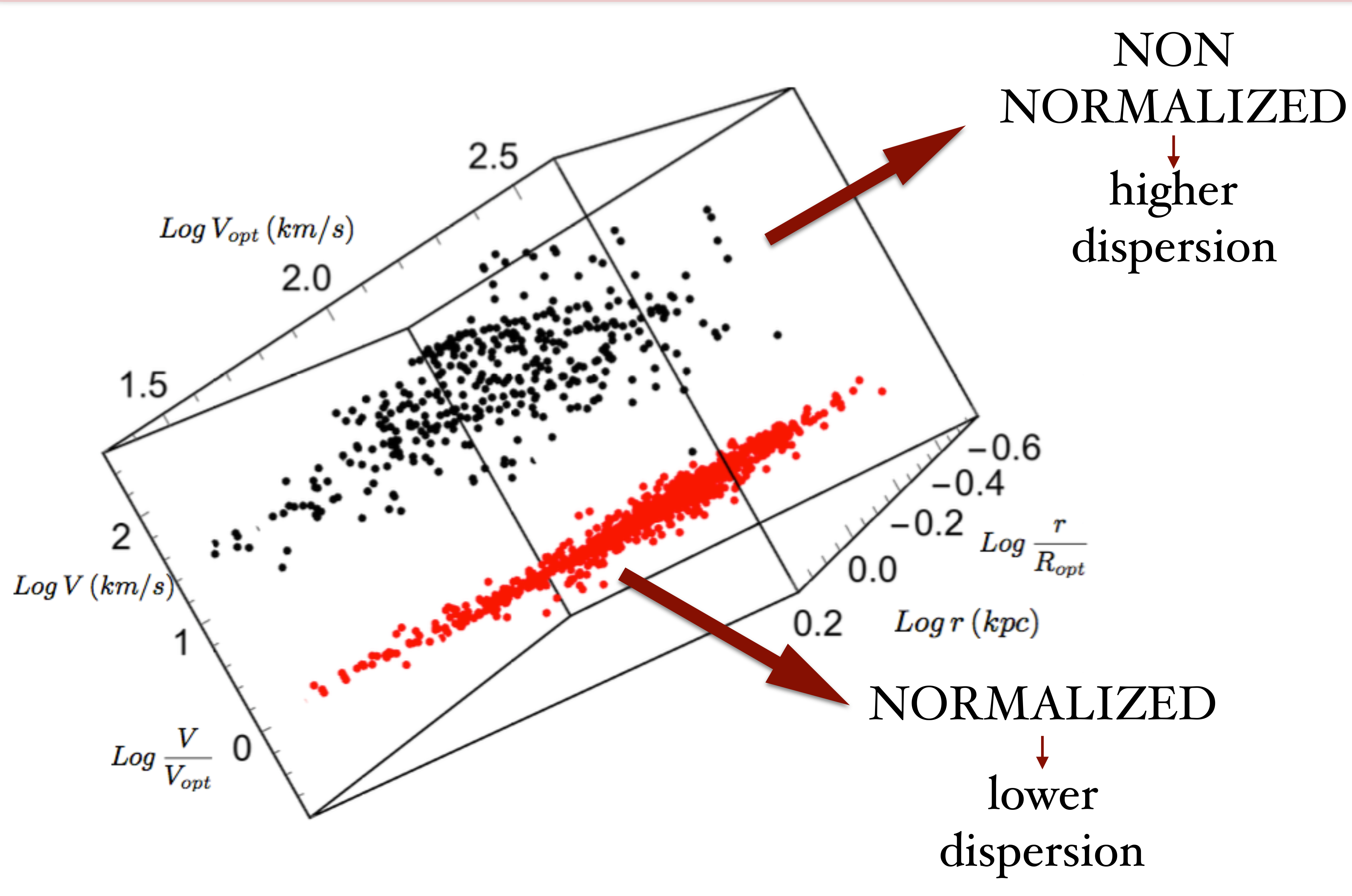}
\caption{The rotation curves of the LSBs sample of \cite{dps} in physical {\it (black)} and normalized units {\it (red)}}
\label{fig:lvlr}
\end{figure}

There is a limited number of recent studies on the RCs of LSB galaxies, although some of these objects appear in
well studied samples of disk systems discussed in the previous sections. In
LSB the 21-cm HI line provides us with the main observational channel probing the gravitational field: radio telescopes 
only now reach sufficient spatial resolution and sensitivity to map small and faint objects like LSB.\footnote{SKA will exponentially increase the amount of available kinematics. }

\cite{dps} applied to LSBs the concept of the stacked analysis of RCs that in spirals has led us to the 
URC. They investigated, in a sample of 72 objects with available 
rotation curves and infrared photometry, the distribution of the baryonic and the dark matter components.
The galaxies were divided in five velocity bins according to their increasing values of $V_{\rm opt}$.
Noticeably, when we plot them in physical units: $\log V(\log r)$, they show a great diversity: objects with a same 
maximum velocity possesses very different RC 
profiles, see Fig.~\ref{fig:lvlr}. Instead, when we adopt the specifically normalized units: $x\equiv r/R_{\rm opt}$ and $v(x)
=V(x)/V(1)$, the rotation curves $\log 
v(\log x)$ of each velocity bin are all alike, see Fig.~\ref{fig:lvlr}, probing, as in spirals, the idea that by stacking
and by coadding diverse RCs, we get a
3D universal profile, i.e., a surface function of $x$ and of one galaxy structural quantity, e.g., $\log V_{\rm opt}$.  The diversity in the RCs is caused by the presence of another structural parameter in the mass distribution
 that the stacking processes and the double-normalization neutralize. From the double-normalized velocities, five coadded RCs have been built: $V_{\rm coadd}(x,V_{\rm opt})$. They are very well fit by the spirals URC velocity
profile $V_{\rm URC}(x;\rho_0, r_0,M_D)$
(see \ref{eq:urc}) see Figs.~5--6 of \cite{dps}.

The resulting URC of LSB galaxies (Fig.~18 of \citealt{dps}) implies that the B-URC halo parameters $\rho_0 $ and $r_0$
connect with $R_D$ and $M_D$ in a way similar to that found in spirals (\citealt{dps}). Moreover, also in these objects we
find: $\rho_0~ r_0 \sim 100\, M_\odot pc^{-2}$

\begin{figure}[htb]
\centering
\includegraphics[width=1.0\textwidth]{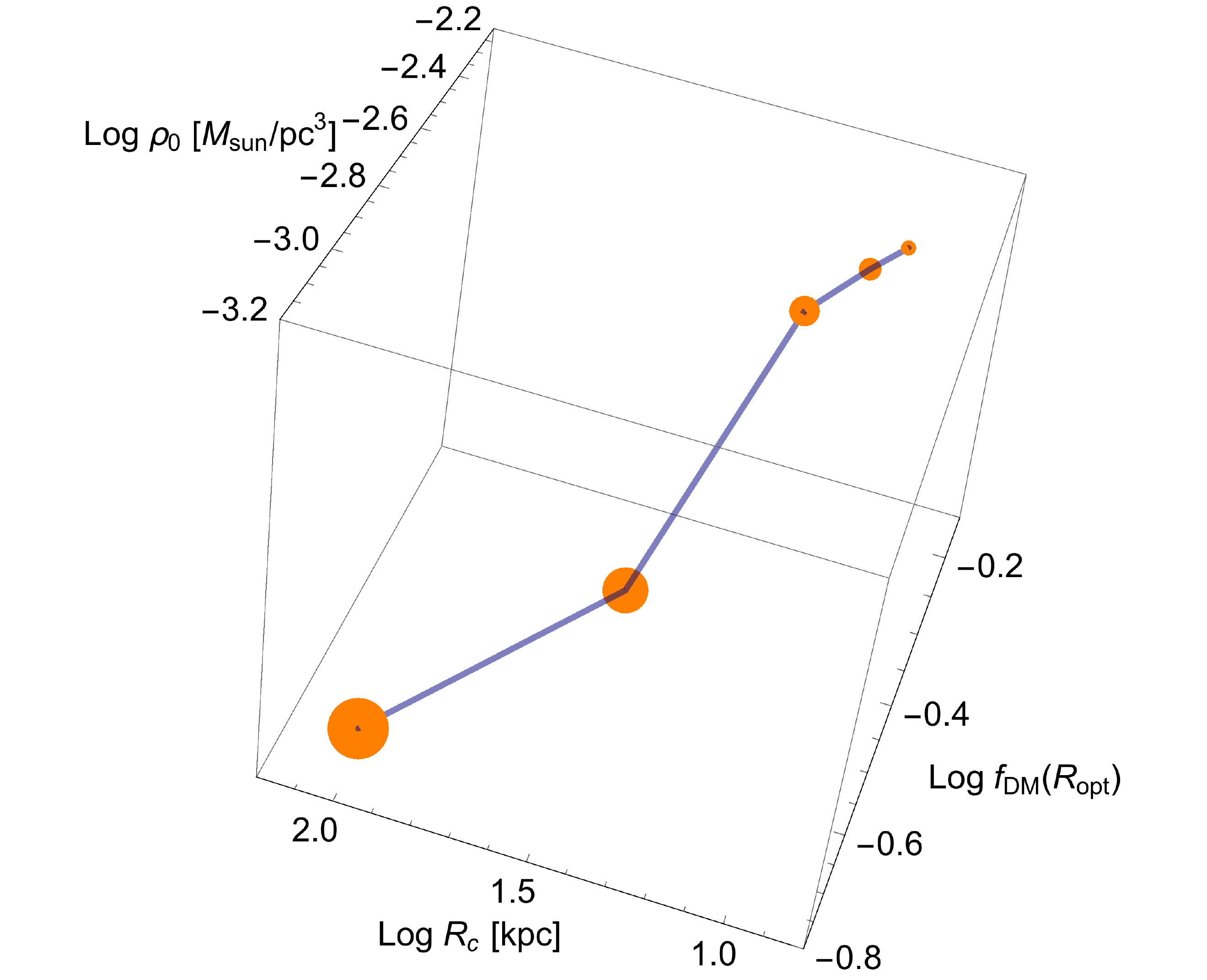}
\caption{The 4D relationship in Fig.~\ref{fig:urcmodel} (bottom, left) for low-surface-brightness galaxies (\citealt{dps}). Legenda: $R_c\equiv r_0$.}
\label{fig:urclsb}
\end{figure}

Remarkably, in LSBs, the URC, expressed in normalized radial units, has {\it two} independent parameters: one, as in
spirals, is the stellar disk or the halo mass, the second is the compactness, either of the dark halo or of the luminous
disk; in fact, a tight correlation between these two quantities emerges (without a plausible physical
explanation) (see Fig.~\ref{fig:comp}).

\subsection{Dwarf disks}

\begin{figure}[htb]
\centering
\includegraphics[width=0.64\textwidth]{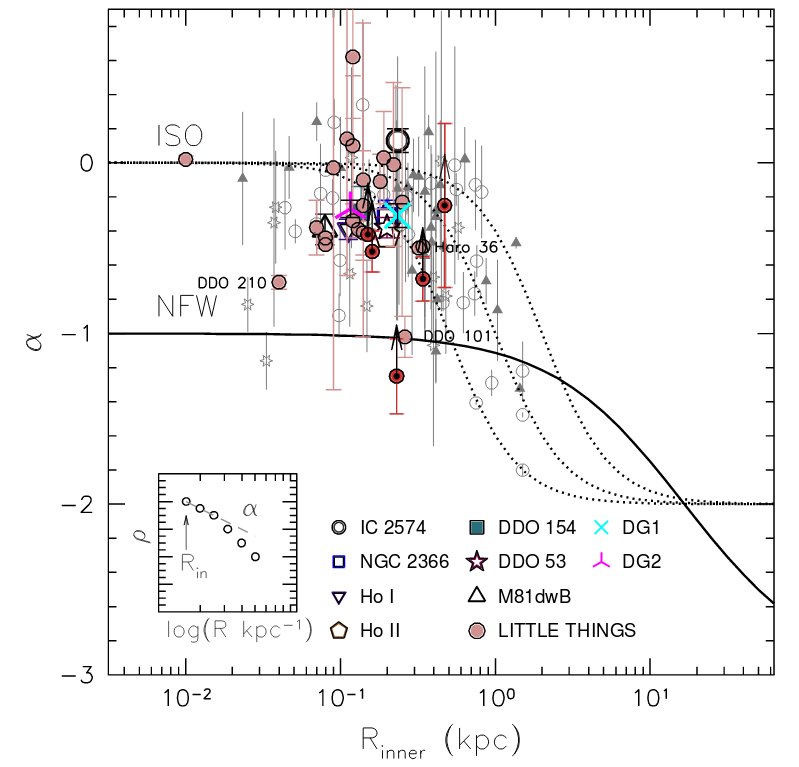}
\caption{The slope $\alpha$ of the DM density: $\rho_{\rm DM}\propto (r/R_{\rm inner})^\alpha $ 
with $R_{\rm inner}$ the innermost radius with velocity measurement. Also shown the predictions for halos of 
mass $10^{10} \, M_\odot$ and a pseudo-isothermal (ISO)
or NFW halo profile. 
Image reproduced with permission from \cite{oel15}, copyright by AAS.}
\label{fig:LITTHI}
\end{figure}

\cite{oel15} have investigated 26 high-resolution rotation curves of dwarf (irregular) disk ({\bf dd}) galaxies from
LITTLE THINGS sample, a high-resolution VLA HI survey of nearby dwarf galaxies. The rotation curves were decomposed
into their baryonic and DM contributions in a very accurate way: in these objects, the first component
is much less important than the second. Generally, the RCs of {\bf dd}s are found to increase with radius out to several
disk length scales. Furthermore, the logarithmic inner slopes $\alpha$ of their DM halo densities are 
very high: $\langle \alpha \rangle =-0.32 \pm 0.24$, in disagreement with the prediction of cusp-like NFW halos 
$\langle \alpha \rangle_{\rm NFW}<-1$ (see Fig.~\ref{fig:LITTHI}). This result is confirmed also by the full mass modelling when it is possible
to accurately perform it. 

\begin{figure}[htb]
\centering
 \includegraphics[width=1.0\textwidth]{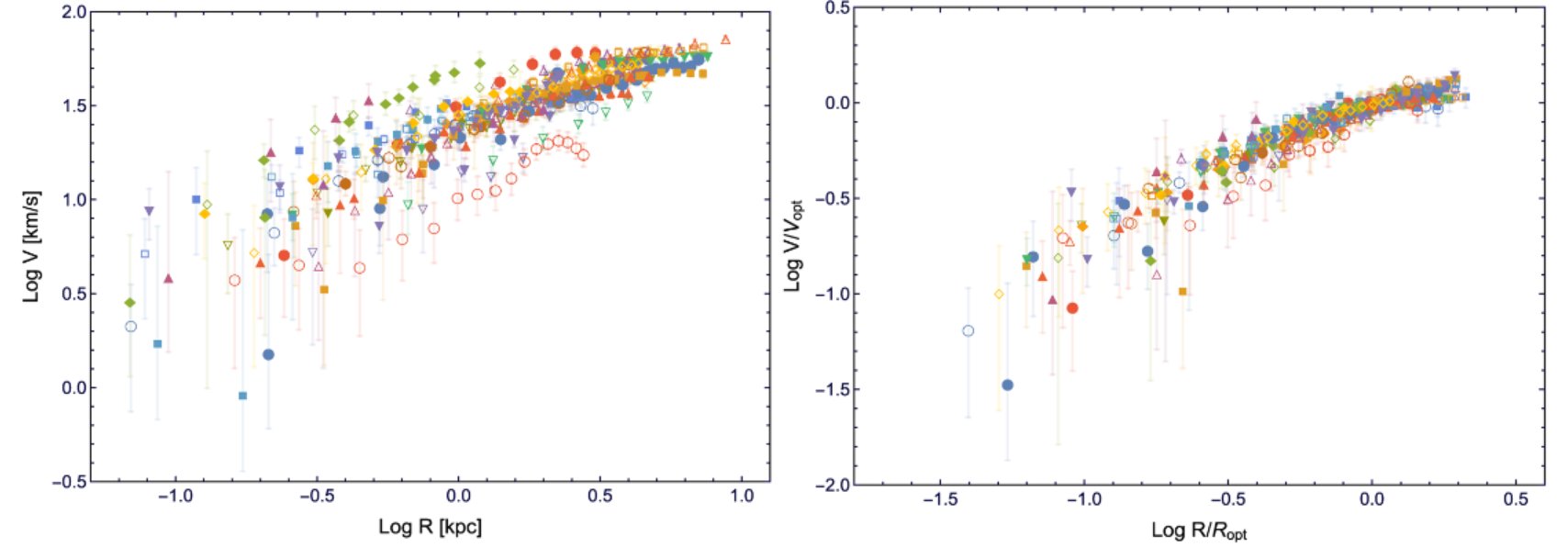}
\caption{{\bf dd}s. The 36 RCs in physical units
{\it (left)} and in two-normalized units {\it (right)}. In \cite{ks17} one finds the $R_{\rm opt} $ vs. $L_K$
relationship whose scatter is responsible for the evident diversity of the various RC profiles when they are  expressed in
physical units. 
Image reproduced with permission from \cite{ks17}, copyright by the authors.}
\label{fig:binorm}
\end{figure}

\cite{ks17} investigated a sample of 36
objects with good quality rotation curve drawn from the Local Volume Sample. They found that, although several objects have a
RC suitable for individual mass modelling, on the whole, the stacked analysis yields very important results. They found that,
despite variations in luminosities of $\sim 2$~dex and, above all, despite a great diversity in their rotation curves profiles
$V(R)$, 
when radii and velocities are normalized by ($R_{\rm opt}$, $V_{\rm opt}$) the RCs look all alike (see
Fig.~\ref{fig:binorm}) and lead to what can be considered as the low-mass continuation
of $V_{\rm coadd}(x, M_I)$, the coadded RCs of spiral galaxies. This finding addresses the ``diversity problem'' (\citealt{O15});
it confirms that dwarf disk galaxies, with the
same maximum circular velocity, exhibit large differences in their inner RC profiles and then, in their inferred
DM densities. However, this pattern disappears when the relevant quantities are expressed in normalized units (see Fig.~\ref{fig:binorm}). The reason is that these galaxies have a large scatter in the luminosity vs.\ size relationship (see
\cite{ks17}) which, exactly as in LSB, gets neutralised by the normalization procedure performed while building the $V_{\rm coadd}$. Of course the issue
itself does not disappear, but it actually thickens and manifests itself as arisen from the strong correlation between the
distribution of dark and
luminous dark matter and from the presence in these objects of an additional structural quantity: the compactness
$C_{\star}$ (see later) belonging to the luminous world, but independent, by construction, of the galaxy luminosity (see \citealt{ks17}).

Let us stress that, differently from
spirals and LSBs, we need just one $V_{\rm opt} $ to represent all {\bf dd}s double-normalized RC's, laying in the range 
$10\ {\rm km/s} <V_{\rm opt}< 80\ {\rm km/s}$: in fact, all their (double normalized) velocity profiles are almost identical. 
The velocity modelling starts from the coadded RC: $V_{\rm coadd}(R/R_{\rm opt}, \langle V_{\rm opt} \rangle)$, with $\langle V_{\rm opt} \rangle =40\ {\rm km/s}$.
As in spirals and LSBs, these data are fitted by the {\bf dd} URC model that includes an exponential Freeman disc, a
B-URC DM halo and a gaseous disks.
The fit is very successful, unlike that relative to the NFW halo+stellar and gaseous disks velocity model (\citealt{ks17}).

These systems are strongly dominated by dark matter halos with cored density profile. The core sizes are
proportional to the corresponding disk length scales: $r_0 = 3\,R_D$, continuing the relationship found in spirals
and extending it 2 dex down in galaxy luminosity (\citealt{ks17}). Also, all the other dark and luminous
structural
properties of the dark and luminous matter, including the stellar/DM compactness $C_\star$ and $C_{\rm DM}$, result amazingly 
correlated (\citealt{ks17}).

All structural relationships established in normal spirals extend down to ``dd'' galaxies, the relevant aspect being that
also those that connect the dark and the luminous world continue, unchanged, in objects where the dark matter 
is, by far, the dominant component.

\section{The distribution of matter in spheroids}

Spheroidal galaxies include the biggest and the smallest galaxies of the Universe. The investigation of their dark matter
component is rather complicated. With respect to spirals, the bulk of stars in ellipticals is much more compact and then it
probes 
much inner and more luminous matter dominated galactic regions than the stellar and HI disks do in spirals. However, the 
 halos of ellipticals are filled with objects, like planetary
nebulae and globular clusters that can be good tracers of the gravitational potential, in spite of their
limited number and totally unknown dynamical state. 
 
 \subsection{The fundamental plane in ellipticals}
 
\begin{figure}[htb]
\includegraphics[width=\textwidth]{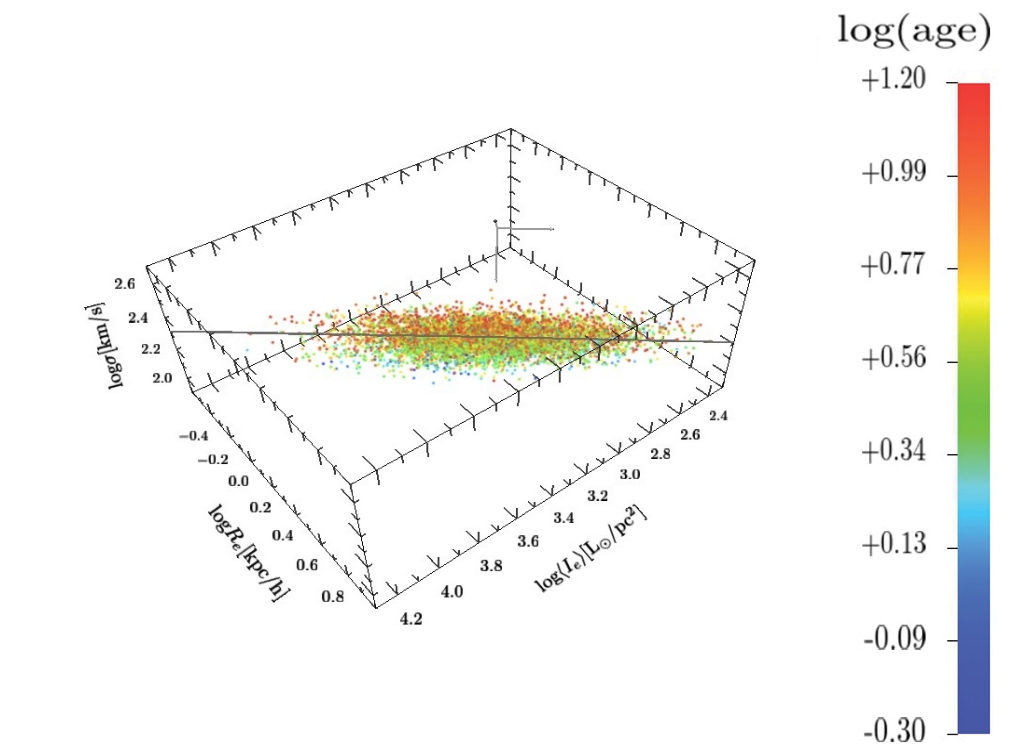}
\caption{The fundamental plane of ellipticals from \cite{HB} in the coordinate system $I_e, R_e, \sigma(1/8 R_e)$. 
Image reproduced with permission from \cite{mag12}, copyright by the authors.}
\label{fig:pianofond}
\end{figure}

The luminous regions of ellipticals show a 3D relationship, known as the fundamental plane, which is usually written as 
\begin{equation}
 \log \frac{R_e}{\rm kpc} = a \log \frac{\sigma}{\rm kms^{-1}} - \frac{b}{2.5}~ \frac{\mu_e}{\rm mags} + c \,,
\end{equation}
where $R_e$ is the effective radius, $\sigma$ is the central velocity 
dispersion (corrected to an aperture of $R_e/8$).
$\mu_e$ and $\log \ I_e$ are the surface brightness and surface luminosity within $R_e$. 
It is worth to remind that for virialized stable objects, all with the same surface profile $I(r/R_e)$ and small amount of
dark matter inside $R_e$, one expects: $R_e = \sigma_0^a/I_e^b $, with $a=2$ and $b=1$. It is well known that the FP has 
different parameters (\citealt{dd, d+, Joel}), e.g., 
$\log R_e= 1.24 \log \sigma_0 - 0.82 \log \langle I \rangle_e$ 
with scatter 0.07~dex in $\log R_e $. As a recent example, \cite{HB} used a sample of about 50,000 early-type galaxies
based
on the
SDSS-DR4/6, photometric and spectroscopic parameters and obtained $a=1.3 \pm 0.05, b= 0.3 \pm 0.05 $ with r.m.s.\ of 0.1~dex
(See Fig.~\ref{fig:pianofond}). 

Moreover, \cite{mag12} investigated the near-infrared FP in $\sim10^4$ early-type galaxies (ETG) included in the 6dF Galaxy
Survey (6dFGS). They fitted the distribution of central velocity dispersions, near-infrared surface brightness and half-light
radii with a three-dimensional Gaussian model that provided an excellent
match to the observed properties. 

The resulting FP reads as: 
$R_{e}\,\propto\,\sigma^{1.52\pm0.03}I_{e}^{-0.89\pm 0.01}$, with a r.m.s.\ of 23\%. The deviation of the FP with respect
to the theoretical predictions, called the tilt of the FP, has been thought to be due a combination of several effects
(\citealt{b03, bo08, HB, gf10} and \cite{zar12} for a review). However, from recent independent and accurate measurements
of the total mass inside $R_e$ by means of stellar
dynamics (\citealt{c06+, tsb}) and strong lensing (\citealt{bbt07, atb}), it is clear that variations among ETGs of the stellar mass to light ratio
$M/L$ are the cause of the tilt. This has clearly emerged in \cite{cap13}: they started with the FP which reads as
\begin{equation}
\log \left(\frac{L}{L_{\odot,r}}\right) = a + b \log \left(\frac{\sigma_e}{130\ {\rm kms}}\right) + c \log \left(\frac{R_e}{2\ {\rm kpc}}
\right) ,
\end{equation} 
$\sigma_e$ and $R_e$ are normalized to the median values found in sample under study. The resulting values of the 
parameters are: $b = 1.25 \pm 0.04 ; c = 0.96 \pm 0.03$ and the r.m.s.\ scatter is 0.1~dex; when 
the galaxy luminosity is replaced by the dynamical mass $ L\times (M/L){\rm dyn}$, obtained by self-consistent JAM modelling (see Sect.~5.5) a smaller r.m.s.\ it is found and the parameters: $b = 1.93 \pm 0.03, c = 0.96\pm 0.02$ acquire the virial values. This confirms
 that a major part of the scatter of the FP is actually due to variations in the $M/L$s values.

Therefore, the fundamental plane of ETGs expresses the properties of the virialized stellar spheroids and, differently from
the
Tully--Fisher in spirals, is not directly related to the properties of DM distribution (inside $R_e$). Finally, this
result lends support to the
idea,
valid in spirals, that the dynamically measured mass is more accurate prior of luminous mass of a galaxy than the
luminosity itself.

 \subsection{The dark matter distribution in ellipticals}

The derivation of the distribution of dark and luminous mass in ellipticals is far more difficult than in disk systems. The
kinematics is more uncertain and the tracers of the gravitational field often do not cover sufficiently well the
crucial region between $1/3 \, R_e$ and $3 \, R_e$ where the system becomes from stellar dominated to DM dominated.

The main issues under investigation are: {\it a)} an
universal power law slope of the \emph{total} density profile: $\rho_{\rm tot}\propto r^{-2}$ and {\it b)}
large variations of the $M_\star/L$ ratio with
mass and other quantities. As regard to the first issue, let us stress that the above density law in ellipticals
and the
case $V(R)={\rm const}$ in spirals are different configurations (see Eq.~\ref{eq:vsigma}). As regard to the second, at fixed galaxy luminosity, the stellar mass-to-light ratios vary in ellipticals much more than in spirals. 

As regard to investigations in early-type galaxies (ETG)s one has to report the several different approaches devised to obtain their
mass distribution.
However, it is fair to stress that it is difficult to make a synthesis of the results obtained so far, being the
situation still in full development. 

Data from the Sloan Lens Advanced Camera for Surveys (SLACS) project (\citealt{bbt}) provided us with the
total matter density profiles for a sample of 73 ETGs with strong lenses and large stellar masses $(M_\star > 10^{11}
M_\odot$ (\citealt{atb}). For each
galaxy the relevant quantities are the Einstein radius $R_E$, its relative enclosed mass, the stellar mass, and
$\sigma_E$ the velocity dispersions at $R_E$. An isotropic mass model was assumed and they found: $(\rho_{\rm tot}(r)
\propto r^{-\gamma})$ with $\left\langle\gamma\right\rangle = 2.08 \pm 0.03$ and with a scatter among galaxies of $\sigma_{\gamma}= 0.16$.

\cite{c+12} determined the total density profile for a sample of 14 ETGs fast-rotators 
(stellar masses $10.2 < \log M_\ast/M_\odot < 11.7$). SLUGGS and ATLAS observations provided the 2D stellar kinematics out to
about to $4 \ R_e$, reaching the region dominated by dark matter and poorly investigated before. They built 
axisymmetric dynamical models based on the Jeans equations solved with a spatially varying anisotropy $\beta$ and a general
density profile for the
dark matter halo. The resulting \emph{total} density profiles were found to be to follow, from $ R_e/10$ to $4 R_e$,  the
the power law:
$\rho_{\rm tot}(r)\propto r^{-\gamma}$
with $\langle\gamma\rangle=2.19\pm 0.03$. This extension of the above power law relationship to regions well
outside $R_{1/2}\simeq R_e$ is far than trivial and likely hides a connection between the dark halo and the
stellar spheroid. 

\cite{T+14} have investigated the central regions ($ r< R_e$) of ETGs by using strong lensing data from SPIDER and
kinematics and photometric data from ATLAS$^{\rm 3D}$. The
analysis extends the range of galaxy stellar mass ($M_\star$) probed by gravitational lensing down to $\sim 10^{10} \, 
M_\odot$. Each galaxy was modeled by two components (dark matter halo + stellar spheroid). 
The following DM halo profiles were considered: NFW, NFW-contracted, and Burkert. The mass-to-light
($M_\star/L$) was normalized to the Chabrier IMF as $M_\star/L=\delta_{\rm IMF}(M_\star/L)_{\rm Chabrier}$ with $\delta_{\rm IMF}$ a free parameter describing the systematically variations of IMF among galaxies. 
They found that, generally: 1) $\delta_{\rm IMF}$ increases 
with galaxy size and mass. 2) $ \alpha(R_e/2) = d\,\log M/d\,\log ~r -3 $ in the most massive
($M_\star \sim 10^{11.5} M_\odot$) or
largest ($R_e \sim 15\ {\rm kpc}$) ETGs reaches the value of $-2$, while in low-mass ($M_\star \sim
10^{10.2} \, M_\odot$) or very small ($R_e \sim 0.5\ {\rm kpc}$) ETGs decreases to the value of $-2.5$. 
As regard to the DM
distribution,
the
result of this work could not reach an explicit preference for a particular profile. 

\cite{ch14} investigated $\sim 2,000$ nearly spherical Sloan Digital Sky Survey (SDSS) ETGs,
at a mean redshift of $\langle z \rangle=0.12$ and assembled mass models based on their aperture, velocity dispersions,
and luminosity
profiles measurements. A two-components 
mass model (i.e., stellar spheroid plus dark halo) successfully fitted, inside $R_{1/2}$, the SDSS aperture velocity
dispersions. As result, they confirmed that, in the region: $0.1\,R_{1/2} < R < R_{1/2}$, the total density (dark halo + stellar
spheroid) exhibits a power
law behavior: $\rho_{\rm tot}(r) \propto r^\gamma$ with $\langle\gamma\rangle = -2.15\pm 0.04$. 

\begin{figure}[htb]
\centering
\includegraphics[width=0.8\textwidth]{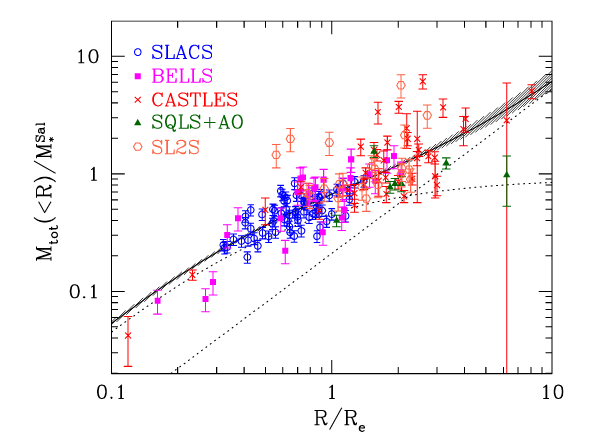}
\caption{Normalized mass of ETGs as function of its normalized radius. 
Also shown: the best-fit mass profile {\it solid line} and the stellar spheroid and the power law DM halo contributions
{\it green and red dotted lines}.
Image reproduced with permission from \cite{og14}, copyright by the authors.}
\label{fig:oguriell}
\end{figure}

\cite{og14} investigated 161 strong gravitational lenses from SLACS and BELLS and a number of strongly lensed quasars. 
They derived the stellar mass $M_*^{\rm Sal}$ for each lensing galaxy by fitting the observed spectral energy distribution 
to a stellar population synthesis model with a Salpeter IMF (\citealt{BC03}). The measurement in these lens galaxies 
of the sizes of their Einstein rings $R_E$ allowed them to build normalized total mass profiles for each
object: $M_{\rm tot}(<R_{E})/M_*^{\rm Sal}$ and to normalize the projected radius
$R$ by the effective luminosity radius $R_e$. Notice that this double-normalization is of the same kind of that performed
in
the {\bf dd} galaxies (\citealt{ks17}). They derived, from each Einstein ring, the relative scaled mass profile $M_{\rm
tot}(<R_E/R_e)/M_*^{\rm Sal}$. These data were fitted by the model
\begin{equation}
\frac{M_{\rm tot}(<R)}{M_*^{\rm Sal}}=A\left(\frac{R}{R_e}\right)^{3+\gamma}.
\end{equation}
 They found $\gamma=-2.11\pm 0.05$. Furthermore, they decomposed the total mass in its dark and
luminous components: a power-law spherical DM
dark halo and a Hernquist spheroid for which, with $y\equiv R/R_e$:
$M_{\rm Her}(y)=M_{\rm sph} y^2/(1.4^2 +y^2)$
\begin{equation}
 \frac{M_{\rm DM}(<R)}{M_\star^{\rm Sal}}=A_{\rm DM}\left(\frac{R}{R_e}\right)^{3+\gamma_{\rm DM}}\,.
\end{equation}
Quasar microlensing measurements break the IMF-stellar mass degeneracy, the DM fraction inside
$R_e$ results: $A_{\rm DM}/A=0.2$ and $\gamma_{\rm DM}=-1.60_{-0.13}^{+0.18}$ that 
implies that DM is distributed in a way shallower than the total matter, as it occurs in disk systems, see Fig.~\ref{fig:oguriell}. 

\cite{pcd17} (see also \citealt{cap13}), by modelling kinematical and photometric data of 258 early-type
galaxies, belonging to the volume-limited ATLAS$^{\rm 3D}$ survey, derived their density profiles and found the usual power
law: $\rho_{\rm tot}(r) = r^\gamma$ with $\gamma = -2.2
\pm 0.2$. Noticeably, however, they did find significant variations of $\gamma$ with $\Sigma_e
$ the surface
brightness inside $R_e$ and $\sigma_e$, in some contrast with previous works.

\begin{figure}[htb]
\centering
\includegraphics[width=1.1\textwidth]{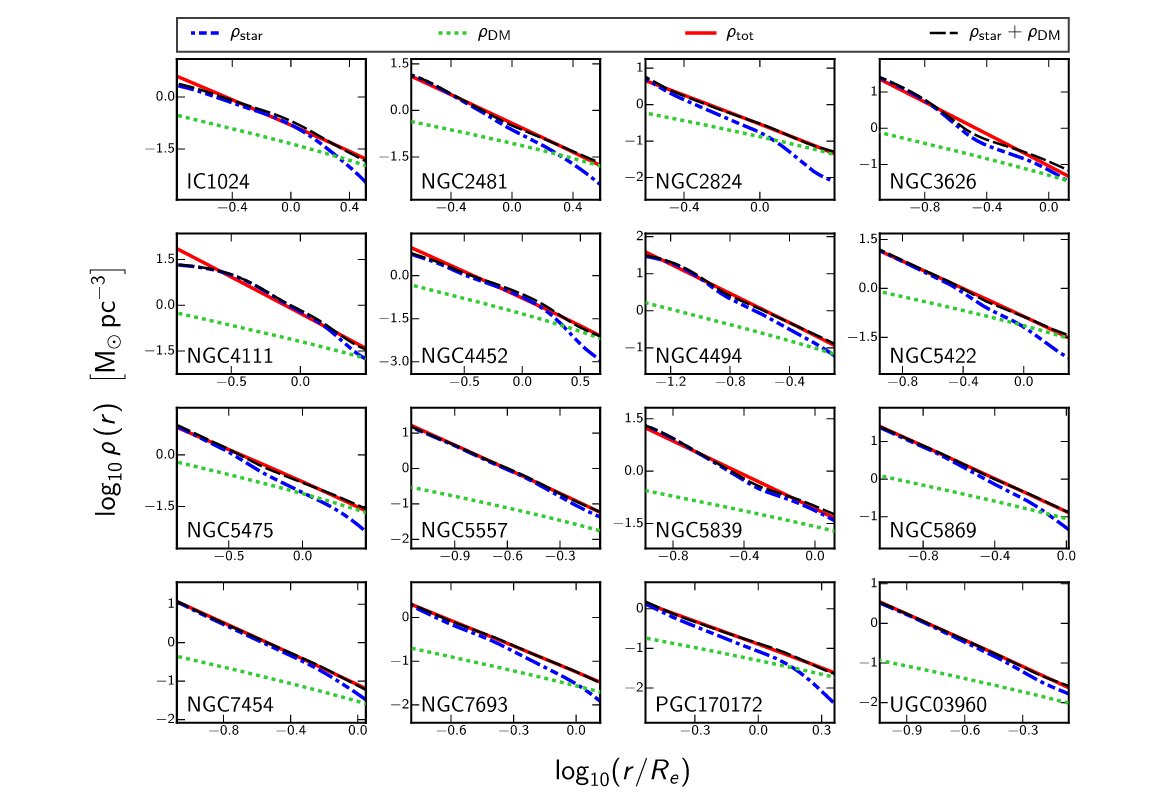}
\caption{The total density profile {\it solid line} and its stellar {\it dotted line} and DM {\it  dot-dashed line}
components for 16 galaxies from the ATLAS survey. 
Image reproduced with permission from \cite{pcd17}, copyright by the authors.}
\label{fig:poci}
\end{figure}

\begin{figure}[htb]
\centering
\includegraphics[width=0.9\textwidth]{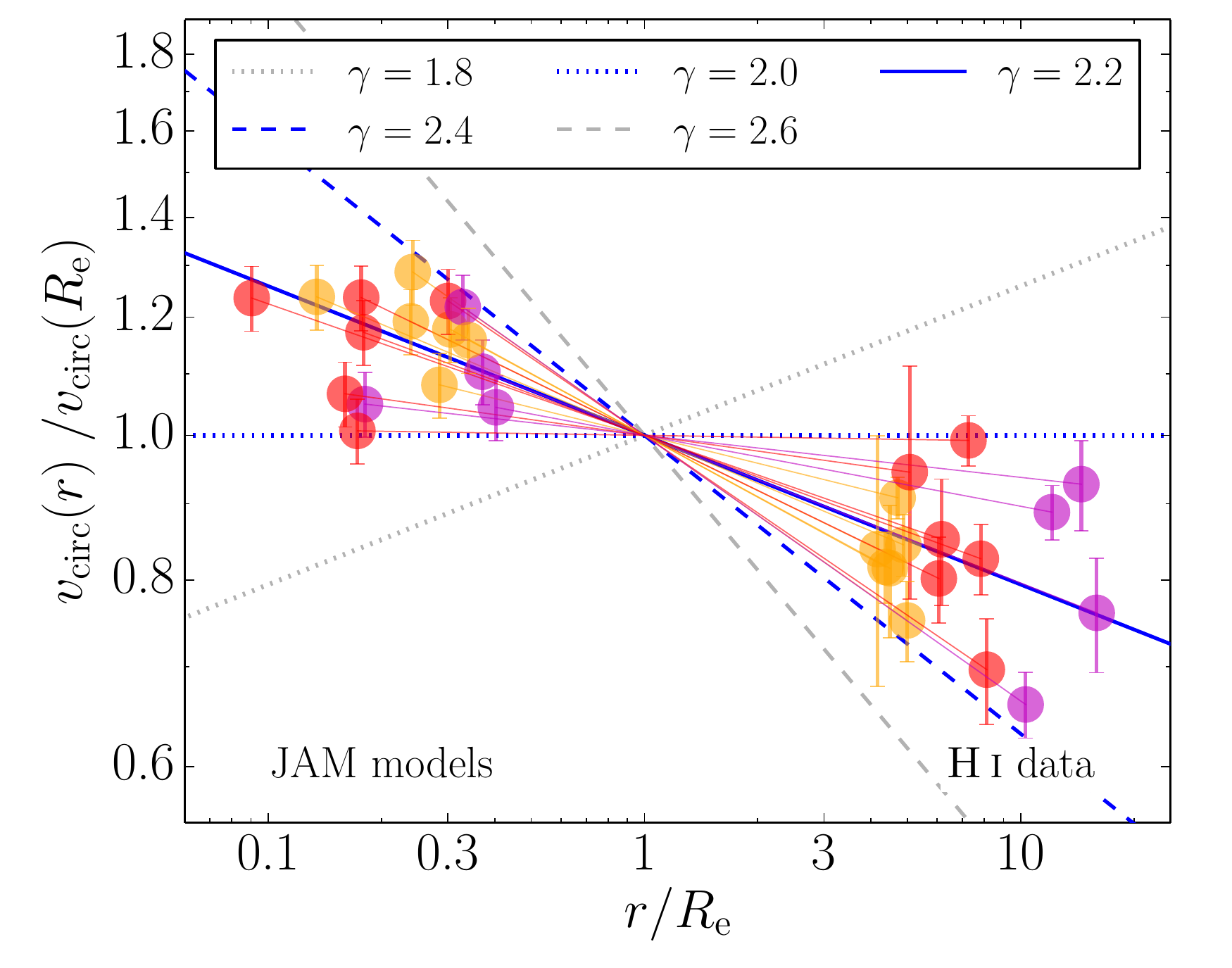}
\caption{Radial profile of the normalized circular velocity for the sample of ellipticals in \cite{Soc16}. Data come from JAM models for $R<R_e$ and from HI 21~cm for $R >R_e$. 
Points and solid lines are coded according to the increasing
$R_{\rm HI} /R_e$ ratio.}
\label{fig:serra}
\end{figure}

\cite{Soc16} investigating a sample of 16 fast-rotator ETGs with HI disks extended out to $\sim 6 \, R_e$
established a tight linear relation between $V_{\rm HI}$ the (flat) circular velocity measured from resolved HI observations in
(external) DM dominated regions (i.e., for $R \gg R_e$) and $\sigma_e$. 
the velocity dispersion measured at $R_e$, i.e., in a luminous matter dominated region:
\begin{equation}
V_{\rm HI} =1.33 \sigma_e \,,
\label{eq:vcvhi}
\end{equation}
with an observed scatter of 12 percent. The tightness of the correlation suggests a strong 
coupling between luminous and dark matter, analogous to the situation in
spirals, in LSBs
and in {\bf dd}s. Eq.~(\ref{eq:vcvhi}) implies a decline in the effective circular velocities 
$V(r)$ from $R_e$ to the outer regions. Such drop is in excellent agreement with the results of \cite{ca15} and, remarkably, 
is similar to that observed in early-type spirals (\citealt{no07}) and in the most luminous late type spirals (\citealt{s7}).
Assuming $\rho_{\rm tot}(r) \propto r^{-\gamma}$, Eq.~(\ref{eq:vcvhi})
implies $<\gamma> = 2.18 \pm 0.03$ across the sample, with a scatter of 0.11 around the average value (see
Fig.~\ref{fig:serra}).

\cite{Ael18} (see also \citealt{Ael16}) used globular cluster kinematics data, primarily from the SLUGGS survey, to measure the
dark matter fraction
$f_{\rm DM}(5\,R_e)$ and the average dark matter density $\rho_{\rm DM}(5\,R_e$ within $5\,R_e$ for 32 nearby ETGs
with
stellar mass log $(M_\star/M_\odot)$ ranging from $10.1$ to $11.8$. 
They found that $f_{\rm DM}(R_e) \sim 0.6$ for 
galaxies with stellar mass lesser than $(M_\star/M_\odot) \sim 10^{11}$. At higher masses, a sudden large 
range of $f_{\rm DM}(R_e)$ values emerges. This seems in contradiction with the total density power law 
$\rho_{\rm tot}\propto r^{-2.1\pm 0.1}$ usually found in other determinations.

\cite{P18} used planetary nebulae (PNe) as tracers of the gravitational field around ellipticals. 
They obtained two-dimensional velocity and velocity dispersion for 33 ETGs. The velocity fields were
reconstructed from the measured PNe velocities. The data extend out from $3\,R_e$ to $13\,R_e$.
The objects show a kinematic transition between the inner luminous matter dominated regions and the outer halo
dominated ones. These transition
radii, in units of $R_e$, anti-correlate with stellar mass, differently from what occurs in spirals. The galaxies appear to have more diverse kinematic properties in their halos
than in their central regions. It is noticeable the fact that 15\% 
of the galaxies in the sample have steeply falling profiles implying that, inside $R_e$, the fraction of dark matter is very
negligible. 

 \begin{figure}
\centering
\includegraphics[width=0.9\textwidth]{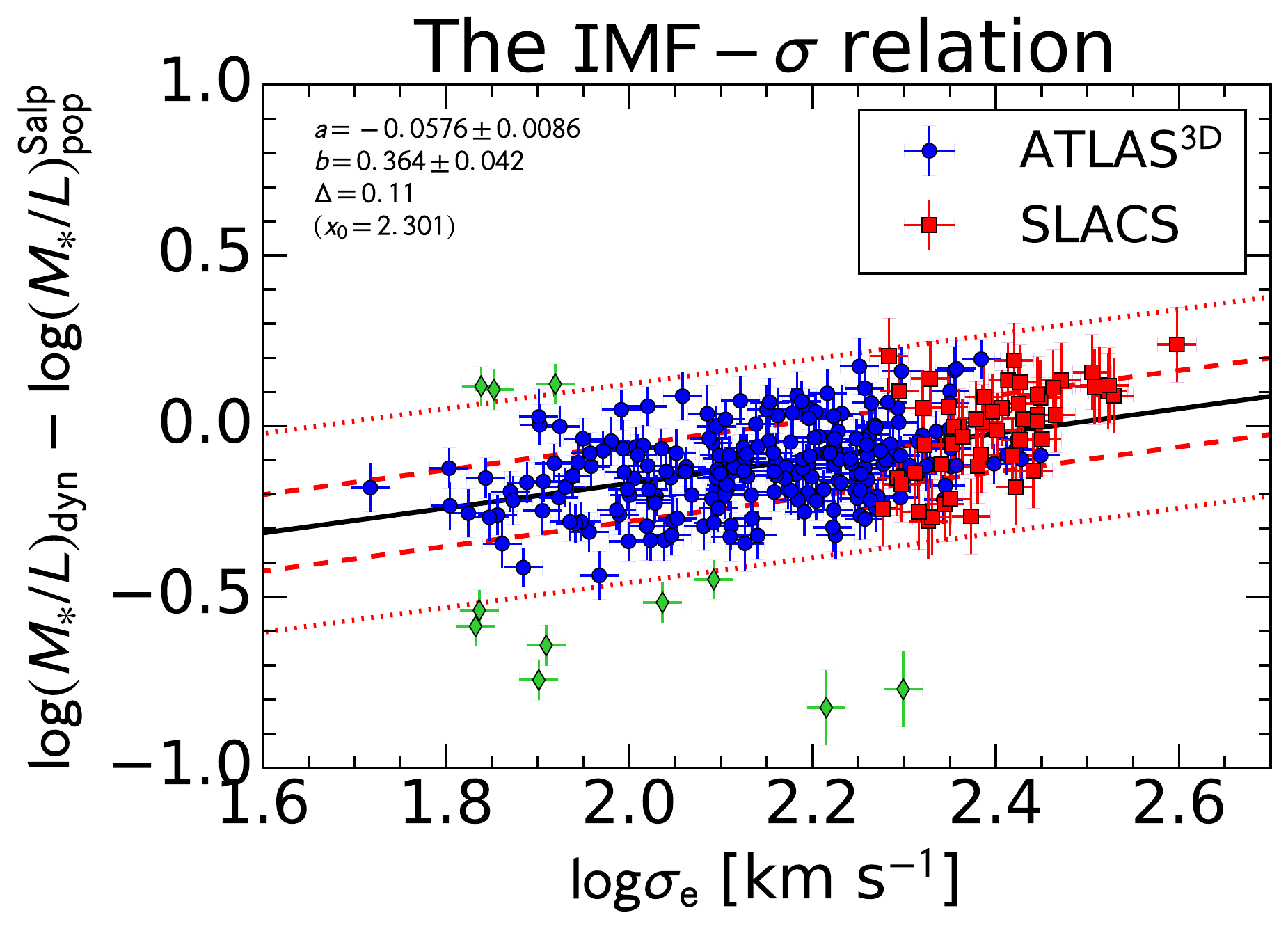}
\caption{The mass discrepancy--$\sigma$ relation, very likely created by systematical variations
of the IMF among ETGs (see \cite{ca16,po15}).}
\label{fig:imfs}
\end{figure}

One important issue of the ETGs is the comparison between the $M/L$'s inferred from their dynamical or strong lensing modelling and those
inferred from the fitting of their spectral energy distributions.\cite{ca16} have investigated it with a large 
sample of objects. The values derived, see Fig.~\ref{fig:imfs}), 
indicate the existence of random variations of the IMF and variations with the galaxy dispersion velocity.  Noticeably, the existence of a non universal initial mass function (IMF) is already present at intermediate redshift (\citealt{Tel17}). 

Evidences that ellipticals have variable IMF them come
Also from their chemical evolution model reproducing the abundance patterns observed in the sample of the Sloan Digital
Sky Survey Data Release 4 (\citealt{MATT1}). The model assumes ellipticals form by fast gas
accretion, and suffer a strong burst of star formation followed by a galactic wind, which quenches star formation. The model
if assumes a fixed initial mass function (IMF) in all galaxies,  fail in simultaneously reproducing the observed trends of chemistry with the
galactic mass; only a varying IMF among ellipticals, leads to an agreement with data.

\begin{figure}[htb]
\centering
\includegraphics[width=0.72\textwidth]{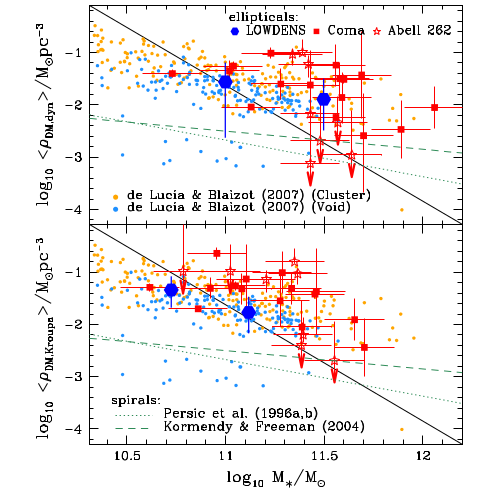}
\caption{The average density inside $2\,R_e$ in NGC 7113 and PGC 67207 (blue hexagons as a function
of their stellar spheroid 
mass $M_\star$ computed a) dynamically {\it (top)} or b) from the photometry  {\it (bottom)}.
 Also shown the values for ETGs in Coma Cluster {\it (red filled)} and in Abell 262 {\it (red open)}. The lines show
the corresponding spirals' relationship. 
Image reproduced with permission from \cite{coel17}, copyright by the authors.}
\label{fig:corsi}
\end{figure}

\cite{coel17} have investigated NGC~7113, and PGC~67207, two bright ETGs in low-density environments. These rare
objects may help us disentangling in ellipticals what is of
pertinence
of the process of their formation and what is inherent to the properties of their dark matter halos. The
surface-brightness distributions and their parameters were derived by K$_S$-ugriz-band two-dimensional photometric
decomposition.
The line-of-sight stellar velocity distributions inside $R_e$ were measured along several position angles. They assumed the BT-URC DM
halo profile (see Eq.~\ref{eq:LOG}). 
The luminous and dark distributions were obtained from the orbit-based axisymmetric dynamical modelling (see Sect.~5.5). The fit model to the data is excellent and implies that these galaxies have a lower content of dark
matter with respect to early-type galaxies living
in high-density environments. Moreover, it is important to notice that their DM density inside $2\,R_e$ is significantly
higher than in similar mass spirals (see Fig.~\ref{fig:corsi}).

 \subsection{DM in dwarf spheroidals}

\begin{figure}[htb]
\centering
\includegraphics[width=0.81\textwidth]{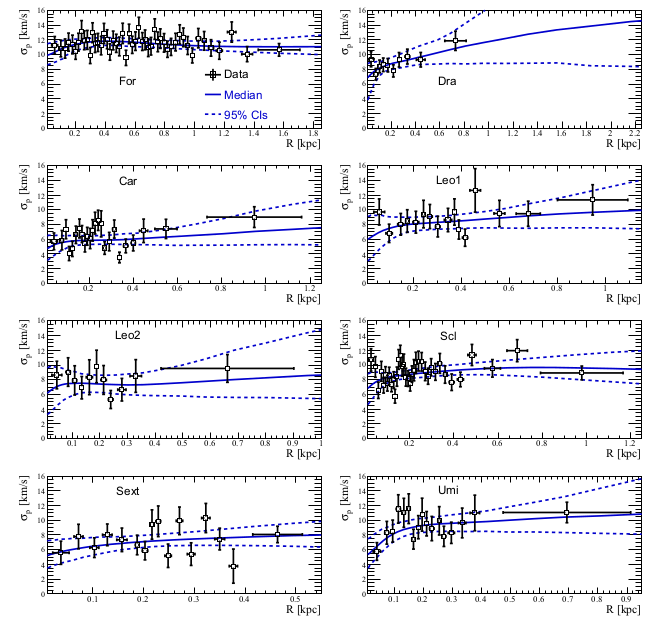}
 
\caption{Line-of-sight dispersion velocities of the ``classical'' dSphs. A large r.m.s.\ is evident. 
Image reproduced with permission from \cite{Bet15}, copyright by the authors.}
\label{fig:bonn}
\end{figure}

Dwarf spheroidal (dSph) galaxies are the smallest and least luminous galaxies in the Universe and provide unique hints on
the nature of DM. They are old, in dynamical equilibrium and with no HI component. They contain a (small) number
of stars, which provide us with tracers of the gravitational field. The very negligible baryonic content that they show
does not affect their mass modelling and it also indicates that this component may have never modified the primordial DM distribution (see
\cite{W13}) Then, by investigating these galaxies, we probe the original structure of the DM
halos (see the review of \citealt{bat13}).

 The stellar component for each dwarf spheroidal galaxy is 
modeled by means of a Plummer density profile with its scale radius $R_e$, see Eq.~(\ref{eq:plu}). The main sample
includes the
 eight larger dSphs of the Milky Way: Carina, Draco, Fornax, Leo~I, Leo~II, Sculptor, Sextans, and Ursa Minor.  The determination of the  DM 
mass profile $M(r)$ requires the velocity dispersion profile along the line-of-sight $\sigma\,{\rm l.o.s.}(r)$ (see Fig.~\ref{fig:bonn}). The very limited number of these galaxies combined with
the large range in the
values of their physical quantities make the stacked analysis approach impossible for investigating the dSphs mass distribution. There are three common
methods that use available observations to infer the DM density profile in dSphs.

\paragraph{Jeans analysis:}

In this approach one feeds Eq.~(\ref{eq:jeans1}) with the values of $\nu_\star(R)$, the stellar density profile, uses a
large number of well determined dispersion velocities  $\sigma_{\rm l.o.s.}(r)$ (\citealt{Waj}) and
assumes a particular anisotropy profile (e.g., as in 
\citealt{Bet15}). Then, through a Monte Carlo analysis, one obtains the free parameters of the DM
density profile $\rho(r)$ and the anisotropy function $\beta$. There are views that this investigation, also when the
tangential velocity dispersions are available, cannot resolve in these objects
the cusp/core issue (\citealt{W09, Sel08, Bet15, Sel18}). The degeneracy in the Jeans equation between the mass and
the anisotropy profiles, combined with a kinematics of limited extension and quality, makes
difficult to determine the density profile by means of this method.

\paragraph{Slope method:}

\cite{WP11} first exploited the fact that in some dSphs there are multiple stellar populations, photometrically and
chemo-dynamically distinct sub-components. They independently trace the (same) gravitational potential. Since
$M(R_e)$, the mass
contained within the effective radius $R_e$ of each component, can be measured independently of their stellar orbital
anisotropies, see Eq.~(\ref{eq:Wolf}) then, we can derive the quantity $\frac{d\,\log M}{d\,\log R}$ 
at different radii without adopting a DM halo profile. The method,
applied to the dSph Fornax and Sculptor, for which two separate stellar sub-components have been disentangled, gives 
$\Delta \log M / \Delta \log r=2.61_{-0.37}^{+0.43}$ and $2.95_{-0.39}^{+0.51}$, respectively, pointing 
 to DM densities that keep an almost constant value within the central few-hundred parsecs of these objects. With the same method, \cite{br13} found that a NFW profile is only marginally allowed in Sculptor.
 
\begin{figure}[htb]
\centering
\includegraphics[width=1.00\textwidth]{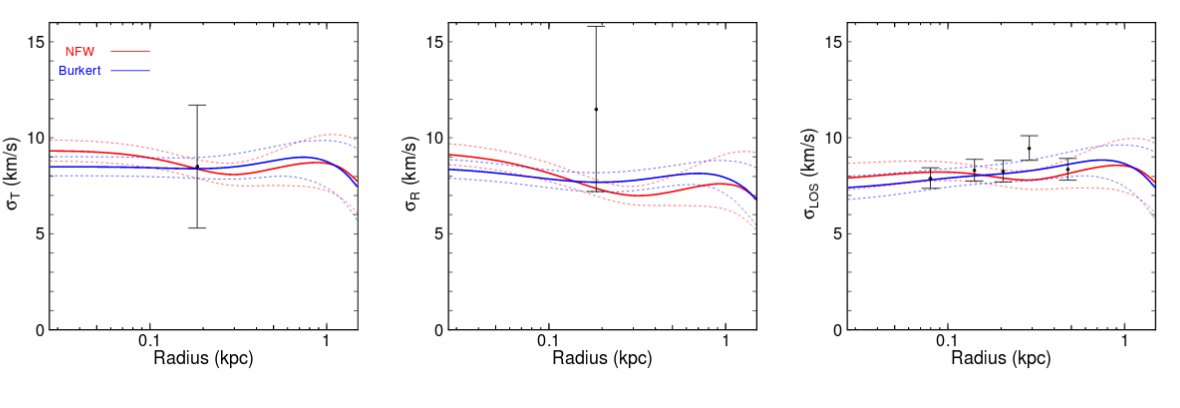}
\caption{Observed versus predicted dispersion velocities form different halo density profiles. 
Image reproduced with permission from \cite{Sel18}, copyright by AAS.}
\label{fig:stri}
\end{figure}

This method has been carefully investigated by \cite{Sel18} in view of determining the level of its
intrinsic bias, see Fig.~\ref{fig:stri} and finding improvements.

\paragraph{Schwarzschild modelling:}

A promising method, based on distribution functions that depends on the action integrals, has
been put forward by \cite{ppn}. This was applied to the Fornax galaxy, finding strong evidence for the presence of a
cored density profile.

\section{The LM/DM universal properties}

One could resume the state of the art of the issue of ``DM in galaxies'', 
by stressing the unexpected scheme shown by the distributions of the dark and luminous matter in galaxies: 
 halo masses, stellar component/baryonic masses, central densities, luminosities, DM density length scales, half-light radii,
and galaxy morphologies are all engaged in a series of relationships, difficult to be understood in a physical sense.
However, since the concurrent view argues that ``galaxy formation is a complex phenomenon which 
could account for the apparently inexplicable observational scenario'', we stress that the above is far
beyond a list of galaxy relationships, but a coherent pattern that can help us in the search of the unknown dark particle.

In disk systems (dwarf disks, low surface brightness galaxies and spirals) when the values of their structural quantities are expressed in
physical units, the stellar component forms a family ruled by three parameters: the disk
length-scale $R_D$ and the magnitude (e.g., $M_I$) and the stellar disk concentration $C_\star$. In the same systems, also the dark
component is represented by a family ruled by three parameters: the core radius $r_0$, the central density $\rho_0$ and
$C_{\rm DM}$ the DM concentration. The two families are closely and mysteriously related: the entanglement is so
deep that it is difficult to
understand which rules which. 

Remarkably, the situation much simplifies when we express the circular velocity $V(r)$\footnote{$V(r)= (r
~d\Phi/dr)^{1/2}$ with $\Phi$ the {\it total} gravitational potential.} in the double-normalized form:
$V(r/R_{\rm opt})/V(R_{\rm opt})$
The profiles of the RCs emerge as a function of just one parameter, at choice among the
above six, plus $V_{\rm opt}$, $M_{\rm vir}$ and the angular momentum for unit mass $j$ (see \citealt{lsd}). Remarkably, this occurs independently on whether a galaxy is dark matter or luminous matter dominated for $R<R_{\rm opt}$. 
The emerging evidence is that structural quantities deeply rooted in the luminous sector, like the disk length scales, 
tightly correlate with structural quantities 
deeply rooted in the dark sector, like the DM halo core radii. 

Let us conclude this section noticing that this scenario is, instead, still under investigation in spheroidal galaxies.

\subsection{The cored distributions of dark matter halos around galaxies}

The current situation is the following: a) in disk
systems
of all {\it morphologies and luminosities} there is strong evidence that the DM halo density profile is very shallow out to the edge of the stellar
distribution $R_{\rm opt}$ b) in
dwarf
spheroidals and in ellipticals, also due to the intrinsic difficulty in these systems to disentangle the actual kinematics
from the 
biased one, the situation is less clear, although, also in these objects, there are several claims of cored DM halo
density profiles. In conclusion, the claim that DM around galaxies have a density distribution well represented by the
cored
B-URC profile is bald, but I believe correct. 

\begin{figure}[htb]
\centering
\includegraphics[width=1.00\textwidth]{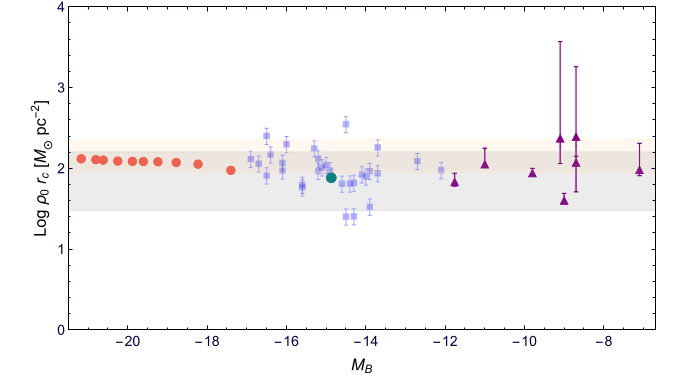}
 
\caption{(Central DM halo density) $\times$ (halo core radius) as a function of a galaxy magnitude. Legenda:
$r_c\equiv r_0$. Data are from the URC of spirals {\it (red circles)}, the scaling relation in \cite{Doet} {\it (orange
area)},
the Milky Way dSphs {\it (purple triangles)} \cite{s12}, the {\bf dd}s {\it (blue squares)} \cite{ks17}. Also shown the relationship by \cite{bu15}:
$\rho_0 ~r_c=75^{+85}_{-45}\,M_{\odot}\ {\rm pc^{-2}}$ {\it (grey area)} (see also \citealt{sma}). 
Image reproduced with permission from \cite{ks17}, copyright by the authors.} 
\label{fig:k4}
\end{figure}

The most intriguing aspect of the DM in galaxies is not that they all possess a universal density profile, but that, 
this latter 
comes with a couple of very unexpected properties. The analysis of rotation curves, dispersion velocities, and weak-lensing
data of 
large samples of dSphs, dwarf irregulars, spirals, and elliptical galaxies, 
found that the product of the DM core radius $r_0$ with the DM central density $\rho_0$ 
is nearly constant in galaxies, i.e., independent of their luminosity (\citealt{Doet}; see also
\citealt{do}). This result, pioneered by \cite{KF04}, is obtained in \cite{Doet}
from the mass models derived from 1) about 1000 coadded RCs of spirals, 2) hundredths individual RCs of 
normal spirals of late and early types 3) galaxy-galaxy weak
lensing signals 4) the inner kinematics of Local Group dwarf spheroidals 5) the RCs of 36 {\bf dd} and 72 LSBs (see \citealt{dps}). 
The relationship reads (see Fig.~\ref{fig:k4}) 
\begin{equation}
\log (r_0 \rho_0) = 2.15 \pm 0.2 \,,
\end{equation}
in units of $\log (M_{\odot}/{\rm pc^2})$. 

This relationship between the two structural quantities of the DM halos is found in galactic systems
spanning
over 14 magnitudes and it exploits mass profiles determined by several independent methods. In the same objects, the
constancy of $\rho_0 r_0$ is in
sharp contrast with the systematically variations, by about 5 orders of magnitude, of all the other DM-related galaxy
quantities, including the central DM density $\rho_0$ and many of the LM-related galaxy properties, as the magnitude. 

\begin{figure}[htb]
\centering
\includegraphics[width=0.9\textwidth]{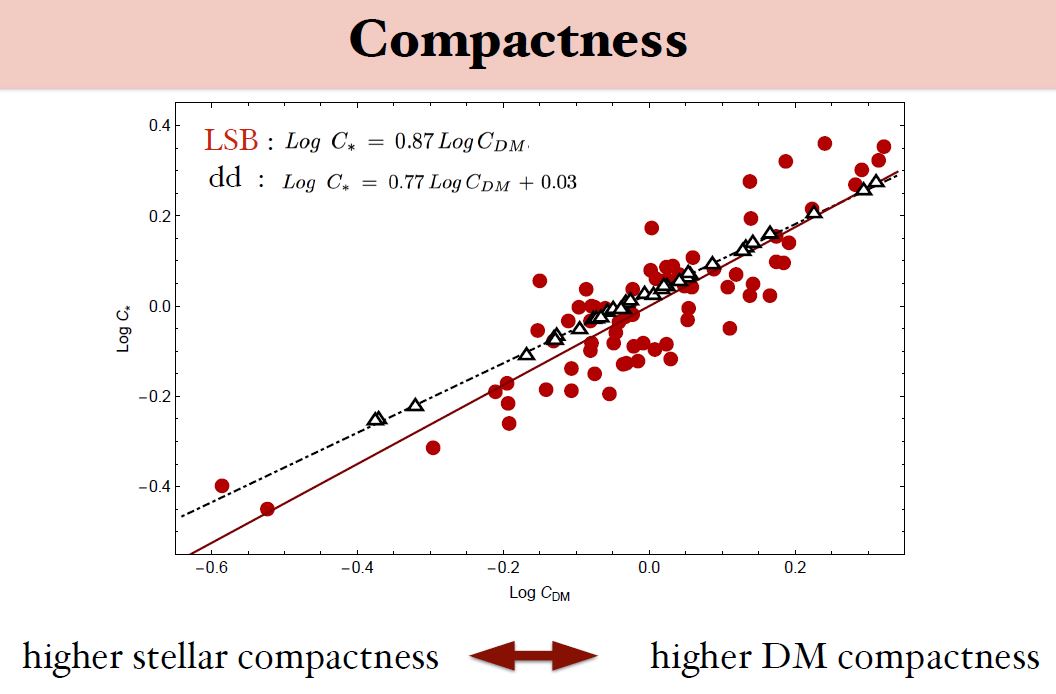}
\caption{The relationships between the compactness of the stellar disks and that of DM halos in {\bf dd} and LSB galaxies
(\citealt{ks17,dps}). Let us set: ${\bf M} $ and ${\bf S}$ for a generic galaxy mass and size. We
perform, in a sample of galaxies, the regression $\log {\bf S} = a + b \log {\bf M}$. For each galaxy $i$ of the sample the
compactness $\log  {\bf C}_i$ 
is defined by 
$\log {\bf C}_i= -\log({\bf S}_i) +a + b \log {\bf M}_i$. Here, for the luminous matter:
${\bf M} \equiv M_\star$, ${\bf S}\equiv R_D$ and ${\bf C}_i\equiv C_\star$ ; for the DM: ${\bf M} \equiv M_{\rm vir}$, ${\bf
S}\equiv r_0$ and ${\bf C}_i\equiv C_{\rm DM}$. }
\label{fig:comp}
\end{figure}

At a higher level there is the correlation between the compactness of the stellar disks and that of the DM halos in
dark matter dominated {\bf dd}s and LSBs (see Fig.~\ref{fig:comp} and the related caption). It is legitimate to
interpret all this as an evidence of the dark and luminous worlds conjuring in galaxies. 

The relationship between the halo mass and the stellar mass located at its center is an important and well investigated one.
It is well known that the mass fraction $\frac{DM}{LM} $ as a function of the halo mass follows a characteristic U-shaped
curve (\citealt{wo10,mos})
for which $M_{\rm vir}/M_\star$ is minimized at the halo mass $M_{\rm vir,break}\approx 3\times 10^{11} M_{\odot}$ and rises at both lower and higher masses.
According
to the URC, the value of  $M_{\rm vir,break}$ corresponds to  $M_{\rm star,break}\sim
1.2\times 10^{10} M_{\odot}$ and to  $L_{\rm break}\sim 5\times 10^9 L_\odot$ in the  $r*$-band luminosity (see also \citealt{lsd}). Outliers of this relationship do exist (\citealt{Be16}), however, here we do not further enter this topic certainly related to the ``galaxy formation process''.

Therefore, the empirical scenario includes six quantities that define a galaxy: three
in the dark sector (halo mass and core radius and DM halo compactness) and three in the luminous sector (stellar/baryonic mass,
half-light radius and stellar disk compactness). They all relate each other but, while some of these relationships lay in the
heart of the DM mystery, others, instead, lay in the
ball-park of the galaxy formation and evolution process.
 

\subsection{The dark-luminous matter coupling 2.0}

In spirals, dwarf disks and LSBs there are extraordinary multiple connections between the dark and the luminous components.
This occurs over many orders of magnitudes in halo masses and over the whole ranges of galaxies morphology and luminosity. The ``standard'' explanation relates to a dynamical evolution of the galaxies, in particular, of their DM halo densities, caused by powerful baryonic feedbacks. Although this scenario is far than being rejected, it seems, however, 
unable to cope with the intriguing wealth of correlations between quantities deep-rooted in opposite dark/luminous
worlds that we have presented in this review. More in detail, while we cannot completely rule out the possibility that astrophysical
phenomena can be responsible for the above intriguing scenario, on the other hand, what emerges in galaxies allow us to propose a shift
of paradigm, according to which, the nature of dark matter is not given to us by convincing theoretical arguments, but must be searched in the various properties of the DM halos and stellar disks. 

In \cite{st17}), it is argued that these new ideas can be justified also by some direct hint: in spirals
the DM pseudo pressure $\rho_{\rm DM}(r) V^2(r)$ reaches a maximum value always close to the
core radius $r_0$ and this maximum takes the same value in all objects, no matter the galaxy mass. 
Moreover, at $r=r_0$ in all disk systems, the quantity $\rho(r) \rho_\star(r)$ takes the same
value. We notice that this density product is proportional to the interaction probability between the ,
the luminous and the dark matter.
This is hardly a coincidence, in that, the quantity like $K_{\rm SA}= \rho_{\rm DM}^2(r) $, which is proportional to the self
interaction of the DM component, is largely varying in galaxies and among galaxies. One can speculate that the structure of the inner parts of the galaxies is driven by a direct interaction between dark
and luminous components on timescales of the order of the age of the Universe. The DM central cusp, outcome of the proto-halo virialization, as time goes by, gets progressively eaten up/absorbed by the
dominant luminous component. The interaction, then, flattens the density of DM and drops the pressure towards the center of
the galaxy and it is likely to leave in inheritance the above galaxy relationships.

\section{Conclusions}

On the fundamental issue of dark matter in galaxies there is a substantial difference between spheroidals and disk systems.
Let us notice that also the latter statement shows that, although we are focused on DM halos,
nonetheless, we must discuss galaxy morphology. And this has been the leitmotiv of this review: the DM
component enters in aspects apparently of pertinence of the luminous matter and vice versa. 

We have started to point out that the luminosity or a reference velocity is the tag that defines the dark and luminous mass
distribution
in galaxies. However, very recent results have proven that in spirals, ``dd'' and LSBs, the universal rotation curve, when 
expressed in physical units, needs two statistically independent 
controlling parameters: the luminosity and the compactness. It must be specified that we are not just flagging some empirical
relationships: we have three structural properties of the stellar discs that enter in close
relation with the three structural properties of the DM halos.

In elliptical galaxies, the situation is still very open. They also 
show regularities in their total mass distributions: its logarithmic derivative from $r=0$ to $r=R_e$ and beyond is very near to 1, despite that in this region the galaxies pass from a totally LM dominated regime to one with a relevant fraction
of dark matter. 
The fundamental plane of ellipticals and S0 entangles two quantities of the luminous world, the luminosity/stellar spheroidal mass, and the half-light ratio and a hybrid one: the dispersion velocity, which is rooted in both luminous and dark worlds.
Universality in the distribution of matter in ellipticals has not been established yet. We believe that this is due to
the
insufficient quality and quantity of proper and useful probes of their gravitational potentials. We also have to notice that
also for these systems there are evidences of cored DM distributions. 

Dwarf spheroidals, despite their limited number, are becoming always more crucial in the investigation of dark matter. Each of
these dark spheres, lying at the lowest mass boundary of the cosmological structures harboring stars, is a wealth of
information on the dark particle. Unfortunately, we can probe their gravitational field only 
very near to their centers, with tracers that provide data that are difficult to be unambiguously interpreted. It is worth
saying, however, that also for this population of galaxies, there are evidences of cored DM halos with
properties
similar to those of the disk systems and ellipticals. 

The non-gravitational nature of DM remains a mystery (\citealt{BerB, desw}). 
 It seems impossible to explain the observational evidences gathered so far in a simple dark
matter framework. In my opinion, they are portals to the new physics that seems to lurk
behind
the phenomenon called ``dark matter''. I think that it will be important to recognize our prejudices and confront them head on, also if this means to end our fascination with the $\Lambda$ CDM Weakly Interacting Massive Particles scenario. 

\section{Future directions} 

As a consequence of the reverse-engineering approach to the mystery of the dark matter in galaxies that I advocate here, 
the future is the past. Namely, I argue that, in the observational properties of galaxies, there is much of the required
information to solve the riddle. Unfortunately, we have recovered only a very small part of it, not because it is difficult or long to do, but because 
we were stuck in a different paradigm where, honestly, all this phenomenology is not so important.

However, the situation is extremely positive because, in the near future, from Gaia to SKA, we will be submerged by an
enormous flux of information, coming from different messengers, on all aspects of galaxies, independently if one believes or not that
this will
lead to a solution of the old mystery of dark matter.

\begin{acknowledgements} 
I thank Francesca Matteucci for motivating me towards the enterprise of writing this review. 
I thank N. Turini, V. Gammaldi, F. Nesti, M. Cobal, A. Bressan, M. Cappellari, G. Danese, A. Lapi, 
 C. Frenk, C. Baccigalupi, A. Pillepich, M.F. de Laurentis, R. Valdarnini and C. di Paolo for very useful discussions. 
I thank Brigitte Greinoecker for help in the process of writing this review.
\end{acknowledgements}


\begin{thebibliography}{}

\bibliographystyle{spbasic} 
 

 

\bibitem[Adams et al.(2014)]{ad14} 
Adams JJ, Simon JD, Fabricius MH, et al (2014) ApJ, 789:63 

\bibitem[Adhikari et al.(2017)]{all}
Adhikari R, Agostini M, Ky NA, et al (2017) JCAP 1:025 
 
\bibitem[Alabi et al.(2016)]{Ael16}
Alabi AB, Forbes DA, Romanowsky AJ, et al (2016) MNRAS, 460,
3838 

\bibitem[Alabi et al.(2018)]{Ael18}
Alabi A, Ferr{\'e}-Mateu A, Romanowsky AJ, et al (2018) ArXiv e-print arXiv:1801.09686 

\bibitem[An and Evans(2011)]{ae11} 
An JH, and Evans NW (2011) MNRAS, 413:1744 

\bibitem[Aprile et al.(2018)]{apr} 
Aprile E, Aalbers J, Agostini F, et al [XENON Collaboration] (2018) PRL 121:111302 

\bibitem[Arcadi et al.(2017)]{amel} 
Arcadi G, Dutra M, Ghosh P et al, EPJC, 78:203 
 
\bibitem[Auger et al.(2010)]{atb}
Auger M W et al (2010) ApJ, 724:511

\bibitem[Bacon et al.(2001)]{bac01} 
Bacon R, Copin Y, Monnet G, et al (2001) MNRAS, 326:23 

\bibitem[Bahcall(1984)]{bah84} 
Bahcall JN (1984) ApJ, 276:169 

\bibitem[Bartelmann and Maturi(2016)]{b16}
Bartelmann M and Maturi M (2016) ArXiv e-print arXiv:1612.06535
 
\bibitem[Battaglia et al.(2013)]{bat13} 
Battaglia G, Helmi A, Breddels M (2013) New Astron Rev, 57:52

\bibitem[Beasley et al.(2016)]{Be16}
Beasley MA, Romanowsky AJ, Pota V, et al (2016) ApJL 819:L20 

\bibitem[Bell et al.(2003)]{Bel3} 
Bell EF, McIntosh DH, Katz N, and Weinberg MD (2003) ApJS 149:289 

\bibitem[Bell and de Jong(2001)]{bdj}
Bell E  and de Jong RS (2001) ApJ, 550:212
 
\bibitem[Bellazzini et al.(2013)]{Bz}
Bellazzini B, Cliche M and Tanedo P (2013) PRD 88:083506

\bibitem[Bernal et al.(2017) ]{Be+17}
Bernal N, Heikinheimo, N Tenkanen T et al, (2017) IJMPA 32:27

\bibitem[Bernardi et al.(2003)]{b03} 
Bernardi M, Sheth RK, Annis J et al (2003) AJ 125:1866 

\bibitem[Bershady et al.(2010a)]{dms1} 
Bershady MA, Verheijen, MAW, Westfall, KB et al (2010a) ApJ 716:234 

\bibitem[Bershady et al.(2010b)]{dms2}
Bershady MA, Verheijen MAW, Swaters RA et al (2010b) ApJ 716:198 

\bibitem[Bertone(2010)]{BerB}
Bertone G (ed) (2010) Particle Dark Matter: Observations, Models and Searches. Cambridge, CUP

\bibitem[Bertone and Hooper(2016)]{BH} 
Bertone G and Hooper D (2016) ArXiv e-prints arXiv:1605.04909
 
\bibitem[Binney and Tremaine(2008)]{BT} 
Binney J and Tremaine S (2008) Galactic Dynamics, Princeton, PUP
 
\bibitem[Bloom et al.(2017)]{bel}
Bloom JV et al MNRAS (2017) 472:1809

\bibitem[Boddy et al.(2014)]{Bo}
Boddy KK, Feng JL, Manoj Kaplinghat M et al (2014) PRD 89:115017
 
\bibitem[Bode et al.(2004)]{bd}
Bode P, Ostriker JP, Turok N (2001) ApJ 556:93

\bibitem[Bottema and Pesta{\~n}a(2015)]{bp15}
Bottema R and Pesta{\~n}a JLG (2015) MNRAS 448:2566 

\bibitem[Boyarsky etal.(2007)]{Boy07} 
Boyarsky A, Nevalainen J, Ruchayskiy O (2007) A\&A 471:51 

\bibitem[Brook et al.(2016)]{BSS} 
Brook CB, Santos-Santos I, and Stinson, G (2016) MNRAS 459:638 

\bibitem[Bolton et al.(2006)]{bbt}
Bolton AS, Burles S, Koopmans LVE et al (2006) ApJ 638:703

\bibitem[Bolton et al.(2008)]{bo08}
Bolton AS et al (2008) ApJ 684:248

\bibitem [Bolton et al.(2007)]{bbt07}
Bolton AS, Burles S, Treu T et al (2007) ApJ 665:105

\bibitem[Bonnivard et al.(2015)]{Bet15}
Bonnivard V et al (2015) MNRAS, 453:849

\bibitem[Bosma(1981a)]{b1}
Bosma A (1981a) AJ 86:1791

\bibitem[Bosma(1981b)]{b2}
Bosma A (1981b) AJ 86:1825

\bibitem[Bothun et al.(1991)]{Bo91}
Bothun GD, Impey CD and Malin DF (1991) ApJ, 376:404 

\bibitem[Breddels et al.(2013)]{br13} 
Breddels MA, Helmi A, van den Bosch RCE et al (2013) MNRAS 433:3173 
 
\bibitem[Bringmann et al.(2016)]{Br}
Bringmann T et al (2016) PRD 94:103529

\bibitem[Brown et al.(2009)]{bet} 
Brown WR, Geller MJ, Kenyon SJ and Diaferio A (2009) ApJ 690:1639

\bibitem[Bruzual and Charlot(2003)]{BC03} 
Bruzual G and Charlot S (2003) MNRAS 344:1000 

\bibitem[Bullock and Boylan-Kolchin(2017)]{BB}
Bullock JS and Boylan-Kolchin M (2017) ARAA 55:343

\bibitem[Burkert(1995)]{bu95}
Burkert A (1995) ApJL 447, L25

\bibitem[Burkert(2015)]{bu15}
Burkert A (2015) ApJ 808:158

\bibitem[Butler(2018)]{But}
Butler J (2018) PoS(ALPS2018)030 

\bibitem[Caldwell and Ostriker(1981)]{cao81} 
Caldwell JAR and Ostriker JP (1981) ApJ 251:61 

\bibitem[Campbell et al.(2017)]{cel17}
Campbell et al (2017) MNRAS 469:2335

\bibitem[Cappellari et al.(2011)]{ca11} 
Cappellari M, Emsellem E, Krajnovi{\'c} D et al (2011) MNRAS 413:813 
 
\bibitem[Cappellari(2012)]{c+12}
Cappellari M et al (2012) Nature 484:485

\bibitem[Cappellari et al.(2015)]{ca15} 
Cappellari M, Romanowsky AJ, Brodie, JP et al (2015) ApJL 804:L21 

\bibitem[Cappellari(2016)]{ca16}
Cappellari M (2016) ARAA 54 597

\bibitem [Cappellari et al.(2013)]{cap13}
Cappellari, M et al (2013) MNRAS 432:1709

\bibitem [Cappellari et al.(2006)]{c06+} 
Cappellari, M et al (2006) MNRAS 366:1126 
 
\bibitem[Carignan and Freeman(1985)]{cf85} 
Carignan, C, and Freeman, KC (1985) ApJ 294:494 
 
\bibitem[Catena and Ullio(2010)]{cu10} 
Catena R, and Ullio P (2010) JCAP 08(2010):004 

\bibitem[Catena and Ullio(2012)]{cu12}
Catena R and Ullio P (2012) JCAP 05(2012):005 
 
\bibitem[Catinella et al.(2006)]{catel06}
Catinella B, Giovanelli R and Haynes MP (2006) ApJ 640:751

\bibitem[Chae(2014)]{ch14} 
Chae K-H (2014) ApJL 788:L15 
 
\bibitem[Coccato et al.(2009)]{cga}
Coccato L, Gerhard O, Arnaboldi M et al MNRAS (2009) 394:1249

\bibitem[Corbelli and Salucci(2000)]{cs} 
Corbelli E and Salucci P (2000) MNRAS 311:441
 
\bibitem[Corsini et al.(2017)]{coel17}
Corsini EM, Wegner GA, Thomas J et al (2017) MNRAS 466:974

\bibitem[Courteau(1997)]{co97}
Courteau S (1997) AJ 114:2402

\bibitem[Cretton et al.(1999)]{cr99} 
Cretton N, de Zeeuw PT, van der Marel RP and Rix H-W (1999) ApJS 124:383 
 
\bibitem[Deason et al.(2012)]{dea12} 
Deason AJ, Belokurov V, Evans NW, An J (2012) MNRAS 424:L44 

\bibitem[de Blok et al.(2001)]{deb01} 
de Blok WJG, McGaugh SS and Rubin VC (2001) AJ 122:2396 

\bibitem[de Blok et al.(2008)]{db08} 
de Blok WJG, Walter F, Brinks E et al (2008) AJ 136:2648

\bibitem[de Blok(2010)]{dB10} 
de Blok WJG (2010) Adv Astron 2010:789293 

\bibitem[De Masi et al.(2018)]{MATT1}
De Masi C, Matteucci F, and Vincenzo F (2018) MNRAS 474:5259 

\bibitem[de Zeeuw et al.(2002)]{dz02} 
de Zeeuw PT, Bureau M, Emsellem E et al (2002) MNRAS 329:513 

\bibitem[Destri et al.(2013)]{dds}
Destri C, de Vega P, Sanchez NG (2013) PRD 88:3512

\bibitem[de Swart et al.(2017)]{desw}
de Swart J, Bertone G, van Dongen J (2017) Nature Astron 1:005
 
\bibitem[de Vega and Sanchez(2017)]{dvs} 
de Vega HJ and Sanchez NG (2017) EPJC 77:1 

\bibitem[Di Cintio et al.(2014)]{dc14} 
Di Cintio A, Brook CB, Dutton AA et al (2014) MNRAS 441:2986

\bibitem[Di Paolo and Salucci(2018)]{dps}
Di Paolo C and Salucci P (2018) ArXiv e-print arXiv:1805.07165 

\bibitem[Di Paolo et al.(2018)]{dnv} 
Di Paolo C, Nesti F and Villante FL (2018) MNRAS 475:5385 

\bibitem[Djorgovski and Davis(1987)]{dd}
Djorgovski S and Davis M (1987) ApJ 313:59

\bibitem[Dodelson and Widrow(1994)]{DW}
Dodelson S and Widrow LM (1994) PRL 72:17

\bibitem[Donato et al.(2009)]{Doet}
Donato F, Gentile G, Salucci P et al (2009) MNRAS 397:1169 

\bibitem[Donato et al.(2004)]{do}
Donato F, Gentile G, Salucci P (2004) MNRAS 353:17

\bibitem [Dressler et al.(1987)]{d+} 
Dressler A, Lynden-Bell D, Burstein D et al (1987) ApJ 313:42 
 
\bibitem [Ellis et al.(2018)]{e18}
Ellis G et al (2018) Found Phys 48:1226

\bibitem[Ettori and Fabian(2006)]{ef}
Ettori S and Fabian AC (2006) MNRAS 369:L42

\bibitem[Evoli et al.(2011)]{e11}
Evoli C, Salucci P, Lapi A, Danese L (2011) ApJ 743:45

\bibitem[Faber and Gallagher(1979)]{FG}
Faber SM and Gallagher JS (1979) ARAA 17:135 
 
\bibitem[Fabricant et al.(1984)]{frg}
Fabricant D, Rybicki G and Gorenstein P (1984) ApJ 286:186

\bibitem[Freeman(1970)]{fre}
Freeman KC (1970) ApJ 160:811

\bibitem[Freese(2017)]{Free} 
Freese K (2017) IJMPD 26:1730012

\bibitem[Gammaldi(2016)]{ga16} 
Gammaldi V (2016) EPJ Web Conf 121:06003 

\bibitem[Gammaldi(2015)]{ga15} 
Gammaldi V (2015) PhD Thesis. UCM Madrid

\bibitem[Garc{\'{\i}}a-Bellido(2017)]{gb} 
Garc{\'{\i}}a-Bellido (2017) J Phys Conf Ser 840:012032 

\bibitem[Gentile et al.(2004)]{ge04}
Gentile G, Salucci P, Klein U, Vergani D and Kalberla P (2004) MNRAS 351:903

\bibitem[Gentile et al.(2005)]{ge05} 
Gentile G, Burkert A, Salucci P et al (2005) ApJ 634:145	  

\bibitem[Genzel et al.(2017)]{GEL} 
Genzel R, Schreiber NMF, {\"U}bler H et al (2017) Nature 543:397 

\bibitem[Graves and Faber(2010)]{gf10} 
Graves GJ and Faber SM (2010) ApJ 717:803 

\bibitem[Gratier et al.(2010)]{G+10} 
Gratier P, Braine J, Rodriguez-Fernandez NJ et al (2010) A\&A 522:A3 

\bibitem[Green(2016)]{grn}
Green AM (2016) PRD 94:063530 

\bibitem[Gondolo(2002)]{go02} 
Gondolo P (2002) PRD 66:103513 

\bibitem[Grillo et al.(2009)]{gg09}
Grillo C, Gobat R, Lombardi M, and Rosati P (2009) A\&A 501:461

\bibitem[Gurovich et al.(2004)]{gel04}
Gurovich S, McGaugh SS, Freeman KC, et al (2004) PASA 21:412 
 
\bibitem[Hessman(2017)]{H17}
Hessman FV (2017) MNRAS 469:1147
 
\bibitem[Hyde and Bernardi(2009)]{HB}
Hyde JB and Bernardi M (2009) MNRAS 396:1171

\bibitem[Hoekstra and Jain(2008)]{hj08}
Hoekstra H and Jain B (2008) Annu Rev Nucl Part Sci 58:99
 
\bibitem[Hui et al.(2016)]{HO}
Hui L, Ostriker JP, S Tremaine and E Witten (2017) PRD 95:043541

\bibitem[Hudson et al.(2015)]{hel} 
Hudson MJ, Gillis BR, Coupon J et al (2015) MNRAS 447:298

\bibitem[Kang et al.(2018)]{Kanel} 
Kang S, Scopel S, Tomar G, and Yoon J.-H (2018) ArXiv e-print arXiv:1805.06113 

\bibitem[Karukes and Salucci(2017)]{ks17}
Karukes EV and Salucci P (2017) MNRAS 465:4703
 
\bibitem[Karukes et al.(2015)]{ks15} 
Karukes EV, Salucci P, and Gentile G (2015) A\&A 578:A13 
 
\bibitem[Kaplinghat et al.(2015)]{ka} 
Kaplinghat M, Linden T, and Yu H-B (2015) PRL 114:211303 
 
\bibitem[Kennedy et al.(2014)]{ke+14} 
Kennedy R, Frenk C, Cole S, and Benson A (2014) MNRAS 442:2487 

\bibitem[Klypin et al.(2010)]{k}
Klypin A, Trujillo-Gomez S, and Primack J (2011) ApJ, 740:102 

\bibitem[Kolb and Turner(1990)]{KT}
Kolb, EW and Turner, MS (1990) The Early Universe, Addison Wesley, New York

\bibitem[Kormendy and Freeman(2004)]{KF04}
Kormendy J and Freeman KC (2004) In: Ryder SD et al (eds) Dark Matter
in Galaxies (IAU S220), San Franciso, ASP, p 377 

\bibitem[Korsaga et al.(2018)]{kal18} 
Korsaga M, Carignan C, Amram P et al (2018) MNRAS 478:50 

\bibitem[Koushiappas and Loeb(2017)]{kl17} 
Koushiappas SM and Loeb A (2017) PRL 119:041102 

\bibitem[Kregel et al.(2002)]{kel}
Kregel M, van der Kruit PC, de Grijs R (2002) MNRAS 334:646

\bibitem[Kusenko(2009)]{Ku}
Kusenko A (2009) Phys Rep 481:1 

\bibitem[Kuzio de Naray et al.(2008)]{kdn} 
Kuzio de Naray R, McGaugh SS and de Blok, WJG (2008) ApJ 676:920
 
\bibitem[Juric et al.(2008)]{ju08} 
Juri{\'c} M and Ivezi{\'c} {\v Z}, Brooks A et al (2008) ApJ 673,864 

\bibitem [Jungman and al.(1996)]{jel96}
Jungman G, Kamionkowski, M Griest K (1996) Phys Rep 267:195

\bibitem[Jorgensen et al.(1996) ]{Joel}
Jorgensen I, Franx M, and Kjaergaard P (1996) MNRAS 280:167

\bibitem[Impey et al.(1988)]{I88} 
Impey C, Bothun G, and Malin D (1988) ApJ 330:634 

\bibitem[Lapi et al.(2018)]{lsd} 
Lapi A, Salucci P and Danese L.(2018) ApJ 859:2
 
\bibitem[Lelli et al.(2016a)]{LMS}
Lelli F, McGaugh SS and Schombert JM (2016a) ApJL 816:L14 

\bibitem[Lelli et al.(2016b)]{lms16}
Lelli F, McGaugh SS and Schombert JM (2016b) AJ 152:157 
 
\bibitem[Li et al.(2017)]{lsrd17}
Li B, Shapiro PR, Rindler-Daller T (2017) PRD 96:063505 

\bibitem[Lisanti(2017)]{L17}
Lisanti M (2017) In: Polchinski J and Vieira P and DeWolfe, O (eds) New Frontiers in Fields and Strings. World Scientific, Singapore, p 399--446
 
\bibitem[Magoulas et al.(2012)]{mag12}
Magoulas C, Springob CM, Colless M et al (2012) MNRAS 427:245 

\bibitem[Maraston(2013)]{mar}
Maraston C (2013) In: Thomas D,  Pasquali A, Ferreras I (eds) The Intriguing Life of Massive Galaxies (IAU S295) Cambridge, CUP, p 272 

\bibitem[Mamon and Lokas(2005)]{ML}
Mamon, G and Lokas EL (2005) MNRAS 363:705

\bibitem[Martinsson et al.(2013)]{Mel13} 
Martinsson T, Verheijen M, Westfall K et al (2013) A\&A 557:131 

\bibitem[Matteucci(2012)]{MATT2}
Matteucci F (2012) Chemical Evolution of Galaxies. Berlin, Heidelberg, Springer 

\bibitem[McGaugh et al.(2000)]{mc00} 
McGaugh SS, Schombert JM, Bothun GD, de Blok WJG (2000) ApJL 533:L99 

\bibitem[McGaugh(2005)]{mgc1}
McGaugh SS, (2005) ApJ 632:859 
 
\bibitem[McMillan(2011)]{mm11}
McMillan PJ (2011) MNRAS 414:2446

\bibitem[Moster et al.(2010)]{mos} 
Moster BP, Somerville, RS, Maulbetsch C et al (2010) ApJ 710:903

\bibitem[Muller et al.(2018)]{mel}
M{\"u}ller O, Pawlowski MS, Jerjen T et al (2018) Science 359:534	

\bibitem[Munshi et al.(2008)]{mv}
Munshi D, Valageas P, van Waerbeke L and Heavens A (2008) Phys Rep 462:67
 
\bibitem[Naab and Ostriker(2017)]{NO}
Naab T and Ostriker JP ARAA (2017) 55:59

\bibitem[Navarro et al.(1997)]{nfw}
Navarro JF, Frenk CS and White SDM (1997) ApJ 490:493

\bibitem[Nesti and Salucci(2013)]{ns13} 
Nesti F and Salucci P (2013) JCAP 7,16

\bibitem[Noordermeer et al.(2007)]{no07} 
Noordermeer E, van der Hulst JM, Sancisi R et al (2007) MNRAS 376:1513 

\bibitem[Oguri et al.(2014)]{og14}
Oguri et al (2014) MNRAS 439 2494

\bibitem[Oh et al.(2008)]{oh}
Oh S-H et al (2008) AJ 136:2761
 
\bibitem[Oh et al.(2011)]{oh11}
Oh S-H, Brook C, Governato F et al (2011) AJ 142:24 

\bibitem[Oh et al.(2015)]{oel15}
Oh S-H, Hunter DA, Brinks E et al (2015) AJ 149:180 

\bibitem[Oman et al.(2015)]{O15}
Oman KA, Navarro JF, Fattahi A et al (2015) MNRAS 452:3650 
 
\bibitem[Honma et al.(2012)]{ho12} 
Honma, M, Nagayama, T, Ando, K, et al (2012) PASJ 64:136 
 
\bibitem[Palunas and Williams(2000)]{pw00} 
Palunas, P and Williams, TB (2000) AJ 120:2884 

\bibitem[Pascale et al.(2018)]{ppn}
Pascale R, Posti L, Nipoti C, Binney J (2018) MNRAS 480:927

\bibitem[Pato and Iocco(2017)]{pi17} 
Pato M and Iocco F (2017) SoftwareX 6:54

\bibitem[Persic and Salucci(1995)]{PS95} 
Persic M and Salucci P (1995) ApJS, 99:501 
	
\bibitem[Persic, Salucci and Stel(1996)]{PSS}
Persic M, Salucci P and Stel F (1996) MNRAS 281:27

\bibitem[Persic and Salucci(1990)]{PS90}
Persic M and Salucci P (1990) MNRAS 245:577 

\bibitem[Persic and Salucci(1991)]{P91}	
Persic M and Salucci P (1991) ApJ 368:60 	

\bibitem[Planck Collaboration et al.(2016)]{P16} 
Planck Collaboration, Ade PAR, Aghanim N et al (2016) A\&A 594:A13 

\bibitem[Plummer(1915)]{pl}
Plummer HC (1915) MNRAS 76:107

\bibitem[Poci et al.(2017)]{pcd17}
Poci A, Cappellari M, McDermid RM (2017) MNRAS 467:1397

\bibitem[Ponomareva et al.(2018)]{pon} 
Ponomareva AA, Verheijen MAW, Papastergis E et al (2018) MNRAS 474:4366

\bibitem[Posacki et al.(2015)]{po15} 
Posacki S, Cappellari M, Treu T et al (2015) MNRAS 446:493

\bibitem[Pulsoni et al.(2017)]{P18} 
Pulsoni C, Gerhard O, Arnaboldi M et al (2017) A\&A 618:A94

\bibitem[Ratnam and Salucci(2000)]{ra} 
Ratnam C and Salucci P (2000) NewA 5:427 

\bibitem[Richards et al.(2015)]{Rich15} 
Richards EE, van Zee L, Barnes KL et al (2015) MNRAS 449:3981 
 
\bibitem[Ringwald(2012)]{Ri}
Ringwald A (2012) Phys Dark Univ 1:116

\bibitem[Roberts(1978)]{R78}
Roberts MS (1978) AJ 83:1026

\bibitem[Roszkowski et al.(2017)]{RST}
Roszkowski L, Sessolo EM, Trojanowski S (2017) Rep Prog Phys 81:066201

\bibitem[Rubin et al.(1980)]{rel80}
Rubin VC, Ford WK Jr and Thonnard N (1980) ApJ 238:471 

\bibitem[Salucci(2001)]{s01} 
Salucci P (2001) MNRAS 320, L1 

\bibitem[Salucci et al(2010)]{sng}	
Salucci P, Nesti F, Gentile G, Frigerio Martins C (2010) A\&A 523:83

\bibitem[Salucci et al.(1993)]{SFP}
Salucci P, Frenk CS, Persic M (1993) MNRAS 262:392

\bibitem[Salucci and Burkert(2000)]{sb00} 
Salucci P and Burkert A (2000) ApJL 537:L9 

\bibitem[Salucci et al.(2007)]{s7}
Salucci P, Lapi A, Tonini C, Gentile G, Yegorova I, Klein U (2007) MNRAS 378:41

\bibitem[Salucci et al.(2008)]{sya} 
Salucci P, Yegorova IA, and Drory N (2008) MNRAS 388:159 

\bibitem[Salucci et al.(2012)]{s12} 
Salucci P, Wilkinson MI, Walker MG et al (2012) MNRAS 420:2034 

\bibitem[Salucci and Turini(2017)]{st17}
Salucci P and Turini N (2017) ArXiv e-print arXiv:1707.01059 
 
\bibitem[Schneider(1996)]{sch} 
Schneider P (1996) MNRAS 283:837
 
\bibitem[Serra et al.(2016)]{Soc16} 
Serra P, Oosterloo T, Cappellari M, den Heijer M, Jozsa GIG (2016) MNRAS 460:1382
 
\bibitem[Shankar et al.(2006)]{sh06}
Shankar F, Lapi A, Salucci P (2006) ApJ 643:14

\bibitem[Shi and Fuller(1999)]{SeF}
Shi X and Fuller GM (1999) PRL 82:2832 
 
\bibitem[Shi et al.(2017)]{din}
Shi D et al (2017) ApJ 846:26

\bibitem[Simon(2005)]{si05} 
Simon JD (2005) PhD Thesis

\bibitem[Sofue(2017)]{sof17} 
Sofue Y (2017) PASJ 69:R1 

\bibitem[Sofue(2013)]{so13} 
Sofue Y (2013) PASJ 65:118 

\bibitem[Somerville and Dave(2015)]{SD} 
Somerville RS and Dave R (2015) ARAA 53:51

\bibitem[Spano et al.(2008)]{sma} 
Spano M, Marcelin M, Amram P et al (2008) MNRAS 383:297 

\bibitem[Spekkens et al.(2005)]{sp05} 
Spekkens K, Giovanelli R, and Haynes MP (2005) AJ 129:2119 

\bibitem[Spergel and Steinhardt(2000)]{spst} 
Spergel DN and Steinhardt PJ (2000) PRL 84:3760

\bibitem[Steigman and Turner(1985)]{ST} 
Steigman S and Turner MS (1985) Nucl Phys B 253:375

\bibitem[Strauss and Willick(1995)]{sw} 
Strauss MJ, Willick JA (1995) Phys Rept 261:271

\bibitem[Strigari et al.(2008)]{Sel08} 
Strigari LE, Bullock JS, Kaplinghat M et al (2008) Nature 454:1096

\bibitem[Strigari et al.(2018)]{Sel18}
Strigari LE, Frenk CS, White SDM (2018) ApJ 860:56

\bibitem[Thomas et al.(2011)]{tsb} 
Thomas J, Saglia RP, Bender R et al (2011) MNRAS 415:545 

\bibitem[Tinsley(1981)]{tin}
Tinsley BM (1981) MNRAS 194:63 

\bibitem[Tiret et al.(2011)]{TS10}
Tiret O, Salucci P, Bernardi M, Maraston C, Pforr J (2011) MNRAS 411:1435

\bibitem[Toloba et al.(2018)]{tl18} 
Toloba E, Lim S, Peng E et al (2018) ApJL 856:L31 

\bibitem[Tortora et al.(2014)]{T+14} 
Tortora C, La Barbera F, Napolitano NR et al (2014) MNRAS 445:115 

\bibitem[Tortora et al.(2018)]{Tel17}
Tortora C, Napolitano NR, Roy N et al (2018) MNRAS 473:969 

\bibitem [Treu(2010)]{T}
Treu T (2010) ARAA 48:87

\bibitem[Tulin et al.(2013)]{Tu}
Tulin S, Yu H, Zurek KM (2013) PRD 87:115007

\bibitem[Tulin and Yu(2017)]{TY}
Tulin S, Yu H (2018) Phys Rep 730:1--57

\bibitem[Tully and Fisher(1977)]{TF} 
Tully RB and Fisher JR (1977) A\&A 54:661 

\bibitem[Turner(2018)]{T18}
Turner MS (2018) Found Phys 48:1261

\bibitem[van Albada et al.(1985)]{vel85} 
van Albada TS, Bahcall JN, Begeman K et al (1985) ApJ 295:305 

\bibitem[van der Kruit and Searle(1981)]{vs81} 
van der Kruit PC and Searle L (1981) A\&A 95:105 

\bibitem[van der Kruit(1988)]{vk88} 
van der Kruit PC (1988) A\&A 192:117 

\bibitem[van der Kruit and Freeman(2011)]{vkf}
van der Kruit PC and Freeman KC (2011) ARAA 49:301--371 
 
\bibitem[van Dokkum et al.(2015)]{vad15} 
van Dokkum PG, Romanowsky AJ, Abraham R et al (2015) ApJL 804:L26 

\bibitem[Verheijen(2001)]{v01} 
Verheijen MAW (2001) ApJ 563:694 

\bibitem[Viel et al.(2005)]{v05} 
Viel M, Branchini E, Cen R et al (2005) MNRAS 360:1110 

\bibitem[Vogelsberger et al.(2014)]{vog14b} 
Vogelsberger M, Genel S, Springel V et al (2014) Nature 509:177

\bibitem[Vogt et al.(2004a)]{vel04} 
Vogt NP, Haynes MP, Herter T, Giovanelli R (2004a) AJ 127:3273

\bibitem[Vogt et al.(2004b)]{vel04b} 
Vogt NP, Haynes MP, Herter T, Giovanelli R (2004b) AJ 127:3325

\bibitem[Walker et al.(2009a)]{Waj}
Walker MG, Mateo M, Olszewski EW (2009a) AJ 137:3100

\bibitem[Walker et al.(2009b)]{W09}
Walker MG, Mateo, M, Olszewski EW et al (2009b) ApJ 704:1274

\bibitem[Walker(2013)]{W13}
Walker M (2013). In: Oswalt TD and Gilmore G (eds) Planets, Stars and Stellar Systems 5. Springer, Dordrecht, p 1039--1089

\bibitem[Walker and Penarrubia(2011)]{WP11}
Walker MG and Penarrubia J (2011) ApJ 742:20 

\bibitem[Wang et al.(2014)]{wael14}
Wang J, Fu J, Aumer M et al (2014) MNRAS 441:2159 

\bibitem[Watkins et al.(2010)]{wa10}
Watkins LL, Evans NW, and An JH (2010) MNRAS 406:264 

\bibitem[Wechsler et al.(2006)]{w06}
Wechsler RH, Zentner AR, Bullock JS et al (2006) ApJ 652:71

\bibitem[Wechsler and Tinker(2018)]{WT} 
Wechsler RH and Tinker JL (2018) ARAA 56:435 

\bibitem[Weinberg(1977)]{W}
Weinberg S (1978) PRL 40:223

\bibitem[Wolf et al.(2010)]{wo10} 
Wolf J, Martinez GD, Bullock JS et al (2010) MNRAS 406:1220

\bibitem[Xue et al.(2008)]{Xet}
Xue XX et al (2008) ApJ 684:1143.

\bibitem[Yegorova and Salucci(2007)]{ys}
Yegorova IA and Salucci P (2007) MNRAS 377:507

\bibitem[Zaritsky(2012)]{zar12} 
Zaritsky D (2012) ISRN Astron Astrophys 2012:189625 

\bibitem[Zavala et al.(2013)]{za}
Zavala J, Vogelsberger M and Walker MG (2013) MNRAS 431, L20 

\bibitem[Zhao(1996)]{Zha} 
Zhao H (1996) MNRAS 278:488 

\bibitem[Zu and Mandelbaum(2015)]{ZM} 
Zu Y and Mandelbaum R (2015) MNRAS 454:1161 


 \end{thebibliography}
\end{document}